\documentclass{IEEEtran} 

\usepackage{graphicx}
\usepackage{tabu}

\usepackage{cuted, ragged2e}

\usepackage{tcolorbox}
\tcbuselibrary{most}

\usepackage{array}
\usepackage{hyperref}
\usepackage{stfloats}
\usepackage{amsmath,amssymb,amsfonts}
\usepackage{bbding} 
\usepackage{textcomp}
\usepackage{xcolor}
\usepackage{booktabs}
\usepackage{multirow}
\usepackage{xspace}
\usepackage{framed}
\usepackage{tablefootnote}
\usepackage[ruled,linesnumbered]{algorithm2e}
\usepackage{subcaption}
\usepackage{bbm}

\usepackage{bm}

\def\eqref#1{equation~(\ref{#1})}

\def\1{\bm{1}}

\def\vx{{\bm{x}}}

\DeclareMathAlphabet{\mathsfit}{\encodingdefault}{\sfdefault}{m}{sl}
\SetMathAlphabet{\mathsfit}{bold}{\encodingdefault}{\sfdefault}{bx}{n}

\DeclareMathOperator*{\argmax}{arg\,max}

\DeclareMathOperator{\sign}{sign}

\hypersetup{hidelinks,
	colorlinks=true,
	allcolors=black,
	pdfstartview=Fit,
	breaklinks=true}

\newcommand{\blue}[1]{\textcolor{blue}{#1}}
\newcommand{\red}[1]{\textcolor{red}{#1}}

\definecolor{xg}{RGB}{189,230,219}

\newcommand{\redbox}[1]{\colorbox{red!20}{#1}}
\newcommand{\greenbox}[1]{\colorbox{green!20}{#1}}

\newcommand{\tool}[1]{\textsc{#1}\xspace}

\newcommand{\dje}{\tool{DeepJudge}}
\newcommand{\deepjudge}{\tool{DeepJudge}}

\newcommand{\norm}[1]{\left\lVert#1\right\rVert}

\newcommand{\yes}[0]{\footnotesize \Checkmark }
\newcommand{\no}[0]{\scriptsize \XSolidBrush}

\newcommand{\commentout}[1]{}
\usepackage{soul}

\begin{document}

\title{Copy, Right? A Testing Framework for Copyright Protection of Deep Learning Models}

\author{
    \IEEEauthorblockN{Jialuo Chen\IEEEauthorrefmark{1}, Jingyi Wang\IEEEauthorrefmark{1}, Tinglan Peng\IEEEauthorrefmark{1}, 
    Youcheng Sun\IEEEauthorrefmark{2}, Peng Cheng\IEEEauthorrefmark{1}, Shouling Ji\IEEEauthorrefmark{1}, \\ 
    Xingjun Ma\IEEEauthorrefmark{3}, Bo Li\IEEEauthorrefmark{4} and Dawn Song\IEEEauthorrefmark{5}
    }

    \vspace{0.8mm}
    \small
    \IEEEauthorblockA{\IEEEauthorrefmark{1}Zhejiang University, \IEEEauthorrefmark{2}University of Manchester, \IEEEauthorrefmark{3}Deakin University, \IEEEauthorrefmark{4}UIUC, \IEEEauthorrefmark{5}UC Berkeley}
    
    \vspace{0.4mm}
    \IEEEauthorblockA{\IEEEauthorrefmark{1}\{chenjialuo, wangjyee, tlpeng\_zju, lunarheart, sji@zju.edu.cn\}, \IEEEauthorrefmark{2}\{youcheng.sun@manchester.ac.uk\}, \\ 
    \IEEEauthorrefmark{3}\{danxjma@gmail.com\}, \IEEEauthorrefmark{4}\{lbo@illinois.edu\}, \IEEEauthorrefmark{5}\{dawnsong@berkeley.edu\}}
    
}

\maketitle

\begin{abstract} 
    Deep learning models, especially those large-scale and high-performance ones, can be very costly to train, demanding a considerable amount of data and computational resources. As a result, deep learning models have become one of the most valuable assets in modern artificial intelligence. Unauthorized duplication or reproduction of deep learning models can lead to copyright infringement and cause huge economic losses to model owners, calling for effective copyright protection techniques. Existing protection techniques are mostly based on watermarking, which embeds an owner-specified watermark into the model. While being able to provide exact ownership verification, these techniques are 1) invasive, i.e., they need to tamper with the training process, which may affect the model utility or introduce new security risks into the model; 2) prone to adaptive attacks that attempt to remove/replace the watermark or adversarially block the retrieval of the watermark; and 3) not robust to the emerging model extraction attacks. Latest fingerprinting work on deep learning models, though being non-invasive, also falls short when facing the diverse and ever-growing attack scenarios.

\let\thefootnote\relax\footnotetext{This paper is to appear in the IEEE Security \& Privacy, May 2022.}

   In this paper, we propose a novel \emph{testing} framework for deep learning copyright protection: \dje. \dje quantitatively tests the similarities between two deep learning models: a victim model and a suspect model. It leverages a diverse set of testing metrics and efficient test case generation algorithms to produce a chain of supporting evidence to help determine \emph{whether a suspect model is a copy of the victim model}. Advantages of \dje include: 1) \emph{non-invasive}, as it works directly on the model and does not tamper with the training process; 2) \emph{efficient}, as it only needs a small set of seed test cases and a quick scan of the two models; 3) \emph{flexible}, i.e., it can easily incorporate new testing metrics or test case generation methods to obtain more confident and robust judgement; and 4) \emph{fairly robust to model extraction attacks and adaptive attacks}. We verify the effectiveness of \dje under three typical copyright infringement scenarios, including model finetuning, pruning and extraction, via extensive experiments on both image classification and speech recognition datasets with a variety of model architectures.
   
\end{abstract}

\section{Introduction}
\label{sec:introduction}
Deep learning models, e.g., deep neural networks (DNNs), have become the standard models for solving many complex real-world problems, such as image recognition \cite{he2016deep}, speech recognition \cite{graves2013speech}, natural language processing \cite{collobert2011natural}, and autonomous driving \cite{chen2015deepdriving}. However, training large-scale DNN models is by no means trivial, which requires not only large-scale datasets but also significant computational resources. The training cost can grow rapidly with task complexity and model capacity. For instance, it can cost \$1.6 million to train a BERT model on Wikipedia and Book corpora (15 GB) \cite{sharir2020cost}.
It is thus of utmost importance to protect DNNs from unauthorized duplication or reproduction.

One concerning fact is that well-trained DNNs are often exposed to the public via remote services (APIs), cloud platforms (e.g., Amazon AWS, Google Cloud and Microsoft Azure), or open-source toolkits like OpenVINO\footnote{\href{https://github.com/openvinotoolkit/open_model_zoo}{https://github.com/openvinotoolkit/open\_model\_zoo}}. It gives rise to adversaries (e.g., a model ``thief'') who attempt to steal the model in stealthy ways, causing copyright infringement and economic losses to the model owners.
Recent studies have shown that stealing a DNN can be done very efficiently without leaving obvious traces \cite{tramer2016stealing,papernot2017practical}. Arguably, unauthorized finetuning or pruning is the most straightforward way of model stealing, if the model parameters are publicly accessible (for research purposes only) or the adversary is an insider. Even when only the API is exposed, the adversary can still exploit advanced \emph{model extraction} techniques \cite{tramer2016stealing,papernot2017practical,orekondy2019knockoff,juuti2019prada,yuan2020attack} to steal most functionalities of the hidden model. These attacks pose serious threats to the copyright of deep learning models, calling for effective protection methods.

A number of defense techniques have been proposed to protect the copyright of DNNs, where DNN watermarking \cite{uchida2017embedding,zhang2018protecting,adi2018turning,darvish2019deepsigns} is one major type of technique. DNN watermarking embeds a secret watermark (e.g., logo or signature) into the model by exploiting the over-parameterization property of DNNs \cite{adi2018turning}. The ownership can then be verified when the same or similar watermark is extracted from a suspect model. The use of watermarks has an obvious advantage, i.e., the owner identity can be embedded and verified exactly, given that the watermark can be fully extracted. However, these methods still suffer from certain weaknesses. Arguably, the most concerning one is that they are \emph{invasive}, i.e., they need to tamper with the training procedure to embed the watermark, which may compromise model utility or introduce new security threats into the model \cite{liu2018fine,wang2019neural,fan2019rethinking,guo2020hidden}.

More recently, DNN fingerprinting \cite{cao2021ipguard,DBLP:conf/iclr/LukasZK21} has been proposed as a non-invasive alternative to watermarking. Lying at the design core of fingerprinting is \emph{uniqueness} --- the unique feature of a DNN model. Specifically, fingerprinting extracts a unique identifier (or fingerprint) from the owner model to differentiate it from other models. The ownership can be claimed if the identifier of the owner model matches with that of a suspect model. However, in the context of deep learning, a single fingerprinting feature/metric can hardly be sufficient or flexible enough to handle all the randomness in DNNs or against different types of model stealing and adaptive attacks (as we will show in our experiments). In other words, there exist many scenarios where a DNN model can easily lose its unique feature or property (i.e., fingerprint).



In this work, we propose a \emph{testing} approach for DNN copyright protection.
Instead of solely relying on one metric, we propose to actively test the ``similarities'' between a victim model and a suspect model from multiple angles. The core idea is to \textbf{1)} carefully construct a set of test cases to comprehensively characterize the victim model, and \textbf{2)} measure how similarly the two models behave on the test cases. Intuitively, if a suspect model is a stolen copy of the victim model, it will behave just like the victim model in certain ways. An extreme case is that the suspect is the exact duplicate of the victim model, and in this case, the two models will behave identically on these test cases. This testing view creates a dilemma for the adversary as better stealing will inevitably lead to higher similarities to the victim model.  We further identify two major challenges for testing-based copyright protection: 1) how to define comprehensive testing metrics to fully characterize the similarities between two models, and 2) how to effectively generate test cases to amplify the similarities. The set of similarity scores can be viewed as a proof obligation that provides a chain of strong evidence to judge a stolen copy.



\begin{figure*}[t]
\centering
\includegraphics[width=0.84\linewidth]{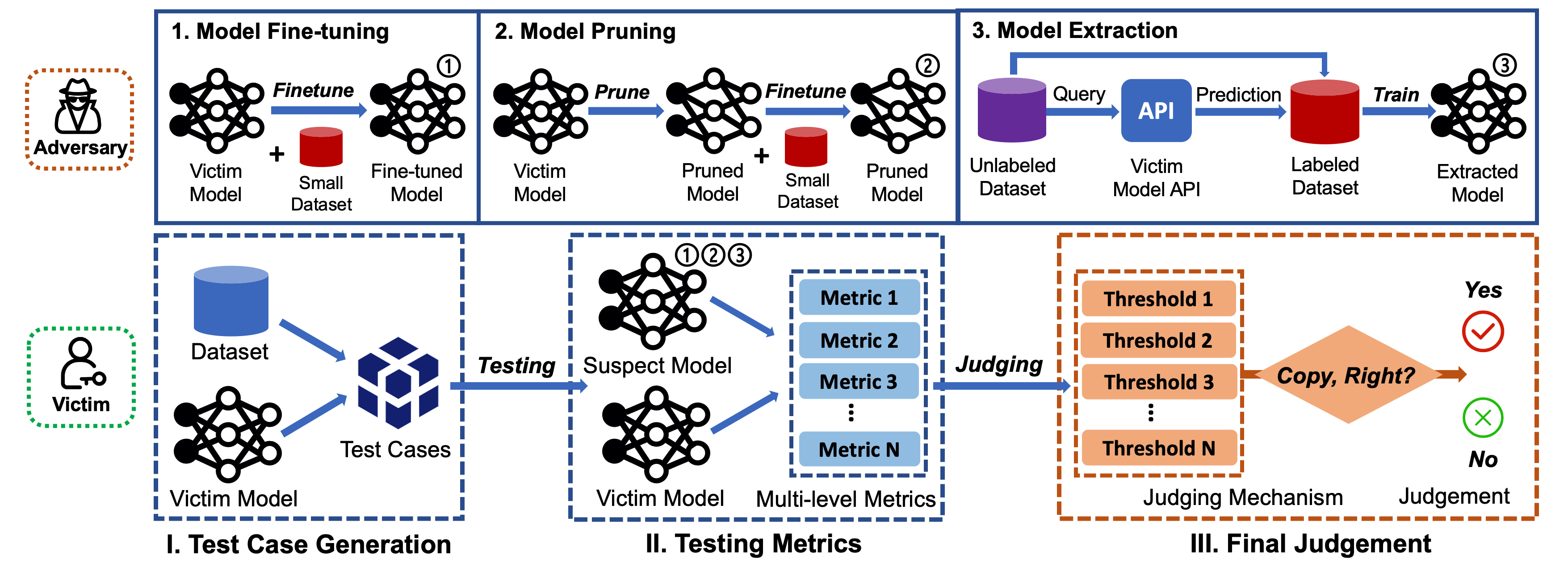}
\caption{The overview of \dje Testing Framework.}
\label{fig:frame}
\end{figure*}

Following the above idea, we design and implement \textsc{DeepJudge}, a novel testing framework for DNN copyright protection. As illustrated in Fig.~\ref{fig:frame}, \dje is composed of three core components. First, we propose a set of multi-level testing metrics to fully characterize a DNN model from different angles. Second, we propose efficient test case generation algorithms to magnify the similarities (or differences) measured by the testing metrics between the two models. Finally, a `yes'/`no' (stolen copy) judgment will be made for the suspect model based on all similarity scores.

The advantages of \textsc{DeepJudge} include \textbf{1) non-invasive}: it works directly on the trained models and does not tamper with the training process; \textbf{2) efficient}: it can be done very efficiently with only a few seed examples and a quick scan of the models; \textbf{3) flexible}: it can easily incorporate new testing metrics or test case generation methods to obtain more evidence and reliable judgement, and can be applied in both white-box and black-box scenarios with different testing metrics; \textbf{4) robust}: it is fairly robust to adaptive attacks such as model extraction and defense-aware attacks. The above advantages make \dje a practical, flexible, and extensible tool for copyright protection of deep learning models.

We have implemented \dje as an open-source self-contained toolkit and evaluated \textsc{DeepJudge} on four benchmark datasets (i.e., MNIST, CIFAR-10, ImageNet and Speech Commands) with different DNN architectures, including both convolutional and recurrent neural networks. The results confirm the effectiveness of \textsc{DeepJudge} in providing strong evidence for identifying the stolen copies of a victim model. \dje is also proven to be more robust to a set of adaptive attacks compared to existing defense techniques.

In summary, our main contributions are:
\begin{itemize}
\item We propose a novel testing framework \dje for copyright protection of deep learning models. \dje determines whether one model is a copy of the other depending on the similarity scores obtained from a comprehensive set of testing metrics and test case generation algorithms.

\item We identify three typical scenarios of model copying including finetuned copy, pruned copy, and extracted copy; define positive and negative suspect models for each scenario; and consider both white-box and black-box protection settings. \dje can produce reliable evidence and judgement to correctly identify the positive suspects across all scenarios and settings.

\item \dje is a self-contained open-source tool for robust copyright protection of deep learning models and a strong complement to existing techniques. \dje  can be flexibly applied in different DNN copyright protection scenarios and is extensible to new testing metrics and test case generation algorithms.

\end{itemize}

\vspace{-2mm}
\section{Background}
\label{sec:background}

\subsection{Deep Neural Network}
\label{sec:background_dnn}
A DNN classifier is a decision function $f: X\rightarrow Y$ mapping an input $\vx \in X$ to a label $y \in Y=\{1, 2, \cdots, C\}$, where $C$ is the total number of classes. It comprises of $L$ layers:
 $\{f^1, f^2, \cdots, f^{L-1}, f^L\},$
where $f^1$ is the input layer, $f^L$ is the probability output layer, 
and $f^2,\cdots,f^{L-1}$ are the hidden layers. Each layer $f^{l}$ can be denoted by a collection of neurons:
$\{n_{l,1}, n_{l,2}, \cdots, n_{l,N_l}\},$
where $N_l$ is the total number of neurons at that layer. Each neuron is a computing unit that computes its output by applying a linear transformation followed by a non-linear operation to its input (i.e., output from the precedent layer). We use $\phi_{l,i}(\vx)$ to denote the function that returns the output of neuron $n_{l,i}$ for a given input $ \vx \in X$. Then, we have the output vector of layer $f^{l} $ ($2 \leq l \leq L$):
$f^{l}(\vx) = \left\langle  \phi_{l,1}(\vx), \phi_{l,2}(\vx), \cdots, \phi_{l,N_l}(\vx)\right\rangle$.
Finally, the output label $f(\vx)$ is computed as $f(\vx)=\argmax f^L(\vx)$. 

\vspace{-2mm}
\subsection{DNN Watermarking}
\label{sec:background_watermarking}
A number of watermarking techniques have been proposed to protect the copyright of DNN models \cite{adi2018turning,darvish2019deepsigns,le2020adversarial,uchida2017embedding,zhang2018protecting,jia2021entangled}. Similar to traditional multimedia watermarking, DNN watermarking works in two steps: \emph{embedding} and \emph{verification}. In the embedding step, the owner embeds a secret watermark (e.g., a signature or a trigger set) into the model during the training process.
Depending on how much knowledge of the model is available in the verification step, existing watermarking methods can be broadly categorized into two classes: a) \textit{white-box} methods for the case when model parameters are available; and b) \textit{black-box} methods when only predictions of the model can be acquired.

White-box watermarking embeds a pre-designed signature (e.g., a string of bits) into the parameter space of the model via certain regularization terms \cite{darvish2019deepsigns,uchida2017embedding}. The ownership could be claimed when the extracted signature from a suspect model is similar to that of the owner model.
Black-box watermarking usually leverages backdoor attacks \cite{gu2017badnets} to implant a watermark pattern into the owner model by training the model with a set of backdoor examples (also known as the trigger set) relabeled to a secret class \cite{le2020adversarial,zhang2018protecting}. The ownership can then be claimed when the defender queries the suspect model for examples attached with the watermark trigger and receives the secret class as predictions.



\vspace{-2mm}
\subsection{DNN Fingerprinting}
\label{sec:background_fingerprinting}
Recently, DNN fingerprinting techniques have been proposed to verify model ownership via two steps: fingerprint \emph{extraction} and \emph{verification}. According to the categorization rule for watermarking, fingerprinting methods \cite{cao2021ipguard, DBLP:conf/iclr/LukasZK21} are all \emph{black-box} techniques. Moreover, they are \emph{non-invasive}, which is in sharp contrast with watermarking techniques. Instead of modifying the training procedure to embed identities, fingerprinting directly extracts a unique feature or property of the owner model as its fingerprint (i.e., a unique identifier). The ownership can then be verified if the fingerprint of the owner model matches with that of the suspect model. For example, IPGuard \cite{cao2021ipguard} leverages data points close to the classification boundary to fingerprint the boundary property of the owner model. A suspect model is determined to be a stolen copy of the owner model if it predicts the same labels for most boundary data points. \cite{DBLP:conf/iclr/LukasZK21} proposes a Conferrable Ensemble Method (CEM) to craft conferrable (a subclass of transferable examples) adversarial examples to fingerprint the overlap between two models' decision boundaries or adversarial subspaces. CEM fingerprinting demonstrates robustness to removal attacks including finetuning, pruning and extraction attacks, except several adapted attacks like adaptive transfer learning and adversarial training \cite{DBLP:conf/iclr/LukasZK21}. It is the closest work to our \dje. However, as a fingerprinting method, CEM targets \emph{uniqueness}, while as a testing framework, our \dje targets \emph{completeness}, i.e., comprehensive characterization of a model with multi-level testing metrics and diverse test case generation methods. Note that CEM fingerprinting can be incorporated into our framework as a black-box metric.

\section{DNN Copyright Threat Model}
\label{sec:threat_model}
We consider a typical attack-defense setting with two parties: the victim and the adversary. Here, the model owner is the victim who trains a DNN model (i.e., the victim model) using private resources. The adversary attempts to steal a copy of the victim model, which 1) mimics its functionality while 2) cannot be easily recognized as a copy.
Following this setting, we identify three common threats to DNN copyright: 1) model finetuning, 2) model pruning, and 3) model extraction. The three threats are illustrated in the top row of Fig.~\ref{fig:frame}.

\vspace{0.5mm}
\noindent\textbf{\textit{Threat 1:} Model Finetuning.} In this case, we assume the adversary has full knowledge of the victim model, including model architecture and parameters, and has a small dataset to finetune the model \cite{adi2018turning,uchida2017embedding}. This occurs, for example, when the victim open-sourced the model for academic purposes only, but the adversary attempts to finetune the model to build commercial products.

\vspace{0.5mm}
\noindent\textbf{\textit{Threat 2:} Model Pruning.} In this case, we also assume the adversary has full knowledge of the victim model's architecture and parameters. Model pruning adversaries first prune the victim model using some pruning methods, then finetune the model using a small set of data \cite{liu2018rethinking,renda2020comparing}.

\vspace{0.5mm}
\noindent\textbf{\textit{Threat 3:} Model Extraction.} In this case, we assume the adversary can only query the victim model for predictions (i.e., the probability vector). The adversary may be aware of the architecture of the victim model but has no knowledge of the training data or model parameters.
The goal of model extraction adversaries is to accurately steal the functionality of the victim model through the prediction API \cite{juuti2019prada,tramer2016stealing,papernot2017practical,orekondy2019knockoff,yuan2020attack}. To achieve this, the adversary first obtains an annotated dataset by querying the victim model for a set of auxiliary samples, then trains a copy of the victim model
on the annotated dataset. The auxiliary samples can be selected from a public dataset \cite{correia2018copycat,orekondy2019knockoff} or synthesized using some adaptive strategies \cite{papernot2017practical,juuti2019prada}.

\section{Testing for DNN Copyright Protection}

\label{sec:methodology}
In this section, we present \dje, the proposed testing framework that produces supporting evidence to determine whether a \emph{suspect} model is a \emph{copy} of a \emph{victim} model. The victim model can be copied by model finetuning, pruning, or extraction, as discussed in Section \ref{sec:threat_model}. We identify the following criteria for a reliable copyright protection method: 
\begin{enumerate}
    \item \textbf{Fidelity.} The protection or ownership verification process should not affect the utility of the owner model.
    \item \textbf{Effectiveness.} The verification should have high precision and recall in identifying stolen model copies.
    \item \textbf{Efficiency.} The verification process should be efficient, e.g., taking much less time than model training.
    \item \textbf{Robustness.} The protection should be resilient to adaptive attacks.
\end{enumerate}

\dje is a testing framework designed to satisfy all the above criteria. 
In the following three subsections, we will first give an overview of \dje, then introduce its multi-level testing metrics and test case generation algorithms.

\subsection{\dje Overview}
\label{subsec:frame}
As illustrated in the bottom row of Fig. \ref{fig:frame}, \dje consists of two components and a final judgement step: i) test case generation, ii) a set of multi-level distance metrics for testing, and iii) a thresholding and voting based judgement mechanism.
Alg.~\ref{alg:overall} depicts the complete procedure of \dje with pseudocode. It takes the victim model $\mathcal{O}$, a suspect model $\mathcal{S}$, and a set of data $\mathcal{D}$ associated with the victim model as inputs and returns the values of the testing metrics as evidence as well as the final judgement. The set of data $\mathcal{D}$ can be provided by the owner from either the training or testing set of the victim model.
At the test case generation step, it selects a set of seeds from the input dataset $\mathcal{D}$ (Line 1) and carefully generates a set of extreme test cases from the seeds (Line 2).
Based on the test cases generated, \dje computes the distance (dissimilarity) scores defined by the testing metrics between the suspect and victim models (Line 3).
The final judgement of whether the suspect is a copy of the victim can be made via a thresholding and voting mechanism according to the dissimilarity scores between the victim and a set of negative suspect models (Line 4).


\begin{algorithm}[t]
\small
    \caption{$\dje(\mathcal{O}, \mathcal{S}, \mathcal{D})$}
    \label{alg:overall}
    \KwIn{owner model $\mathcal{O}$, suspect model $\mathcal{S}$, data set $\mathcal{D}$ }
    \KwOut{judgement $\mathcal{J}$, evidence $\mathcal{E}$}
    \SetKwFunction{SelectSeeds}{SelectSeeds}
    \SetKwFunction{GenTestCase}{GenerateTestCases}
    \SetKwFunction{Metrics}{ComputeMetrics}
    \SetKwFunction{Judging}{Judging}
    \SetKwFunction{Evidence}{Evidence}
    \tcp{Test case generation (Section \ref{sec:test-gen})}
    $Seeds$ $\leftarrow$ \SelectSeeds$(\mathcal{O}, \mathcal{D})$ 

    ${T}$ $\leftarrow$ \GenTestCase$(\mathcal{O}, Seeds)$

    \tcp{Testing metrics (Section \ref{sec:metrics})}
    $\mathcal{E}$ $\leftarrow$ \Metrics$(\mathcal{O}, \mathcal{S}, {T})$ 

    \tcp{Final judgement (Section \ref{subsec:verification})}
    $\mathcal{J}$ $\leftarrow$ \Judging$(\mathcal{E})$  \tcp{Copy, Right? Yes or No.}

    \Return $\mathcal{J}, \mathcal{E}$
\end{algorithm}


\subsection{Multi-level Testing Metrics}
\label{sec:metrics}
We first introduce the testing metrics for two different settings respectively: white-box and black-box. \emph{1) White-box Setting:} In this setting, \dje has full access to the internals (i.e., intermediate layer outputs) and the final probability vectors of the suspect model $\mathcal{S}$. 
    \emph{2) Black-box Setting:} In this setting, \dje can only query the suspect model $\mathcal{S}$ to obtain the probability vectors or the predicted labels.
In both settings, we assume the model owner is willing to provide full access to the victim model $\mathcal{O}$, including the training and test datasets, and the training details if necessary.

\begin{table}[t]
    \renewcommand\arraystretch{1.2}
    \centering
    \caption{Proposed multi-level testing metrics.}\label{tab:metrics}
    \begin{tabu}{lll}
    \tabucline[1pt]{-}
    \textbf{Level}                         & \textbf{Metric}                           & \textbf{Defense Setting} \\ \tabucline[1pt]{-}
    \emph{Property-level}                & Robustness Distance (RobD)  & Black-box   \\ \hline
    \multirow{2}{*}{\emph{Neuron-level}} & Neuron Output Distance (NOD)
                                  & White-box  \\ 
    & Neuron Activation Distance (NAD)   & White-box \\ \hline
    \multirow{3}{*}{\emph{Layer-level}} 
    & Layer Outputs Distance (LOD)      & White-box   \\
    & Layer Activation Distance (LAD)   & White-box   \\ 
    & Jensen-Shanon Distance (JSD)    & Black-box   \\\tabucline[1pt]{-}
    \end{tabu}
\end{table}


The proposed testing metrics are summarized in Table~\ref{tab:metrics}, with their suitable defense settings highlighted in the last column. \dje advocates evidence-based ownership verification of DNNs via multi-level testing metrics that complement each other to produce more reliable judgement. 

\subsubsection{Property-level metrics}
There is an abundant set of model properties that could be used to characterize the similarities between two models, such as the adversarial robustness property \cite{fawzi2017robustness,carlini2017towards,cao2021ipguard,DBLP:conf/iclr/LukasZK21} and the fairness property \cite{mehrabi2019survey}. Here, we consider the former and define the \emph{robustness distance} to measure the adversarial robustness discrepancy between two models on the same set of test cases.
We will test more properties in our future work.

Denote the function represented by the victim model $\mathcal{O}$ by $f$, given an input $\vx_i$ and its ground truth label $y_i$, an adversarial example ${\vx'_i}$ can be crafted by slightly perturbing $\vx_i$ towards maximizing the classification error of $f$. This process is known as the adversarial attack, and $f(\vx'_i) \neq y_i$ indicates a successful attack. Adversarial examples can be generated using any existing adversarial attack methods such as FGSM \cite{goodfellow2014explaining} and PGD \cite{madry2017towards}.
Given a set of test cases, we can obtain its adversarial version $T=\{\vx'_1, \vx'_2, \cdots\}$, where $\vx'_i$ denotes the adversarial example of $\vx_i$.
The robustness property of model $f$ can then be defined as its accuracy on $T$:
$$Rob(f,T) =\frac{1}{|T|}\sum_{i=1}^{|T|} (f(\vx'_{i})) = y_i). $$

\noindent \textbf{Robustness Distance (RobD)}. 
Let $\hat{f}$ be the suspect model, we define the robustness distance between $f$ and $\hat{f}$ by the absolute difference between the two models' robustness:
$$RobD(f, \hat{f}, T) = |Rob(\hat{f}, T) - Rob(f,T)|.$$
The intuition behind \emph{RobD} is that model robustness is closely related to the decision boundary learned by the model through its unique optimization process, and should be considered as a type of fingerprint of the model. \emph{RobD} requires minimal knowledge of the model (only its output labels).

\subsubsection{Neuron-level metrics}
\label{subsubsec:neuron-level}

Neuron-level metrics are suitable for white-box testing scenarios where the internal layers' output of the model is accessible.
Intuitively, the output of each neuron in a model follows its own statistical distribution, and the neuron outputs in different models should vary. Motivated by this, \dje uses the output status of neurons to capture the difference between two models and defines the following two neuron-level metrics \emph{NOD} and \emph{NAD}. 

\noindent \textbf{Neuron Output Distance (NOD)}. 
For a particular neuron $n_{l,i}$ with $l$ being the layer index and $i$ being the neuron index within the layer, we denote the neuron output function of the owner's victim model and the suspect copy model by $\phi_{l,i}$ and $ \hat{\phi}_{l,i}$ respectively. \emph{NOD} measures the average neuron output difference between the two models over a given set  $T=\{\vx_1, \vx_2, \cdots\}$ of test cases:
$$NOD(\phi_{l,i}, \hat{\phi}_{l,i}, T)= \frac{1}{|T|} \sum_{\vx\in T} |\phi_{l,i}(\vx)-\hat{\phi}_{l,i}(\vx)|.$$ 

\noindent \textbf{Neuron Activation Distance (NAD)}.
Inspired by the Neuron Coverage \cite{pei2017deepxplore} for testing deep learning models, \emph{NAD} measures the difference in activation status (`activated' vs. `not activated') between the neurons of two models. Specifically, for a given test case $\vx \in T$, the neuron $n_{l,i}$ is determined to be `activated' if its output value $\phi_{l,i}(\vx)$ is larger than a pre-specified threshold.
The \emph{NAD} between the two models with respect to neuron $n_{l,i}$ can then be calculated as:
$$NAD(\phi_{l,i},\hat{\phi}_{l,i},T) = \frac{1}{|T|}\sum_{\vx\in T}|S({\phi_{l,i}(\vx)})-S({\hat{\phi}_{l,i}(\vx)})|,$$
where the step function $S(\phi_{l,i}(\vx))$ returns $1$ if $\phi_{l,i}(\vx)$ is greater than a certain threshold, $0$ otherwise.

\subsubsection{Layer-level metrics}
The layer-wise metrics in \dje take into account the output values of the entire layer in a DNN model. 
Compared with neuron-level metrics, layer-level metrics provide a full-scale view of the intermediate layer output difference between two models.

\vspace{0.5mm}
\noindent \textbf{Layer Output Distance (LOD)}. Given a layer index $l$, let $f^l$ and $\hat{f}^l$ represent the layer output functions of the victim model and the suspect model, respectively. \emph{LOD} measures the $L^p$-norm distance between the two models' layer outputs:
$$LOD(f^l,\hat{f}^l,T) =\frac{1}{|T|} \sum_{\vx\in T}||f^l(\vx)-\hat{f}^l(\vx)||_p,$$
where $||\cdot||_p$ denotes the $L^p$-norm ($p=2$ in our experiments).

\vspace{0.5mm}
\noindent \textbf{Layer Activation Distance (LAD)}. \emph{LAD} measures the average \emph{NAD} of all neurons within the same layer:
$$LAD(f^l,\hat{f}^l,T)=\frac{1}{|N_l|} \sum_{i=1}^{|N_l|} NAD(\phi_{l,i},\hat{\phi}_{l,i},T),$$
where $N_l$ is the total number of neurons at the $l$-th layer, and $\phi_{l,i}$ and $\hat{\phi}_{l,i}$ are the neuron output functions from $f^l$ and $\hat{f}^l$.

\vspace{0.5mm}
\noindent \textbf{Jensen-Shanon Distance (JSD)}. JSD \cite{fuglede2004jensen} is a metric that measures the similarly of two probability distributions, and a small \emph{JSD} value implies the two distributions are very similar. 
Let $f^L$ and $\hat{f}^L$ denote the output functions (output layer) of the victim model and the suspect model, respectively. Here, we apply \emph{JSD} to the output layer as follows:
$$ JSD(f^L,\hat{f}^L,T)=\frac{1}{2|T|} \sum_{\vx\in T} K(f^L(\vx),q)+K(\hat{f}^L(\vx),q),$$
where $q=(f^L(\vx)+\hat{f}^L(\vx))/2$ and $K(\cdot,\cdot)$ is the Kullback-Leibler divergence.
\emph{JSD} quantifies the similarity between two models' output distributions, and is particularly more powerful against model extraction attacks where the suspect model is extracted based on the probability vectors (distributions) returned by the victim model.

\subsection{Test Case Generation}
\label{sec:test-gen}

To fully exercise the testing metrics defined above, we need to magnify the similarities between a positive suspect and the victim model, while minimizing the similarities of a negative suspect to the victim model.
In \dje, this is achieved by smart test case generation methods. Meanwhile, test case generation should respect the model accessibility in different defense settings, i.e., black-box vs. white-box.

\subsubsection{Black-box setting} 
\label{subsubsec:bound}

\begin{figure}[t]
    \centering
    \includegraphics[width=0.7\linewidth]{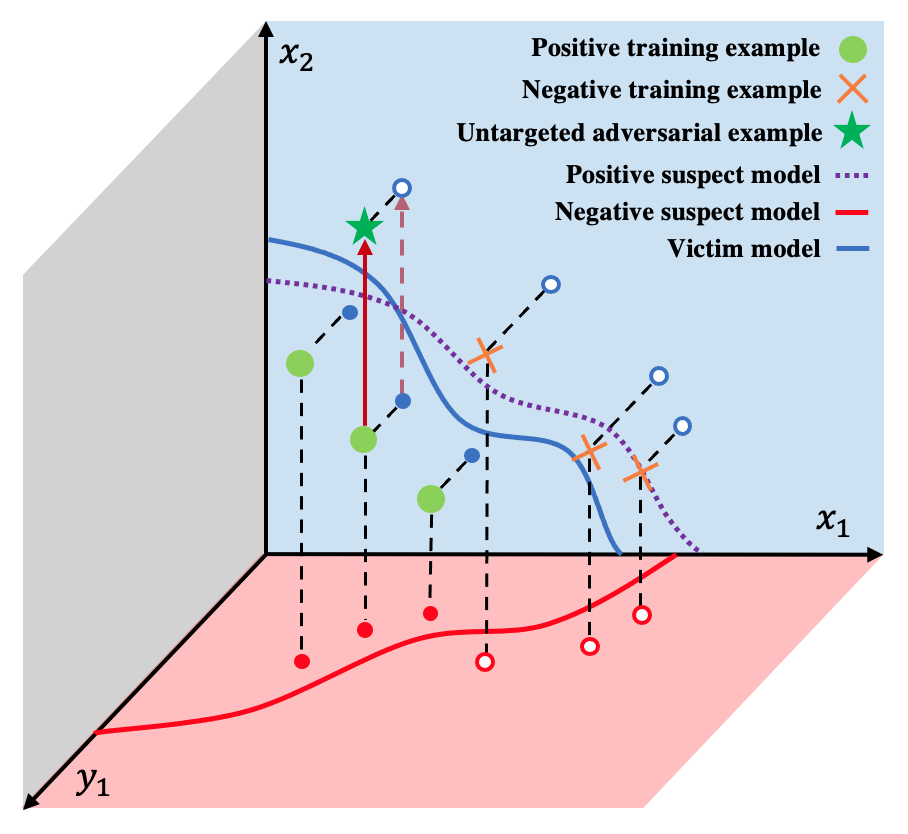}
    \caption{\dje uses adversarial examples of the victim model to probe the difference in models' decision boundary.}
    \label{fig:adv}
\end{figure}

When only the input and output of a suspect model are accessible, we populate the test set $T$ using adversarial inputs generated by existing adversarial attack methods on the victim model.
We consider three widely used adversarial attack methods, including Fast Gradient Sign Method (FGSM) \cite{goodfellow2014explaining}, Projected Gradient Descent (PGD) \cite{madry2017towards}, and Carlini \& Wagner’s (CW) attack \cite{carlini2017towards}, where FGSM and PGD are $L^{\infty}$-bounded adversarial methods, and CW is an $L^{2}$-bounded attack method. This gives us more diverse test cases with both $L^{\infty}$- and $L^{2}$-norm perturbed adversarial test cases. The detailed description and exact parameters used for adversarial test case generation are provided in Appendix \ref{subsec:generation}.

Fig.~\ref{fig:adv} illustrates the rationale behind using adversarial examples as test cases.
Finetuned and pruned model copies are directly derived from the victim model, thus they share similar decision boundaries (purple line) as the victim model. However, the negative suspect models are trained from scratch on different data or with different initializations, thus having minimum or no overlapping with the victim model's decision boundary. By subverting the model's predictions, adversarial examples cross the decision boundary from one side to the other (we use untargeted adversarial examples). Although the extracted models by model extraction attacks are trained from scratch by the adversary, the training relies on the probability vectors returned by the victim model, which contains information about the decision boundary. This implies that the extracted model will gradually mimic the decision boundary of the victim model. From this perspective, the decision boundary (or robustness) based testing imposes a dilemma to model extraction adversaries: the better the extraction, the more similar the extracted model to the victim model, and the easier it to be identified by our decision boundary based testing.

\subsubsection{White-box setting}
\begin{figure}[t]
    \centering
    \includegraphics[width=0.83\linewidth]{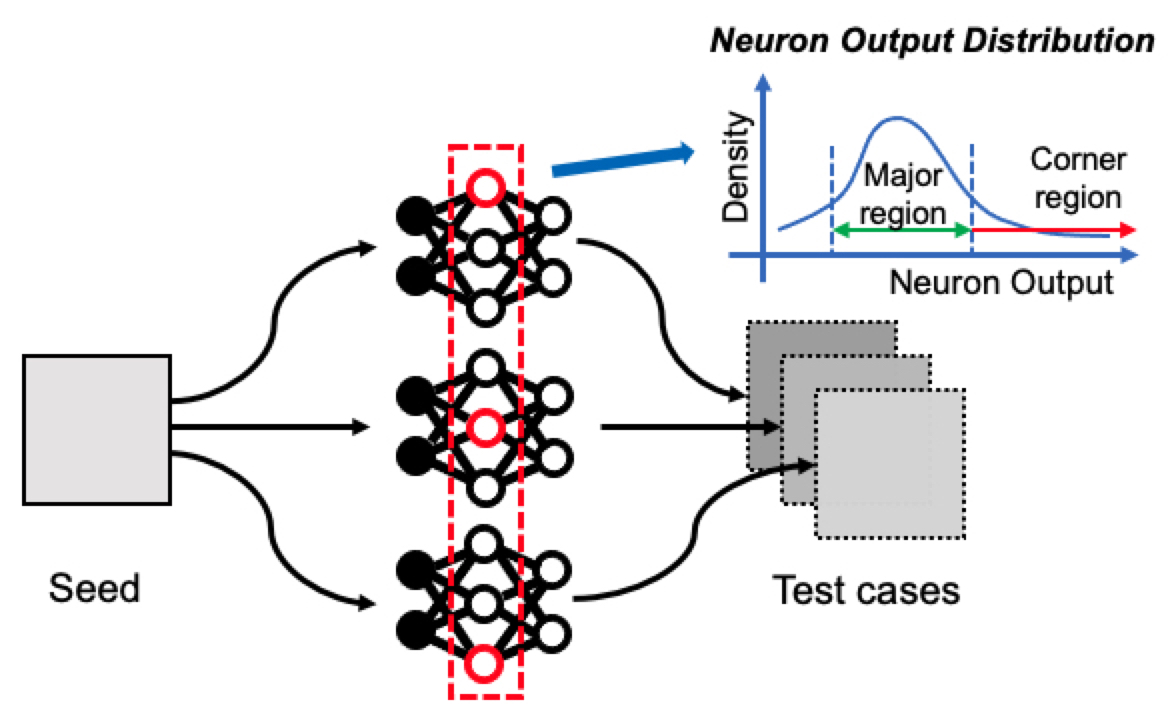}
    \caption{\dje tests each neuron and generates a test case to explore the corner region of its output distribution.}
    \label{fig:testing}
\end{figure}

In this case, the internals of the suspect model are accessible, thus a more fine-grained approach for test case generation becomes feasible. As shown in Fig.~\ref{fig:testing}, given a seed input and a specified layer, \dje generates one test case for each neuron, and the corner case of the neuron's activation is of our particular interest.

The test generation algorithm is described in Alg. \ref{alg:neuron}. It takes the owner's victim model $\mathcal{O}$ and a set of selected seeds $Seeds$ as input, and returns the set of generated test cases $T$. $T$ is initialized to be empty (Line 1). 
The main content of the algorithm is a nested loop (Lines 2-13), in which
for each neuron $n_{l,i}$ required by the metrics in Section \ref{sec:metrics}, the algorithm searches for an input that activates the neuron's output $\phi_{l,i}(\vx')$ more than a threshold value. At each outer loop iteration, an input is sampled from the $Seeds$ (Lines 3-4). It is then iteratively perturbed in the inner loop following the neuron activation's gradient update (Lines 6-7), until an input $\vx'$ that can satisfy the threshold condition is found and is added into the test suite $T$ (Lines 8-11) or when the maximum number of iterations is reached. The parameter $lr$ (Line 7) is used to control that the search space of the input with perturbation is close enough to its seed input.  Finally, the generated test suite $T$ is returned.


We discuss how to configure the specific threshold $k$ for a neuron $n_{l,i}$ used in Alg.~\ref{alg:neuron}. Since the statistics may vary across different neurons, we pre-compute the threshold $k$ based on the training data and the owner model, which is the maximum value (upper bound) of the corresponding neuron output over all training samples.
The final threshold value is then adjusted by a hyper-parameter $m$ to be used in Alg.~\ref{alg:neuron} for more reasonable and adaptive thresholds.
Note that the thresholds for all interested neurons can be calculated once by populating layer by layer across the model.

\begin{algorithm}[t]
\small
    \caption{GenerateTestCases($\mathcal{O}$, $Seeds$)}
    \label{alg:neuron}
    \KwIn{owner model $\mathcal{O}$, a set of seed inputs $Seeds$} 
    \KwOut{test suite $T$}
    Initialize test suite $T\leftarrow\emptyset$
    
    \For{each neuron $n_{l,i}$}{
    
        Sample a seed input $\vx\leftarrow$ $Seeds.choice()$
    
        $\vx'\leftarrow copy(\vx)$ 
        
        \For{$iter = 1$ \KwTo $iters$}{            


            Calculate gradients $grads \leftarrow \frac{\nabla \phi_{l,i}(\vx')}{\nabla \vx'}$
            
            Perturb input $\vx' \leftarrow \vx' + lr \cdot grads$

            \If{$\phi_{l,i}(\vx') > $ threshold $k$}{
                Add new test case $T\leftarrow T\cup \{\vx'\}$\\
                \textbf{break}
            }
        }
    }
    \Return $T$
\end{algorithm}




  
\subsection{Final Judgement}
\label{subsec:verification}
The judgment mechanism of \deepjudge has two steps: thresholding and voting. The thresholding step determines a proper threshold for each testing metric based on the statistics of a set of negative suspect models (see Section \ref{subsubsec:negative} for more details). The voting step examines a suspect model against each testing metric, and gives it a positive vote if its distance to the victim model is \emph{lower} than the threshold of that metric. The lower a measured metric value, the more likely the suspect model is a copy of the victim, according to this metric. The final judgment can then be made based on the votes: \emph{the suspect model will be identified as a positive suspect if it receives more positive votes, and a negative suspect otherwise}.

For each testing metric $\lambda$, we set the threshold adaptively using an $\varepsilon$-difference strategy. Specifically, we use one-tailed T-test to calculate the lower bound $LB_\lambda$ based on the statistics of the negative suspect models at the 99\% confidence level. If the measured difference $\lambda(\mathcal{O},\mathcal{S},T)$ is lower than $LB_\lambda$, $\mathcal{S}$ will be a copy of $\mathcal{O}$ with high probability. The threshold for each metric $\lambda$ is defined as: $\tau_\lambda = \alpha_{\lambda} \cdot LB_\lambda$, where $\alpha_\lambda$ is a user-specified relaxing parameter controlling the sensitivity of the judgement. As $\alpha_\lambda$ decreases, the false positive rate (the possibility of misclassifying a negative suspect as a stolen copy) will also increase. We empirically set $\alpha=0.9$ for black-box metrics and $\alpha=0.6$ for white-box metrics respectively, depending on the negative statistics.

\dje makes the final judgement by voting below:
\begin{equation*}
    p_{copy}(\mathcal{O},\mathcal{S},T) = \frac{1}{|\Lambda|}\sum_{\lambda \in \Lambda}\mathbbm{1}\big(\lambda(\mathcal{O},\mathcal{S},T) \leq \tau_\lambda\big),
\end{equation*}
where $\mathbbm{1}$ in the indicator function and $\Lambda$ denotes the set of \dje testing metrics, i.e., \{\emph{RobD}, \emph{NOD}, \emph{NAD}, \emph{LOD}, \emph{LAD}, \emph{JSD}\}. Note that, depending on the defense setting (white-box vs. black-box), only a subset of the testing metrics can be applied and the averaging can only be applied on the available metrics. \dje identifies a positive suspect copy if $p_{copy}$ is larger than 0.5 and a negative one otherwise. Arguably, voting is the most straightforward way of making the final judgement. While this simple voting strategy works reasonably well in our experiments, we believe more advanced judgement rules can be developed for diverse real-world protection scenarios.



\subsection{\deepjudge vs. Watermarking \& Fingerprinting}
\label{subsec:comp}


\begin{table*}[t]
\small
    \renewcommand\arraystretch{1.2}
    \centering

    \footnotesize
    \caption{A comparison of different copyright protection methods.} \label{tab:compare}
    \scalebox{1.03}{
    \begin{tabu}{cc|c|cc|ccc}
    \tabucline[1pt]{-}
    \multirow{2}{*}{\textbf{Method}}                                      & \multirow{2}{*}{\textbf{Type}} & \multirow{2}{*}{\begin{tabular}[c]{@{}c@{}}\textbf{Non-invasive} \end{tabular}} 
    & \multicolumn{2}{c|}{\textbf{Evaluated Settings}} & \multicolumn{3}{c}{\textbf{Evaluated Attacks}}                              \\ \cline{4-8} 
                                                               &                         &                                                                               & \emph{Black-box}           & \emph{White-box}           & \emph{Finetuning}         & \emph{Pruning}             & \emph{Extraction}   \\ \tabucline[1pt]{-} 
    Uchida et al. \cite{uchida2017embedding}   & \emph{Watermarking}            & \no                                                            & \no  & \yes & \yes & \yes & \no  \\ \hline
    Merrer et al. \cite{le2020adversarial}     & \emph{Watermarking}            & \no                                                            & \yes & \no  & \yes & \yes  & \no  \\ \hline
    Adi et al. \cite{adi2018turning}           & \emph{Watermarking}            & \no                                                            & \yes & \no  & \yes & \no  & \no \\ \hline
    Zhang et al. \cite{zhang2018protecting}    & \emph{Watermarking}            & \no                                                            & \yes & \no  & \yes & \yes & \no   \\ \hline
    Darvish et al. \cite{darvish2019deepsigns} & \emph{Watermarking}            & \no                                                            & \yes & \yes & \yes & \yes & \no  \\ \hline
    Jia et al. \cite{jia2021entangled} & \emph{Watermarking}            & \no                                                            & \yes & \no & \yes & \yes & \yes  \\ \hline
    Cao et al. \cite{cao2021ipguard}  & \emph{Fingerprinting}       & \yes                                                            & \yes & \no  & \yes & \yes & \no    \\ \hline
    Lukas et al. \cite{DBLP:conf/iclr/LukasZK21} & \emph{Fingerprinting}       & \yes                                                            & \yes & \no  & \yes & \yes & \yes  \\ \hline
    \textbf{DeepJudge (Ours)}                 & \emph{Testing}                 & \yes                                                                           & \yes & \yes & \yes & \yes & \yes  \\ \tabucline[1pt]{-}
    \end{tabu}
    }
    \end{table*}

Here, we briefly discuss why our testing approach is more favorable in certain settings and how it complements existing defense techniques. Table \ref{tab:compare} summarizes the differences of \dje to existing watermarking and fingerprinting methods, from three aspects: 1) whether the method is non-invasive (i.e., independent of model training); 2) whether it is particularly designed for or evaluated in different defense settings (i.e., white-box vs. black-box); and 3) whether the method is evaluated against different attacks (i.e., finetuning, pruning and extraction). \dje is training-independent, able to be flexibly applied in either white-box or black-box settings, and evaluated (also proven to be robust) against all three types of common copyright attacks including model finetuning, pruning and extraction, with empirical evaluations and comparisons deferred to Section~\ref{subsec:comparison}. 

Watermarking is invasive (training-dependent), whereas fingerprinting and testing are non-invasive (training independent). The effectiveness of watermarking depends on how well the owner model memorizes the watermark and how robust the memorization is to different attacks. While watermarking can be robust to finetuning or pruning attacks \cite{uchida2017embedding, zhang2018protecting}, it is particularly vulnerable to the emerging model extraction attack (see Section \ref{subsubsec:fail}). This is because model extraction attacks only extract the key functionality of the model, however, watermarks are often task-irrelevant. Despite the above weaknesses, watermarking is the only technique that can embed the owner identity/signature into the model, which is beyond the functionalities of fingerprinting or testing.



Fingerprinting shares certain similarities with testing. However, they differ in their goals. Fingerprinting aims for ``uniqueness'', i.e., a unique fingerprint of the model, while testing aims for ``completeness'', i.e., to test as many dimensions as possible to characterize not only the unique but also the common properties of the model. Arguably, effective fingerprints are also valid black-box testing metrics.
But as a testing framework, our \dje is not restricted to a particular metric or test case generation method. Our experiments in Section~\ref{sec:adaptive_exp} show that a single metric or fingerprint is not sufficient to handle the diverse and adaptive model stealing attacks.
In Section~\ref{sec:adaptive_attack2}, we will also show that our \dje can survive those adaptive attacks that break fingerprinting by dynamically changing the test case generation strategy. 
We anticipate a long-running arms race in deep learning copyright protection between model owners and adversaries, where watermarking, fingerprinting and testing methods are all important for a comprehensive defense.

\section{Experiments}
\label{sec:evaluation}

We have implemented \dje as a self-contained toolkit in Python\footnote{The tool and all the data in the experiment are publicly available via \url{https://github.com/Testing4AI/DeepJudge}}.
In the following, 
we first evaluate the performance of \dje against model finetuning and model pruning (Section \ref{subsec:tuning}), which are two threat scenarios extensively studied by watermarking methods \cite{adi2018turning,darvish2019deepsigns}. We then examine \dje against more challenging model extraction attacks in Section \ref{subsec:extraction_exp}.
Finally, we test the robustness of \dje under adaptive attacks in Section \ref{sec:adaptive_exp}.
Overall, we evaluated \dje with 11 attack methods, 3 baselines, and over 300 deep learning models trained on 4 datasets.

\subsection{Experimental Setup}\label{subsec:setup}
\subsubsection{Datasets \& Victim Models}
We run the experiments on three image classification datasets (i.e., MNIST \cite{lecun2010mnist}, CIFAR-10 \cite{krizhevsky2009learning} and ImageNet \cite{russakovsky2015imagenet}) and one audio recognition dataset (i.e., SpeechCommands \cite{warden2018speech}). The models used for the four datasets are summarized in Table~\ref{tab:setup}, including three convolutional architectures and one recurrent neural network. 
For each dataset, we divide the training data into two subsets. The first subset (50\% of the training examples) is used to train the victim model. More detailed experimental settings can be found in Appendix \ref{subsec:datasets}.

\subsubsection{Positive suspect models}
Positive suspect models are derived from the victim models via finetuning, pruning, or model extraction. These models are considered as stolen copies of the owner's victim model. \dje should provide evidence for the victim to claim ownership.

\subsubsection{Negative suspect models} 
\label{subsubsec:negative}
Negative suspect models have the same architecture as the victim models but are trained independently using either the remaining 50\% of training data or the same data but with different random initializations. The negative suspect models serve as the control group to show that \dje will not claim wrong ownership. These models are also used to compute the testing thresholds ($\tau$). The same training pipeline and the setting are used to train the negative suspect models. Specifically, ``Neg-1'' are trained with different random initializations while ``Neg-2'' are trained using a separate dataset (the other 50\% of training samples).

\subsubsection{Seed selection}
\label{subsubsec:seed}
Seed selection prepares the $Seeds$ examples used to generate the test cases. 
Here, we apply the sampling strategy used in DeepGini \cite{feng2020deepgini} to select a set of high-confidence seeds from the test dataset (details are in Appendix \ref{subsec:gini}). The intuition is that high-confidence seeds are well-learned by the victim model, thus carrying more unique features of the victim model. More adaptive seed selection strategies are explored in the adaptive attack section \ref{subsubsec:advtrain}.

\subsubsection{Adversarial example generation} We use three classic attacks including FGSM~\cite{goodfellow2014explaining}, PGD~\cite{madry2017towards} and CW \cite{carlini2017towards} to generate adversarial test cases as introduced in Section \ref{subsubsec:bound}.

\begin{table}[t]\centering
   \renewcommand\arraystretch{1.15}
   \footnotesize
   \caption{Datasets and victim models.} \label{tab:setup}
   \begin{tabu}{ccccc}
   \tabucline[1pt]{-}
   \multicolumn{1}{c}{\textbf{Dataset}} & \textbf{Type}  & \textbf{Model}    & \textbf{\#Params} & \multicolumn{1}{c}{\textbf{Accuracy}} \\ \tabucline[1pt]{-}
   \multicolumn{1}{c}{MNIST}    & Image  & LeNet-5      & 107.8\,K      & \multicolumn{1}{c}{98.5\%}   \\ \hline
   \multicolumn{1}{c}{CIFAR-10} & Image  & ResNet-20     & 274.4\,K    & \multicolumn{1}{c}{84.8\%}   \\ \hline 
   \multicolumn{1}{c}{ImageNet} & Image  & VGG-16       & 33.65\,M    & \multicolumn{1}{c}{74.4\%}     \\ \hline
   \multicolumn{1}{c}{SpeechCommands} & Audio & LSTM(128) & 132.4\,K  & \multicolumn{1}{c}{94.9\%} \\ 
   \tabucline[1pt]{-}
   \multicolumn{4}{l}{\#Params: number of parameters}
    
   \end{tabu}
   \end{table}

\subsection{Defending Against Model Finetuning \& Pruning}
\label{subsec:tuning}
As model finetuning and pruning threats are similar in processing the victim model (see Section \ref{sec:threat_model}), we discuss them together here. These two are also the most extensively studied threats in prior watermarking works \cite{adi2018turning,uchida2017embedding}.

\subsubsection{Attack strategies} Given a victim model and a small set of data in the same task domain, we consider the following four commonly used model finetuning \& pruning strategies:
   \textbf{a) Finetune the last layer (FT-LL).} Update the parameters of the last layer while freezing all other layers.
   \textbf{b) Finetune all layers (FT-AL).} Update the parameters of the entire model.
   \textbf{c) Retrain all layers (RT-AL).} Re-initialize the parameters of the last layer then update the parameters of the entire model.
   \textbf{d) Parameter pruning (P-r\%).} 
   Prune $r$ percentage of the parameters that have the smallest absolute values, then finetune the pruned model to restore the accuracy. We test both low ($r$=$20\%$) and high ($r$=$60\%$) pruning rates.
Typical data-augmentations are also used to strengthen the attacks. More details of these attacks are in Appendix \ref{subsec:aug}.



\begin{table*}[t]
   \renewcommand\arraystretch{1.2}
   \setlength\tabcolsep{4pt}
   \footnotesize
   \centering
      \caption{Performance of \dje against model finetuning and pruning attacks in the \textbf{black-box setting}. PGD \cite{madry2017towards} is used to generate the adversarial test cases. ACC is the validation accuracy. For each metric, the values below (indicating `copy') or above (indicating `not copy') the threshold $\tau_\lambda$ (the last row) are highlighted in \redbox{red} (copy alert) and \greenbox{green} (no alert), respectively.  `\textbf{Yes} (2/2)': two of the metrics vote for `copy' ($p_{copy}=100\%$); `\textbf{No} (0/2)': none of the metrics vote for  `copy' ($p_{copy}=0\%$).} \label{tab:black-box}
\begin{tabu}{c|c|cccc|cccc}
\tabucline[1pt]{-}
\multicolumn{2}{c|}{\multirow{2}{*}{\textbf{Model Type}}}                                                                   & \multicolumn{4}{c|}{\textbf{MNIST}} & \multicolumn{4}{c}{\textbf{CIFAR-10}}            \\ \cline{3-10} 
\multicolumn{2}{c|}{}                                                                                              & ACC    & \emph{RobD}    & \emph{JSD} & \textbf{Copy?}    & ACC  & \emph{RobD}  & \emph{JSD} &  \textbf{Copy?}\\ \tabucline[1pt]{-}
\multicolumn{2}{c|}{Victim Model}                                                                                  & 98.5\% & --  & --  & --  & 84.8\%      & --   & --  & --     \\ \tabucline[1pt]{-}
\multicolumn{1}{c|}{\multirow{5}{*}{\begin{tabular}[c]{@{}c@{}}Positive\\ Suspect\\ Models\end{tabular}}} & FT-LL   &  98.8$\pm$0.0\%     & \redbox{0.019$\pm$0.003}   &    \redbox{0.016$\pm$0.002}  & \textbf{Yes} (2/2)  & 82.1$\pm$0.1\% & \redbox{0.000$\pm$0.000}       & \redbox{0.002$\pm$0.001}  & \textbf{Yes} (2/2) \\ \cline{2-10} 
\multicolumn{1}{c|}{}                                                                                     & FT-AL   &  98.7$\pm$0.1\%      & \redbox{0.045$\pm$0.016}    &   \redbox{0.033$\pm$0.010}   & \textbf{Yes} (2/2) & 79.9$\pm$1.4\% & \redbox{0.192$\pm$0.028} & \redbox{0.162$\pm$0.014}  & \textbf{Yes} (2/2)\\ \cline{2-10} 
\multicolumn{1}{c|}{}                                                                                     & RT-AL   &  98.4$\pm$0.2\%      & \redbox{0.298$\pm$0.039}        &  \redbox{0.151$\pm$0.017}   & \textbf{Yes} (2/2)  & 79.4$\pm$0.8\% & \redbox{0.237$\pm$0.055} & \redbox{0.197$\pm$0.027} & \textbf{Yes} (2/2)\\ \cline{2-10} 
\multicolumn{1}{c|}{}                                                                                     & P-20\% &    98.7$\pm$0.1\%    & \redbox{0.058$\pm$0.014}        &    \redbox{0.035$\pm$0.009}   & \textbf{Yes} (2/2) & 81.7$\pm$0.2\% & \redbox{0.155$\pm$0.032} & \redbox{0.128$\pm$0.018} & \textbf{Yes} (2/2)\\ \cline{2-10} 
\multicolumn{1}{c|}{}                                                                                     & P-60\% &    98.6$\pm$0.1\%    & \redbox{0.172$\pm$0.024}        &    \redbox{0.097$\pm$0.010}  & \textbf{Yes} (2/2)  & 81.1$\pm$0.6\% & \redbox{0.318$\pm$0.036} &  \redbox{0.233$\pm$0.019} & \textbf{Yes} (2/2)\\ \tabucline[1pt]{-}

\multicolumn{1}{c|}{\multirow{3}{*}{\begin{tabular}[c]{@{}c@{}}Negative\\ Suspect\\ Models\end{tabular}}} & Neg-1  & 98.4$\pm$0.3\%        &    \greenbox{0.968$\pm$0.014}     &   \greenbox{0.614$\pm$0.016}  & \textbf{No} (0/2)   & 84.2$\pm$0.6\% & \greenbox{0.920$\pm$0.021} & \greenbox{0.603$\pm$0.016} & \textbf{No} (0/2)\\ \cline{2-10} 
\multicolumn{1}{c|}{}                                                                                     & Neg-2  & 98.3$\pm$0.2\%        &    \greenbox{0.949$\pm$0.029}     &   \greenbox{0.600$\pm$0.020}   & \textbf{No} (0/2)  & 84.9$\pm$0.5\% & \greenbox{0.926$\pm$0.030} & \greenbox{0.615$\pm$0.021} & \textbf{No} (0/2)\\ \cline{2-10} 
\multicolumn{1}{c|}{}                                                                                     & \textbf{$\tau_\lambda$} & --      & \textbf{0.852}        &  \textbf{0.538}      & --      & --      & \textbf{0.816} & \textbf{0.537} & --\\ \tabucline[1pt]{-}
\end{tabu}

\vspace{1mm}

\begin{tabu}{c|c|cccc|cccc}
\tabucline[1pt]{-}
\multicolumn{2}{c|}{\multirow{2}{*}{\textbf{Model Type}}}                                                                   & \multicolumn{4}{c|}{\textbf{ImageNet}} & \multicolumn{4}{c}{\textbf{SpeechCommands}}    \\ \cline{3-10} 
\multicolumn{2}{c|}{}                                                                                              & ACC    & \emph{RobD}    & \emph{JSD}  & \textbf{Copy?}   & ACC  & \emph{RobD}  & \emph{JSD} & \textbf{Copy?}\\ \tabucline[1pt]{-}
\multicolumn{2}{c|}{Victim model}                                                                                  & 74.4\% & --       & --   & --    & 94.9\%      & --           & --      & --     \\ \tabucline[1pt]{-}

\multicolumn{1}{c|}{\multirow{5}{*}{\begin{tabular}[c]{@{}c@{}}Positive\\ Suspect\\ Models\end{tabular}}} & FT-LL   & 73.2$\pm$0.4\%       & \redbox{0.034$\pm$0.007}   &  \redbox{0.009$\pm$0.003}   & \textbf{Yes} (2/2)   & 95.2$\pm$0.1\% &  \redbox{0.104$\pm$0.007} &  \redbox{0.036$\pm$0.006} & \textbf{Yes} (2/2)\\ \cline{2-10} 
\multicolumn{1}{c|}{}                                                                                     & FT-AL   & 70.8$\pm$0.9\%       &  \redbox{0.073$\pm$0.011}  &  \redbox{0.043$\pm$0.011}     & \textbf{Yes} (2/2)    & 95.8$\pm$0.3\% &  \redbox{0.326$\pm$0.024} &   \redbox{0.155$\pm$0.014} & \textbf{Yes} (2/2)\\ \cline{2-10} 
\multicolumn{1}{c|}{}                                                                                     & RT-AL   &  53.3$\pm$0.8\%      &  \redbox{0.192$\pm$0.008}  & \redbox{0.251$\pm$0.015}    & \textbf{Yes} (2/2)    & 94.3$\pm$0.3\% &  \redbox{0.445$\pm$0.019} &  \redbox{0.231$\pm$0.016} & \textbf{Yes} (2/2)\\ \cline{2-10} 
\multicolumn{1}{c|}{}                                                                                     & P-20\% &    69.7$\pm$1.1\%    &  \redbox{0.106$\pm$0.010}       & \redbox{0.064$\pm$0.003}     & \textbf{Yes} (2/2)    & 95.4$\pm$0.2\% &    \redbox{0.310$\pm$0.026} &  \redbox{0.152$\pm$0.013}  & \textbf{Yes} (2/2)\\ \cline{2-10} 
\multicolumn{1}{c|}{}                                                                                     & P-60\% &   68.8$\pm$1.0\%     &   \redbox{0.161$\pm$0.017}      &  \redbox{0.091$\pm$0.004}     & \textbf{Yes} (2/2)  & 95.0$\pm$0.5\% &    \redbox{0.437$\pm$0.030}  &  \redbox{0.215$\pm$0.013} & \textbf{Yes} (2/2)\\ \tabucline[1pt]{-}

\multicolumn{1}{c|}{\multirow{3}{*}{\begin{tabular}[c]{@{}c@{}}Negative\\ Suspect\\ Models\end{tabular}}} & Neg-1  & 74.2$\pm$0.3\%        &  \greenbox{0.737$\pm$0.007}    & \greenbox{0.395$\pm$0.006}    & \textbf{No} (0/2)   & 94.9$\pm$0.7\% &   \greenbox{0.819$\pm$0.025}  &   \greenbox{0.456$\pm$0.014} & \textbf{No} (0/2) \\ \cline{2-10} 
\multicolumn{1}{c|}{}                                                                                     & Neg-2  &  73.9$\pm$0.5\%       &  \greenbox{0.760$\pm$0.010}     &  \greenbox{0.429$\pm$0.004}   & \textbf{No} (0/2)   & 94.5$\pm$0.8\% &   \greenbox{0.832$\pm$0.024}  &   \greenbox{0.472$\pm$0.012}  & \textbf{No} (0/2) \\ \cline{2-10} 
\multicolumn{1}{c|}{}                                                                                     & \textbf{$\tau_\lambda$} & --      &  \textbf{0.659}   &  \textbf{0.356}   & --      & --    & \textbf{0.727} &  \textbf{0.405} & --\\ \tabucline[1pt]{-}
\end{tabu}

\end{table*}

\begin{table*}[t]
   \renewcommand\arraystretch{1.2}
   \setlength\tabcolsep{2.1pt}
   \centering
   \footnotesize
   \caption{Performance of \dje against model finetuning and pruning attacks in the \textbf{white-box setting}. Algorithm~\ref{alg:neuron} is used to generate the test cases. For each metric, the values below (indicating `copy') or above (indicating `not copy') the threshold $\tau_\lambda$ (the last row) are highlighted in \redbox{red} (copy alert) and \greenbox{green} (no alert) respectively. `\textbf{Yes} (4/4)': all 4 metrics vote for  `copy' ($p_{copy}=100\%$); `\textbf{No} (0/4)': none of the metrics vote for  `copy' ($p_{copy}=0\%$).} \label{tab:white-box}
   
   \begin{tabu}{cc|ccccc|ccccc}
\tabucline[1pt]{-}
\multicolumn{2}{c|}{\multirow{2}{*}{\textbf{Model Type}}}                                                                   & \multicolumn{5}{c|}{\textbf{MNIST}}            & \multicolumn{5}{c}{\textbf{CIFAR-10}}                   \\ \cline{3-12} 
\multicolumn{2}{c|}{}                                                                                              & \emph{NOD}     & \emph{NAD}     & \emph{LOD}     & \emph{LAD}  & \textbf{Copy?}  & \emph{NOD}       & \emph{NAD}        & \emph{LOD}        & \emph{LAD}  &  \textbf{Copy?}   \\ \tabucline[1pt]{-}
\multicolumn{1}{c|}{\multirow{5}{*}{\begin{tabular}[c]{@{}c@{}}Positive\\ Suspect\\ Models\end{tabular}}} & FT-LL   &   \redbox{0.00$\pm$0.00}      &    \redbox{0.00$\pm$0.00}     &   \redbox{0.00$\pm$0.00}    &     \redbox{0.00$\pm$0.00}   & \textbf{Yes} (4/4)  & \redbox{0.00$\pm$0.00}    & \redbox{0.00$\pm$0.00}  & \redbox{0.00$\pm$0.00}      & \redbox{0.00$\pm$0.00}   & \textbf{Yes} (4/4) \\ \cline{2-12} 
\multicolumn{1}{c|}{}                                                                                     & FT-AL   &  \redbox{0.08$\pm$0.01}       &   \redbox{0.23$\pm$0.21}      &    \redbox{0.32$\pm$0.03}    &   \redbox{0.82$\pm$0.16}   & \textbf{Yes} (4/4)   & \redbox{0.15$\pm$0.02} & \redbox{0.30$\pm$0.12}  & \redbox{0.74$\pm$0.07}  & \redbox{0.21$\pm$0.04} & \textbf{Yes} (4/4) \\ \cline{2-12} 
\multicolumn{1}{c|}{}                                                                                     & RT-AL   &  \redbox{0.31$\pm$0.02}       &    \redbox{0.37$\pm$0.20}     &   \redbox{0.97$\pm$0.04}       & \redbox{1.27$\pm$0.29}   & \textbf{Yes}  (4/4)   & \redbox{0.18$\pm$0.02} & \redbox{0.26$\pm$0.10}  & \redbox{0.78$\pm$0.03}  & \redbox{0.22$\pm$0.02}  & \textbf{Yes} (4/4) \\ \cline{2-12} 
\multicolumn{1}{c|}{}                                                                                     & P-20\% &    \redbox{0.10$\pm$0.01}     &     \redbox{0.16$\pm$0.12}    &  \redbox{0.36$\pm$0.03}       &  \redbox{0.79$\pm$0.15}   & \textbf{Yes} (4/4)   & \redbox{0.28$\pm$0.03} & \redbox{0.32$\pm$0.09}  & \redbox{0.77$\pm$0.06}  & \redbox{0.24$\pm$0.02}  & \textbf{Yes} (4/4) \\ \cline{2-12} 
\multicolumn{1}{c|}{}                                                                                     & P-60\% &    \redbox{0.11$\pm$0.01}     &  \redbox{0.82$\pm$0.26}    &  \redbox{0.43$\pm$0.03}       &  \redbox{1.16$\pm$0.08}   & \textbf{Yes}  (4/4)  & \redbox{0.62$\pm$0.03} & \redbox{1.65$\pm$0.34}  & \redbox{2.80$\pm$0.21}  & \redbox{0.93$\pm$0.10} & \textbf{Yes} (4/4) \\ \tabucline[1pt]{-}

\multicolumn{1}{c|}{\multirow{3}{*}{\begin{tabular}[c]{@{}c@{}}Negative\\ Suspect\\ Models\end{tabular}}} & Neg-1  &    \greenbox{0.77$\pm$0.07}     &    \greenbox{11.46$\pm$1.14}    & \greenbox{1.73$\pm$0.06}        &  \greenbox{6.42$\pm$0.84} & \textbf{No}  (0/4)    & \greenbox{3.09$\pm$0.30} & \greenbox{10.94$\pm$1.74} & \greenbox{11.85$\pm$1.01} & \greenbox{5.41$\pm$0.67} & \textbf{No} (0/4) \\ \cline{2-12} 
\multicolumn{1}{c|}{}                                                                                     & Neg-2  &    \greenbox{0.79$\pm$0.08}     &  \greenbox{12.28$\pm$1.50}        & \greenbox{1.78$\pm$0.13}     &    \greenbox{6.37$\pm$0.47}  & \textbf{No} (0/4)  & \greenbox{3.21$\pm$0.18} & \greenbox{11.09$\pm$0.71} & \greenbox{12.60$\pm$1.33} & \greenbox{5.37$\pm$0.72} & \textbf{No} (0/4) \\ \cline{2-12} 
\multicolumn{1}{c|}{}                                                                                     & $\tau_\lambda$      & \textbf{0.45}   &  \textbf{6.74}   &  \textbf{1.03}  &   \textbf{3.65}  & --    & \textbf{1.79} & \textbf{6.14} & \textbf{6.89} & \textbf{3.01}  & -- \\ \tabucline[1pt]{-}
\end{tabu}

\vspace{1mm}
\begin{tabu}{cc|ccccc|ccccc}
\tabucline[1pt]{-}
\multicolumn{2}{c|}{\multirow{2}{*}{\textbf{Model Type}}}                                                                   & \multicolumn{5}{c|}{\textbf{ImageNet}}            & \multicolumn{5}{c}{\textbf{SpeechCommands}}                   \\ \cline{3-12} 
\multicolumn{2}{c|}{}                                                                                              & \emph{NOD}     & \emph{NAD}     & \emph{LOD}     & \emph{LAD}  & \textbf{Copy?}   & \emph{NOD}       & \emph{NAD}        & \emph{LOD}        & \emph{LAD}   &\textbf{Copy?}     \\ \tabucline[1pt]{-}
\multicolumn{1}{c|}{\multirow{5}{*}{\begin{tabular}[c]{@{}c@{}}Positive\\ Suspect\\ Models\end{tabular}}} & FT-LL   &    \redbox{0.00$\pm$0.00}     &    \redbox{0.00$\pm$0.00}     &    \redbox{0.00$\pm$0.00}     &    \redbox{0.00$\pm$0.00}  & \textbf{Yes} (4/4)  & \redbox{0.000$\pm$0.000}     & \redbox{0.00$\pm$0.00}      & \redbox{0.00$\pm$0.00}      & \redbox{0.000$\pm$0.00}   & \textbf{Yes}  (4/4) \\ \cline{2-12} 
\multicolumn{1}{c|}{}                                                                                     & FT-AL   &  \redbox{0.02$\pm$0.01}       &    \redbox{0.18$\pm$0.09}    &    \redbox{0.16$\pm$0.05}     &  \redbox{0.58$\pm$0.13}    & \textbf{Yes} (4/4)   &  \redbox{0.037$\pm$0.003} & \redbox{0.05$\pm$0.02}  & \redbox{0.42$\pm$0.02}  & \redbox{12.82$\pm$1.00} & \textbf{Yes} (4/4) \\ \cline{2-12} 
\multicolumn{1}{c|}{}                                                                                     & RT-AL   &  \redbox{0.03$\pm$0.00}   & \redbox{0.30$\pm$0.07}    &    \redbox{0.25$\pm$0.03}  &  \redbox{0.78$\pm$0.05}    & \textbf{Yes} (4/4)    & \redbox{0.055$\pm$0.003} & \redbox{0.25$\pm$0.31}  & \redbox{0.64$\pm$0.08}  & \redbox{21.64$\pm$2.47} & \textbf{Yes} (4/4) \\ \cline{2-12} 
\multicolumn{1}{c|}{}                                                                                     & P-20\% &   \redbox{0.11$\pm$0.01}  & \redbox{0.83$\pm$0.06}        &    \redbox{0.76$\pm$0.01}     &  \redbox{1.67$\pm$0.22}   & \textbf{Yes} (4/4)    & \redbox{0.038$\pm$0.002} & \redbox{0.03$\pm$0.02}   & \redbox{0.44$\pm$0.02}  & \redbox{14.57$\pm$3.12} & \textbf{Yes} (4/4)  \\ \cline{2-12} 
\multicolumn{1}{c|}{}                                                                                     & P-60\% &   \redbox{0.77$\pm$0.01}    & \redbox{3.09$\pm$0.12}   & \redbox{3.41$\pm$0.03}   &  \redbox{6.63$\pm$0.23}    & \textbf{Yes} (4/4)   & \redbox{0.094$\pm$0.004} & \redbox{0.45$\pm$0.32}  & \redbox{0.67$\pm$0.04}  & \redbox{20.58$\pm$3.44} & \textbf{Yes} (4/4)  \\ \tabucline[1pt]{-}

\multicolumn{1}{c|}{\multirow{3}{*}{\begin{tabular}[c]{@{}c@{}}Negative\\ Suspect\\ Models\end{tabular}}} & Neg-1  &  \greenbox{6.55$\pm$0.78}         &  \greenbox{32.18$\pm$2.97}        &  \greenbox{35.03$\pm$3.13}       & \greenbox{30.32$\pm$1.91}   & \textbf{No} (0/4)     & \greenbox{0.488$\pm$0.013} &  \greenbox{39.61$\pm$9.74}  &  \greenbox{2.82$\pm$0.08} & \greenbox{64.32$\pm$2.42} & \textbf{No} (0/4) \\ \cline{2-12} 
\multicolumn{1}{c|}{}                                                                                     & Neg-2  &  \greenbox{6.25$\pm$0.39}  &  \greenbox{30.04$\pm$2.44}  & \greenbox{44.21$\pm$3.11}        &  \greenbox{29.58$\pm$0.86}   & \textbf{No} (0/4)   & \greenbox{0.480$\pm$0.012} & \greenbox{34.84$\pm$6.07}   & \greenbox{2.79$\pm$0.09} &  \greenbox{62.69$\pm$1.75} & \textbf{No} (0/4) \\ \cline{2-12} 
\multicolumn{1}{c|}{}                                                                                     & $\tau_\lambda$      &   \textbf{3.48}     &   \textbf{17.17}  &  \textbf{20.74}  &    \textbf{17.20}  & --     & \textbf{0.286} & \textbf{19.77} & \textbf{1.66} & \textbf{37.48}  & -- \\ \tabucline[1pt]{-}
\end{tabu}
\end{table*}

\subsubsection{Effectiveness of  \dje} The results are presented separately for black-box vs. white-box settings.

\vspace{0.5mm}
\noindent\textbf{Black-box Testing.}
In this setting, only the output probabilities of the suspect model are accessible. Here, \dje uses the two black-box metrics: \emph{RobD and JSD}. For both metrics, the smaller the value, the more similar the suspect model is to the victim model. Table~\ref{tab:black-box} reports the results of \dje on the four datasets. Note that we randomly repeat the experiment 6 times for each finetuning or pruning attack and 12 times for independent training (as more negative suspect models will result in a more accurate judging threshold).
Then, we report the average and standard deviation (in the form of $a\pm b$) in each entry of Table~\ref{tab:black-box}. Clearly, all positive suspect models are more similar to the victim model with significantly smaller \emph{RobD} and \emph{JSD} values than negative suspect models. Specifically, a low \emph{RobD} value indicates that the adversarial examples generated on the victim model have a high transferability to the suspect model, i.e., its decision boundary is closer to the victim model. In contrast, the \emph{RobD} values of the negative suspect models are much larger than that of the positives, which matches our intuition in Fig.~\ref{fig:adv}. 

\begin{figure}[t]
   \centering
   \includegraphics[width=0.45\linewidth]{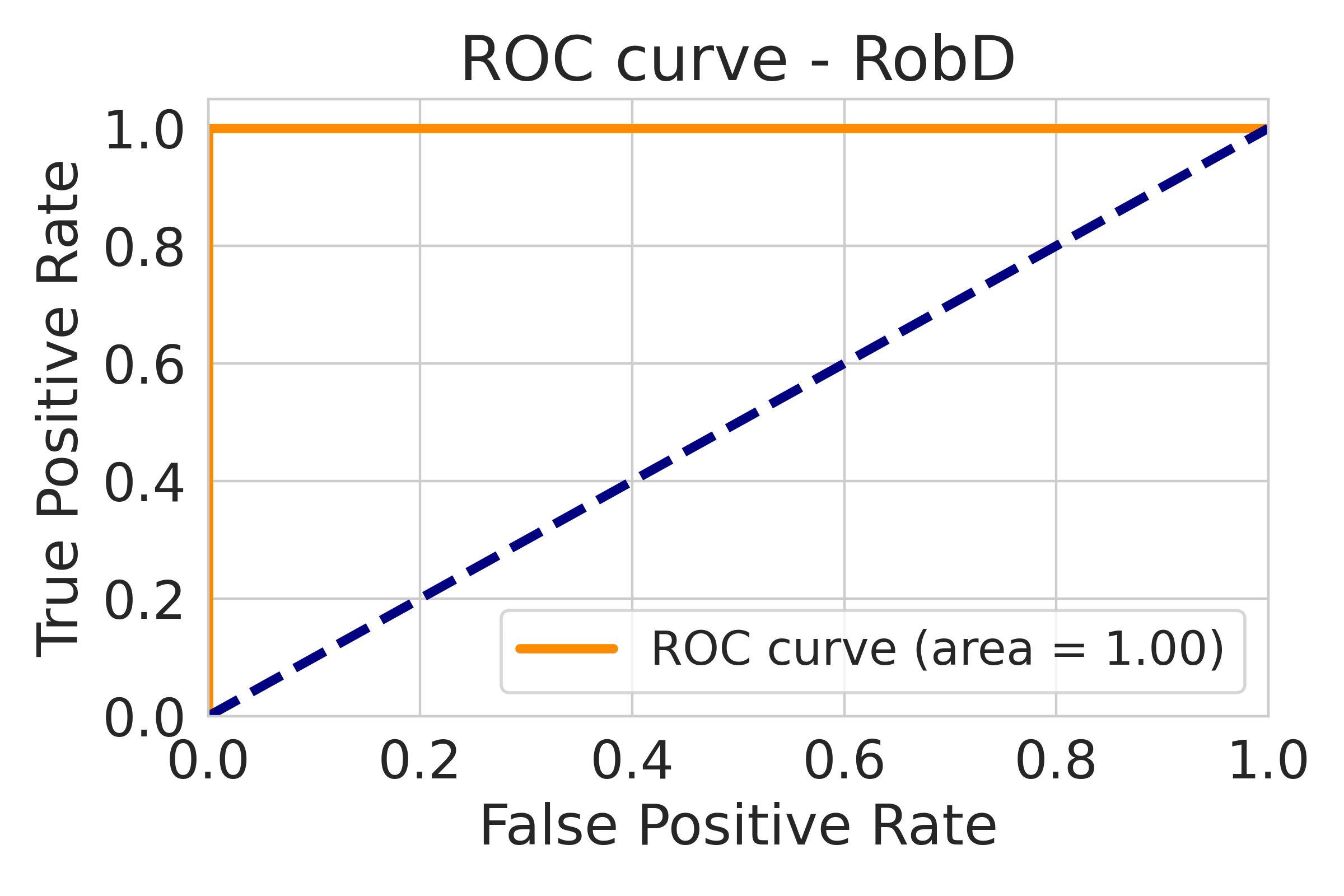}
   \hspace{2mm}
   \includegraphics[width=0.45\linewidth]{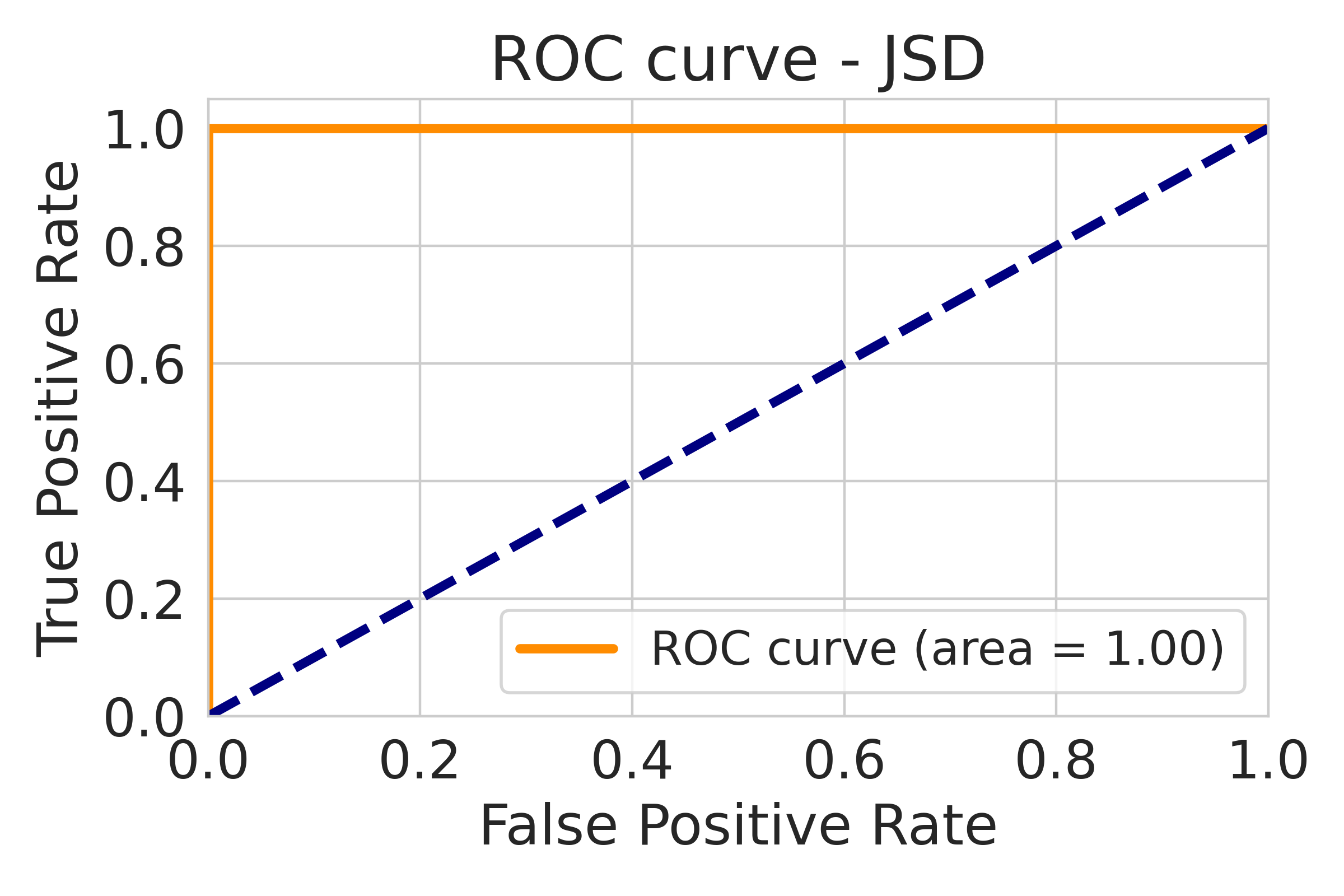}
   \caption{The detection ROC curves of metrics \emph{RobD} and \emph{JSD} on CIFAR-10 suspect models, and $AUC=1$ for both metrics.}
   \label{fig:ROC}
\end{figure}

To further confirm the effectiveness of the proposed metrics, we show the ROC curve for a total of 54 models (30 positive suspect models and 24 negative suspect models) for \emph{RobD} and \emph{JSD} in Figure \ref{fig:ROC}. The AUC values are 1 for both metrics. Note that we omit the plots for the following white-box testing as the AUC values for all metrics are also 1.

\vspace{0.5mm}
\noindent\textbf{White-box Testing.}
In this setting, all intermediate-layer outputs of the suspect model are accessible. \dje can thus use the four white-box metrics (i.e., \emph{NOD, NAD, LOD, and LAD}) to test the models. Table~\ref{tab:white-box} reports the results on the four datasets. Similar to the two black-box metrics, the smaller the white-box metrics, the more likely the suspect model is a stolen copy. As shown in Table~\ref{tab:white-box}, there is a fundamental difference between the two sets (positive vs. negative) of suspect models according to each of the four metrics. That is, the two sets of models are completely separable, leading to highly accurate detection of the positive copies.
It is not surprising as white-box testing can collect more fine-grained information from the suspect models.
In both the black-box and white-box settings, the voting in \dje overwhelmingly supports the correct final judgement (the `Copy?' column). 

\begin{figure*}[t]
   \centering
   \includegraphics[width=0.4\linewidth]{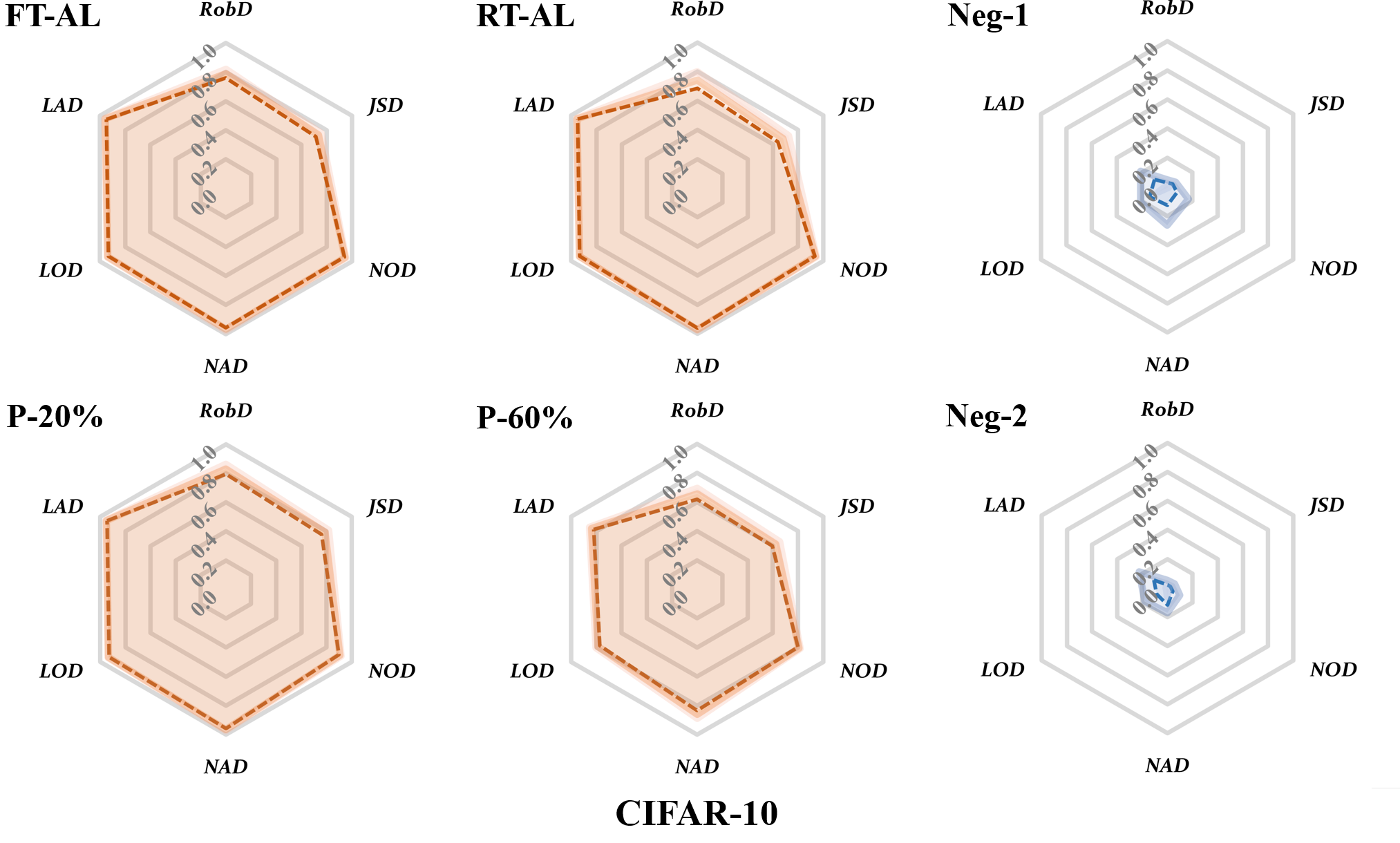}
   \hspace{8mm}
   \includegraphics[width=0.4\linewidth]{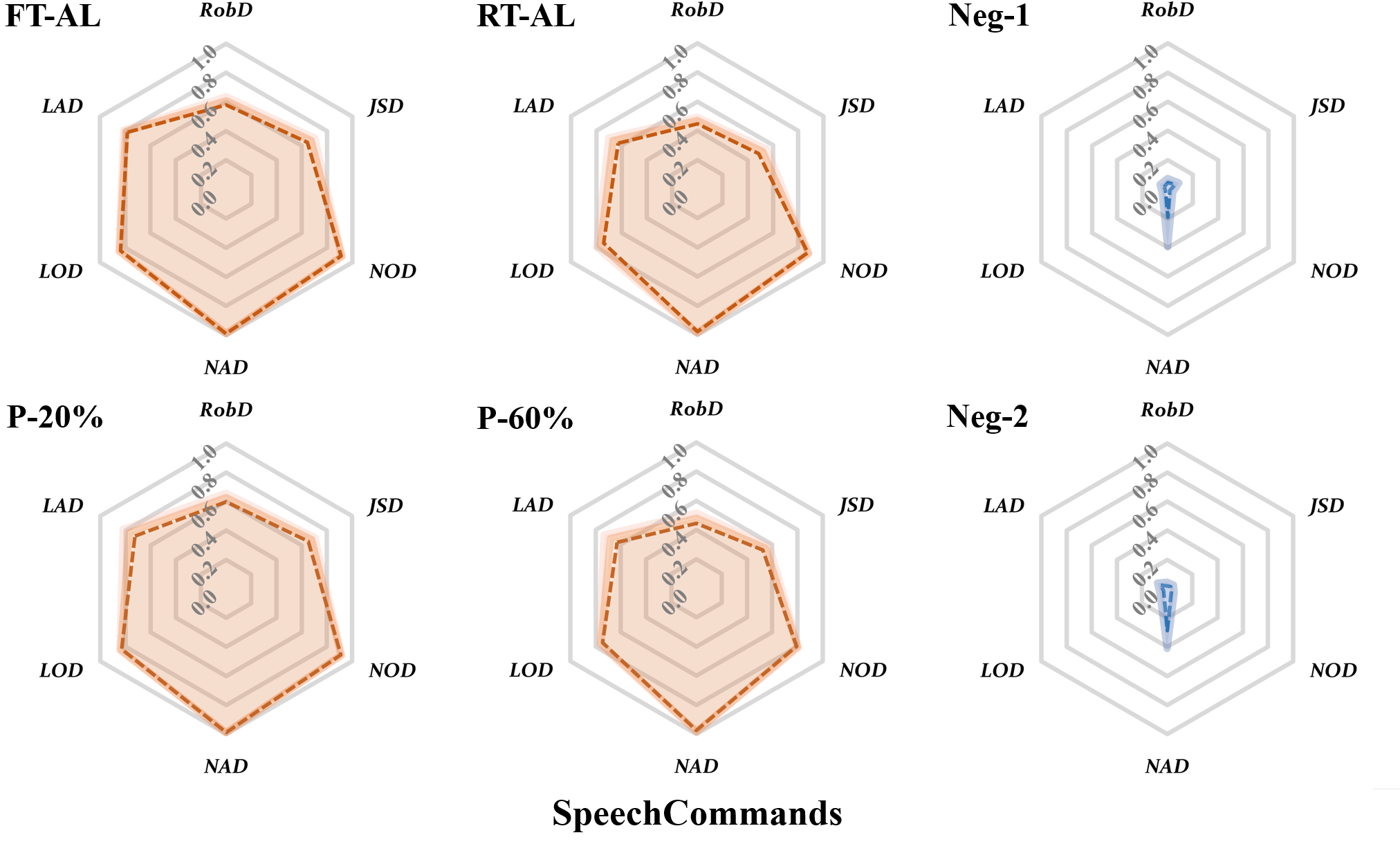}
  \setlength{\abovecaptionskip}{3pt}
   \caption{Similarities of different suspect models to the victim model on CIFAR-10 (left 3 columns) and SpeechCommands (right 3 columns). We use the \textcolor{orange}{orange} line for the positive suspect models and the \blue{blue} line for negatives. Each dimension of the radar chart corresponds to a \emph{similarity score} given by one \dje metric. The similarity score is computed by first normalizing the metric, e.g., $RobD$, to $[0,1]$ then taking $1-RobD$.}
   \label{fig:radars}
\end{figure*}

\noindent\textbf{Combined Visualization.} To better understand the power of \dje, we combine the black-box and white-box testing results for each suspect model into a single radar chart in Fig.~\ref{fig:radars}. Each dimension of the radar chart corresponds to a \emph{similarity score} given by one testing metric. For better visual effect,
we normalize the values of the testing metrics into the range $[0,1]$, and the larger the normalized value, the more similar the suspect model to the victim. Thus, the filled area could be viewed as the \emph{accumulated supporting evidence} by \dje metrics for determining whether the suspect model is a stolen copy. Clearly, \dje is able to accurately distinguish positive suspects from negative ones.
Among the positive suspect models, the areas of RT-AL and P-60\% are noticeably smaller than the other two, meaning they are harder to detect. This is because these two attacks make the most parameter modifications to the victim model. Comparing the metrics, activation-based metrics (e.g., \emph{NAD}) demonstrate better performance than output-based metrics (e.g., \emph{NOD}), while white-box metrics are stronger than black-box metrics, especially against strong attacks like RT-AL. In Appendix~\ref{subsec:generation}, we also analyze the influencing factors including adversarial test case generation and layer selection (for computing the testing metrics) via several calibration experiments. An analysis of how different levels of finetuning or pruning affect \dje is presented in Appendix~\ref{discuss}.

\vspace{0.5mm}
\noindent\textbf{Time Cost of \dje.}
The time cost of generating test cases using 1k seeds is provided in appendix Table \ref{tab:time}. For the black-box setting, we report the cost of PGD-based generation, while for the white-box setting, we report that of Algorithm~\ref{alg:neuron}. It shows that the time cost of white-box generation is slightly higher but is still very efficient in practice. The maximum time cost occurs on the SpeechCommands dataset for white-box generation, which is $\sim1.2$ hours. This time cost is regarded as \emph{efficient} since test case generation is a \emph{one-time effort}, and the additional time cost of scanning a suspect model with the test cases is almost negligible.

\begin{tcolorbox}[fonttitle = \bfseries]
  \textbf{Remark 1:} \textsc{DeepJudge} is effective and efficient in identifying finetuning and pruning copies.
\end{tcolorbox}



\subsubsection{Comparison with existing techniques} 
\label{subsec:comparison}
We compare \dje with three state-of-the-art copyright defense methods against model finetuning and pruning attacks. More details of these defense methods can be found in Appendix~\ref{subsec:watermarking}.

\begin{figure*}[]
 \centering
 \includegraphics[width=0.4\linewidth]{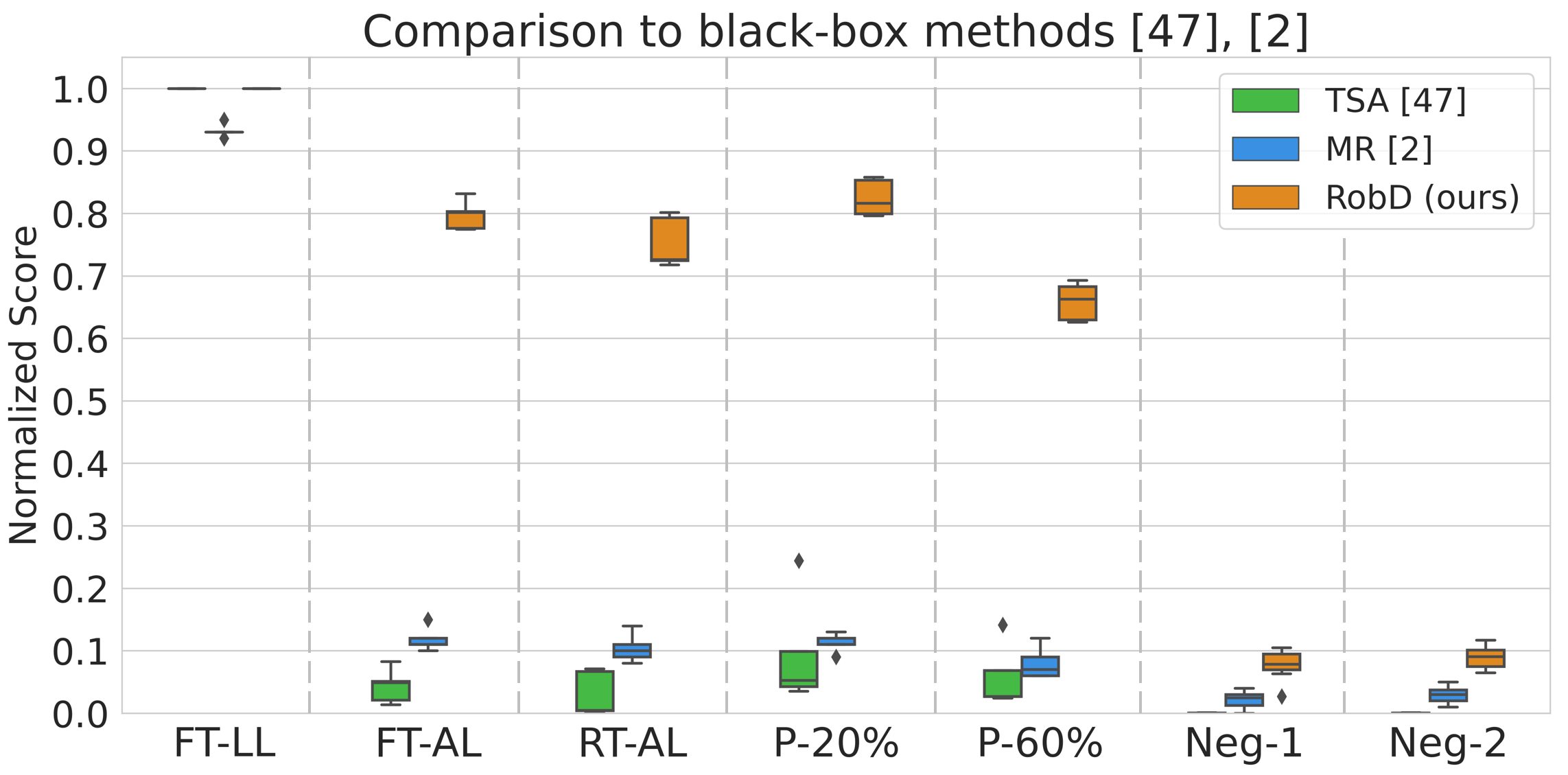}
 \hspace{8mm}
 \includegraphics[width=0.4\linewidth]{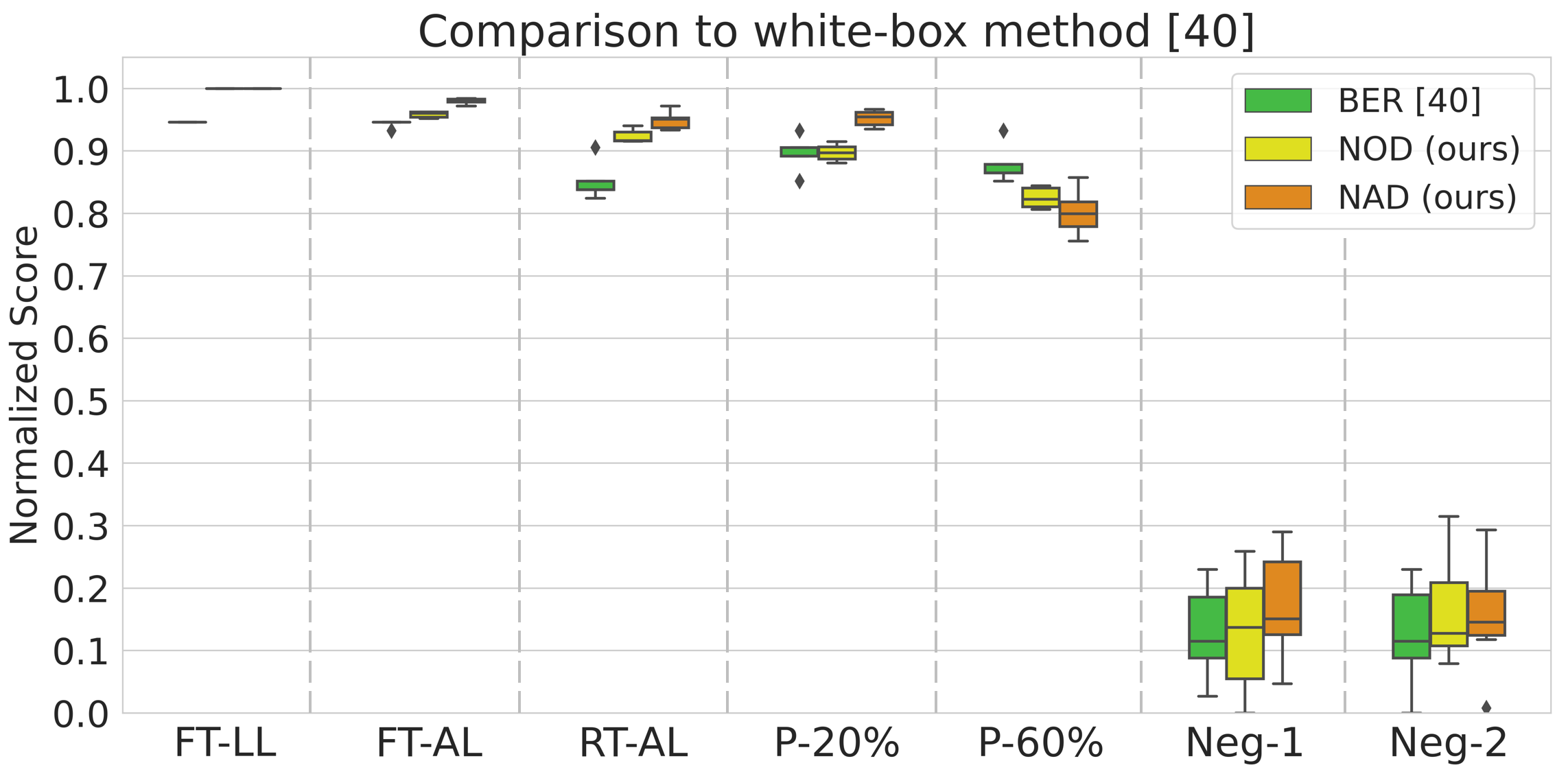}
 \caption{\dje vs. three state-of-the-art copyright defense methods. \emph{Left}: a comparison with \textbf{two black-box methods} \cite{zhang2018protecting, cao2021ipguard}; \emph{Right}: a comparison with \textbf{one white-box method} \cite{uchida2017embedding}. The results are normalized into $[0,1]$ for better visualization. The higher the normalized value, the better the identification of a positive suspect model.}
 \label{fig:baseline}
\end{figure*}

\vspace{0.5mm}
\noindent\textbf{Black-box: Comparison to Watermarking and Fingerprinting}. DNNWatermarking \cite{zhang2018protecting} is a black-box watermarking method based on backdoors, and IPGuard \cite{cao2021ipguard} is a black-box fingerprinting method based on targeted adversarial attacks. Here, we compare these two baselines with \dje in the black-box setting. For DNNWatermarking, we train the watermarked model (i.e., victim model) using additionally patched samples from scratch to embed the watermarks, and the \emph{TSA} (Trigger Set Accuracy) of the suspect model is calculated for ownership verification. 
IPGuard first generates targeted adversarial examples for the watermarked model then calculates the \emph{MR} (Matching Rate) (between the victim and the suspect) for verification. 
For \dje, we only apply the \emph{RobD} (robustness distance) metric here for a fair comparison. 

The left subfigure of Fig.~\ref{fig:baseline} visualizes the results. \dje demonstrates the best overall performance in this black-box setting. DNNWatermarking and IPGuard fail to identify the positive suspect models duplicated by FT-AL, RT-AL, P-20\% and P-60\%. Their scores (\emph{TSA} and \emph{MR}) drop drastically against these four attacks. This basically means that the embedded watermarks are completely removed, or the fingerprint can no longer be verified. While for the \emph{RobD} metric of \dje, the gap remains huge between the negative and positive suspects, demonstrating much better effectiveness to diverse finetuning and pruning attacks. 



\vspace{0.5mm}
\noindent \textbf{White-box: Comparison to Watermarking}. EmbeddingWatermark \cite{uchida2017embedding} is a white-box watermarking method based on signatures. It requires access to model parameters for signature extraction. We train the victim model with the embedding regularizer \cite{uchida2017embedding} from scratch to embed a 128-bits signature. The \emph{BER} (Bit Error Rate) is calculated and used to measure the verification performance. The right subfigure of Fig.~\ref{fig:baseline} visualizes the comparison results to two white-box \dje metrics \emph{NOD} and \emph{NAD}. 
The three metrics demonstrate a comparable performance with \emph{NAD} wins on 4 out of the 5 positive suspects. Note that the huge gap between the positives and negatives indicates that all metrics can correctly identify the positive suspects.
Here, a single metric of \dje was able to achieve the same level of protection as EmbeddingWatermark.

\begin{tcolorbox}[fonttitle = \bfseries]
  \textbf{Remark 2:} Compared to state-of-the-art defense methods, \dje performs better in the black-box setting and comparably in the white-box setting against model finetuning and pruning attacks, while not tampering with model training.
\end{tcolorbox}



\begin{table*}[t]
   \scriptsize
   \caption{Performance of \dje against model extraction attacks in the \textbf{black-box setting}. PGD \cite{madry2017towards} is used to generate adversarial test cases. ACC is the validation accuracy. For each metric, the values below (indicating `copy') or above (indicating `not copy') the threshold $\tau_\lambda$ (the last row) are highlighted in \redbox{red} (copy alert) and \greenbox{green} (no alert), respectively. `\textbf{Yes} (2/2)': two of the metrics vote for positive ($p_{copy}=100\%$); `\textbf{No} (0/2)': none of the metrics vote for positive ($p_{copy}=0\%$). See more details about the 3 extraction attacks in Appendix~\ref{subsec:extraction}.}
   \label{tab:steal}
   \centering
   \setlength\tabcolsep{1.2pt}
   \renewcommand\arraystretch{1.2}
   \begin{tabu}{cc|cccc|cccc|cccc} 
\tabucline[1pt]{-}
\multicolumn{2}{c|}{\multirow{2}{*}{\textbf{Model Type}}}                                                                     & \multicolumn{4}{c|}{\textbf{MNIST}}  & \multicolumn{4}{c|}{\textbf{CIFAR-10}} & \multicolumn{4}{c}{\textbf{SpeechCommands}} \\ \cline{3-14} 
\multicolumn{2}{c|}{}                                                                                                & ACC     & \emph{RobD}    & \emph{JSD} &\textbf{Copy?}  & ACC   & \emph{RobD}     & \emph{JSD}  &\textbf{Copy?}   & ACC     & \emph{RobD}    & \emph{JSD}  &\textbf{Copy?}   \\\tabucline[1pt]{-}

\multicolumn{1}{c|}{\multirow{3}{*}{\begin{tabular}[c]{@{}c@{}}Positive\\ Suspect\\ Models\end{tabular}}} & JBA      &    83.6$\pm$1.7\%    &  \greenbox{0.866$\pm$0.034}  &  \greenbox{0.596$\pm$0.006} & \textbf{No} (0/2)&   40.3$\pm$1.5\%      &    \redbox{0.497$\pm$0.044}      &    \greenbox{0.541$\pm$0.015} & \textbf{No} (1/2)   &    40.1$\pm$1.7\%     &  \redbox{0.381$\pm$0.030}       & \greenbox{0.470$\pm$0.011}  & \textbf{No} (1/2)      \\ \cline{2-14} 
\multicolumn{1}{c|}{}                                                                                     & Knock &   94.8$\pm$0.6\%  &    \redbox{0.491$\pm$0.032}   & \redbox{0.273$\pm$0.021}   & \textbf{Yes} (2/2)    &      74.4$\pm$1.0\%    &  \redbox{0.715$\pm$0.018}     &  \redbox{0.436$\pm$0.019}     & \textbf{Yes} (2/2)   &    86.6$\pm$0.5\%     &   \redbox{0.618$\pm$0.012}      &   \redbox{0.303$\pm$0.007}  & \textbf{Yes} (2/2)    \\ \cline{2-14} 
\multicolumn{1}{c|}{}                                                                                     & 
ESA  & 88.7$\pm$2.5\%    &   \redbox{0.175$\pm$0.056}      &  \redbox{0.141$\pm$0.042}  & \textbf{Yes} (2/2)     &  67.1$\pm$1.9\%        &  \redbox{0.144$\pm$0.031}    &   \redbox{0.249$\pm$0.033}   & \textbf{Yes} (2/2)     & $\times$  & $\times$   & $\times$    & --   \\ \tabucline[1pt]{-}

\multicolumn{1}{c|}{\multirow{3}{*}{\begin{tabular}[c]{@{}c@{}}Negative\\ Suspect\\ Models\end{tabular}}} & Neg-1    &  98.4$\pm$0.3\%       &  \greenbox{0.968$\pm$0.014}       &     \greenbox{0.614$\pm$0.016}  & \textbf{No} (0/2) & 84.2$\pm$0.6\%    &  \greenbox{0.920$\pm$0.021}        &  \greenbox{0.603$\pm$0.016}    & \textbf{No} (0/2)  &    94.9$\pm$0.7\%     &  \greenbox{0.817$\pm$0.025}       & \greenbox{0.456$\pm$0.014}  & \textbf{No} (0/2)     \\ \cline{2-14} 
\multicolumn{1}{c|}{}                                                                                     & Neg-2    &  98.3$\pm$0.2\%       &  \greenbox{0.949$\pm$0.029}       &     \greenbox{0.600$\pm$0.020} & \textbf{No} (0/2)  & 84.9$\pm$0.5\%    &   \greenbox{0.926$\pm$0.030}      &  \greenbox{0.615$\pm$0.021}    & \textbf{No} (0/2) &     94.5$\pm$0.8\%    &   \greenbox{0.832$\pm$0.024}      &   \greenbox{0.472$\pm$0.012}  & \textbf{No} (0/2)  \\ \cline{2-14} 
\multicolumn{1}{c|}{}                                                                                     & $\tau_\lambda$    & --      &    \textbf{0.852} &    \textbf{0.538}  & --   & --  &      \textbf{0.816}    &  \textbf{0.537}  & --     & --       &   \textbf{0.727}      &    \textbf{0.405}  & --    \\ \tabucline[1pt]{-}
\end{tabu}
\end{table*}

\subsection{Defending Against Model Extraction}
\label{subsec:extraction_exp}  
Model extraction (also known as model stealing) is considered to be a more challenging threat to DNN copyright. In this part, we evaluate \deepjudge against model extraction attacks, which has not been thoroughly studied in prior work.

\subsubsection{Attack strategies}
We consider model extraction with two different types of supporting data: auxiliary or synthetic (see Section~\ref{sec:threat_model}). We consider the following state-of-the-art model extraction attacks: \textbf{a) JBA (Jacobian-Based Augmentation \cite{papernot2017practical})} samples a set of seeds from the test dataset, then applies Jacobian-based data augmentation to synthesize more data from the seeds. \textbf{b) Knockoff (Knockoff Nets \cite{orekondy2019knockoff})} works with an auxiliary dataset that shares similar attributes as the original training data used to train the victim model. \textbf{c) ESA (ES Attack \cite{yuan2020attack})} requires no additional data but a huge amount of queries. ESA utilizes an adaptive gradient-based optimization algorithm to synthesize data from random noise. ESA could be applied in scenarios where it is hard to access the task domain data, such as personal health data. With the extracted data, the adversary can train a new model from scratch, assuming knowledge of the victim model's architecture. The new model is considered as a successful stealing if its performance matches with the victim model.

\subsubsection{Failure of watermarking}
\label{subsubsec:fail}
Our experiments in Section \ref{subsec:tuning} show the effectiveness and robustness of watermarking to finetuning and pruning attacks. Unfortunately, here we show that the embedded watermarks can be removed by model extraction attacks. We show the results of DNNWatermarking and EmbeddingWatermark in Fig.~\ref{fig:failure}.
The extracted models by different extraction attacks all differ greatly from the victim model according to either TSA (from DNNWatermarking) or BER (from EmbeddingWatermark). 
For example, the TSA value for the victim model is 100\%, however, the TSA values for the three extracted copies are all below 1\%.
This basically means that the original watermarks are all erased in the extracted models. It will inevitably lead to failed ownership claims. This is somewhat not too surprising as watermarks are task-irrelevant contents and not the focus of model extraction.

\subsubsection{Effectiveness of \dje} 
\label{subsec:valid_extraction}
Table~\ref{tab:steal} summarizes the results of \dje, which successfully identifies all positive suspect models, except when the stolen copies (by JBA) have extremely poor performance with 15\%, 44\% and 55\% lower accuracy than the corresponding victim model. We note that model extraction does not always work, and poorly performed extractions are less likely to pose a real threat. We also observe that \deepjudge works better when the extraction is better, which therefore counters the ultimate perfect matching goal of model extraction attacks.

Compared to model finetuning or pruning, the average \emph{RobD} and \emph{JSD} values on extracted models are relatively larger, meaning that the decision boundaries of extracted models are more different from that of the victim model. The reason is that extracted models are often trained from a random point, while finetuning only slightly shifts the original boundary of the victim model, as depicted in Fig.~\ref{fig:adv}.
As such, model extraction is more stealthy and more challenging for ownership verification. 
Nonetheless, the two metrics \emph{RobD} and \emph{JSD}, can still reveal the unique similarities (smaller values) of the extracted models to the victim model: the better the extraction (higher accuracy of the extracted model), the lower the \emph{RobD} and \emph{JSD} values. This indicates that the extracted model behaves more similarly to the victim as its decision boundary gradually approaching that of the victim, and also highlights the unique advantage of \dje against model extraction attacks. Note that JBA attack can only extract 50\% of the original accuracy on either CIFAR-10 or SpeechCommands, which should not be considered as successful extractions.



\begin{figure}[t]
   \centering
   \includegraphics[width=0.49\linewidth]{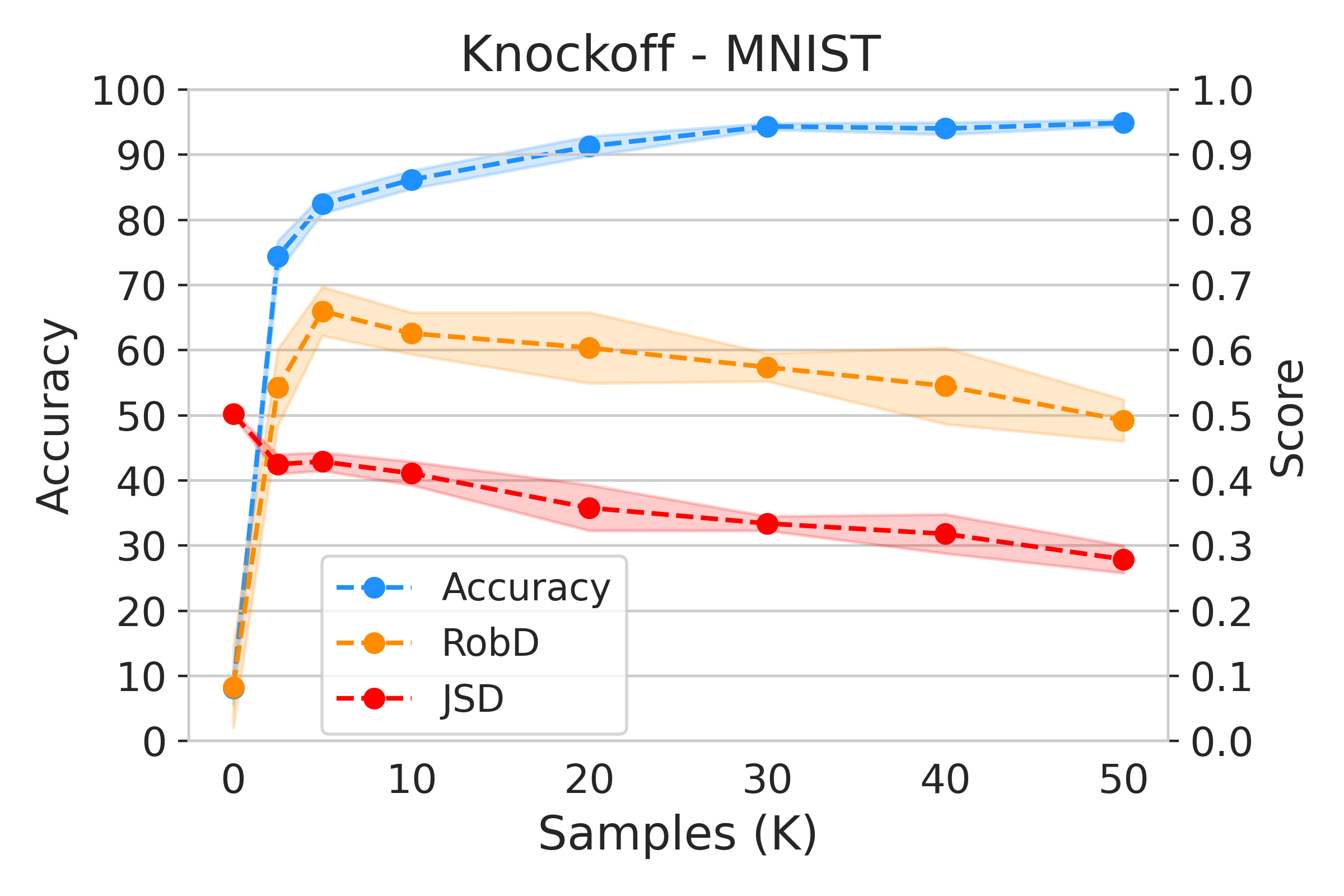}
   \includegraphics[width=0.49\linewidth]{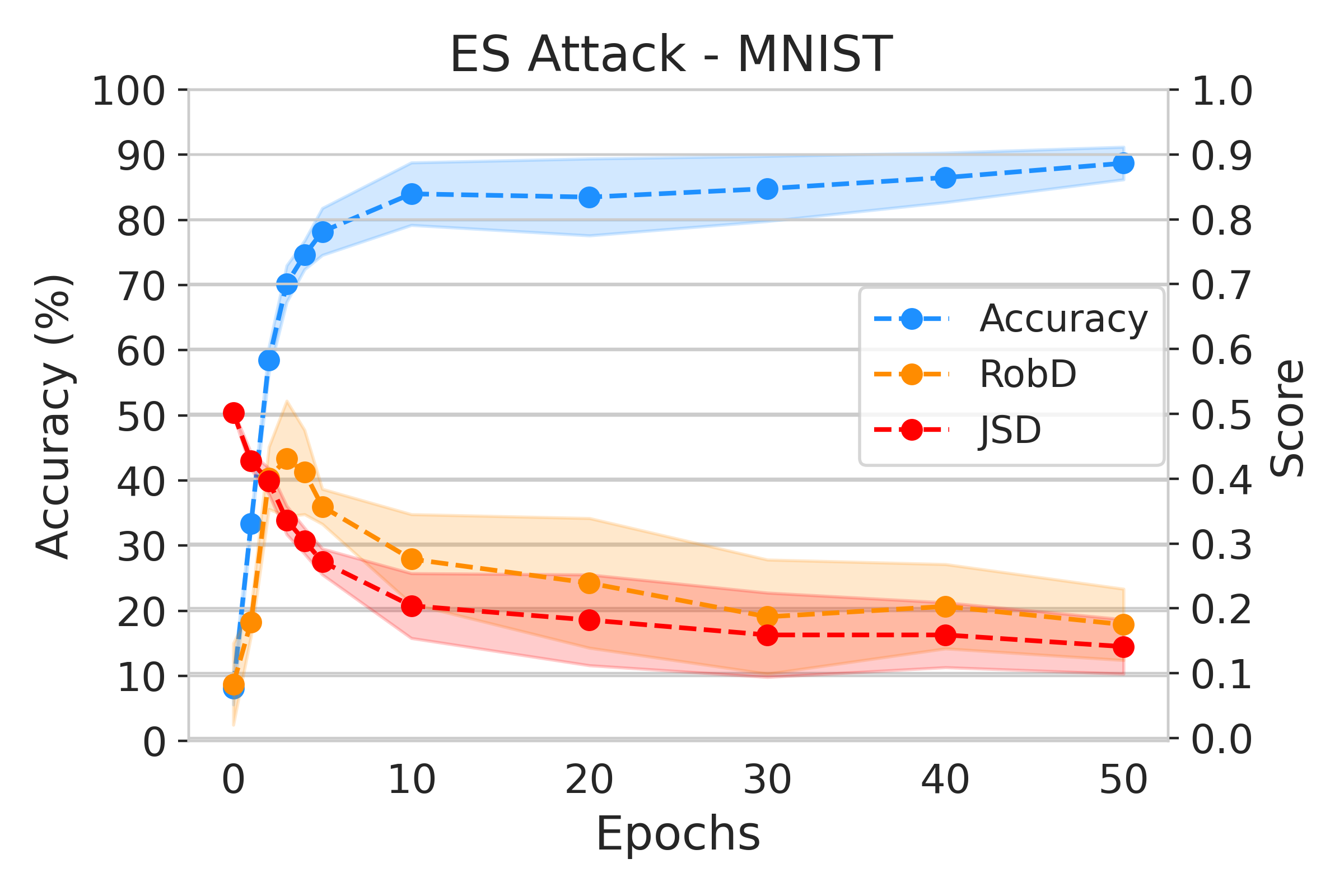}
   \includegraphics[width=0.49\linewidth]{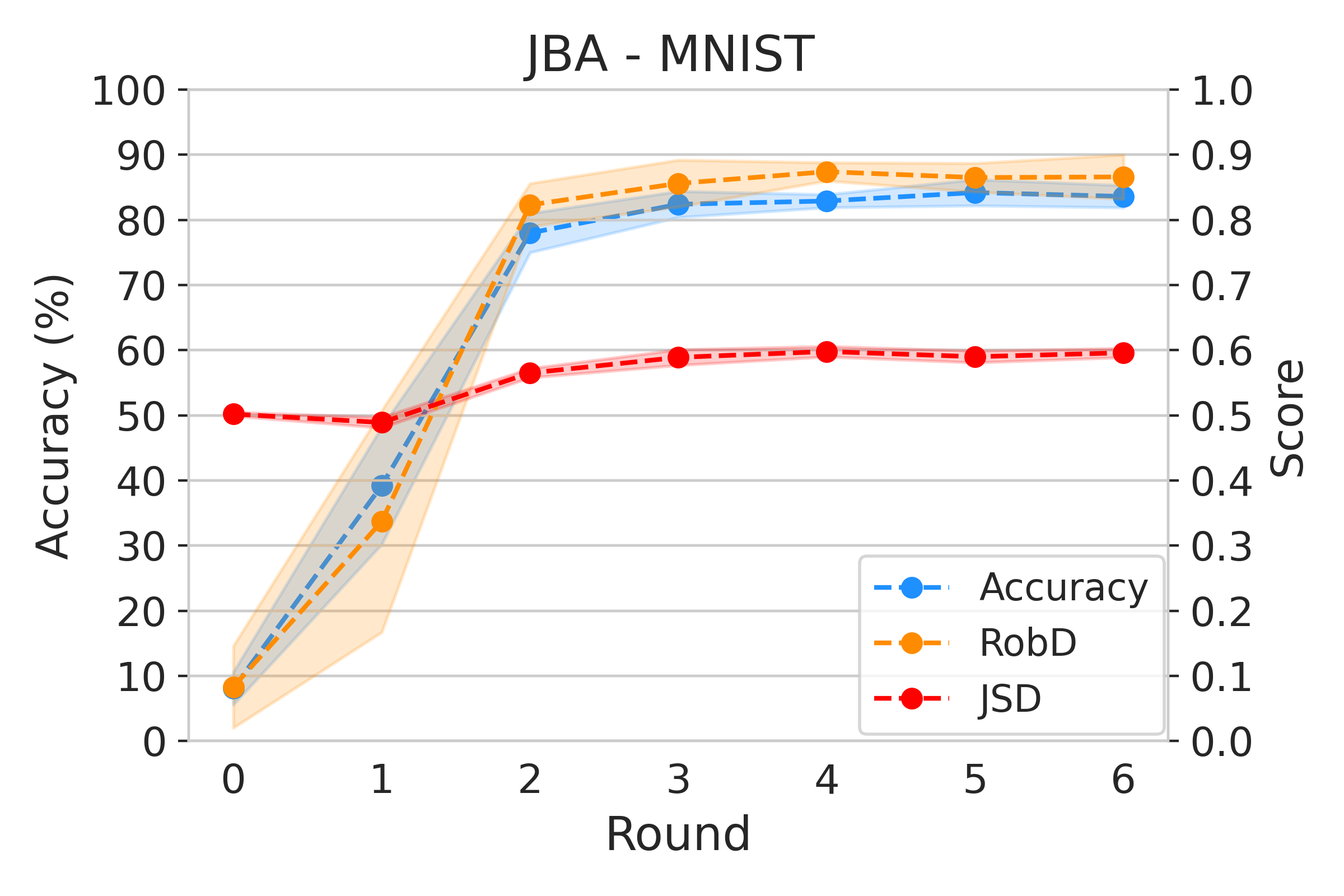}
   \includegraphics[width=0.49\linewidth]{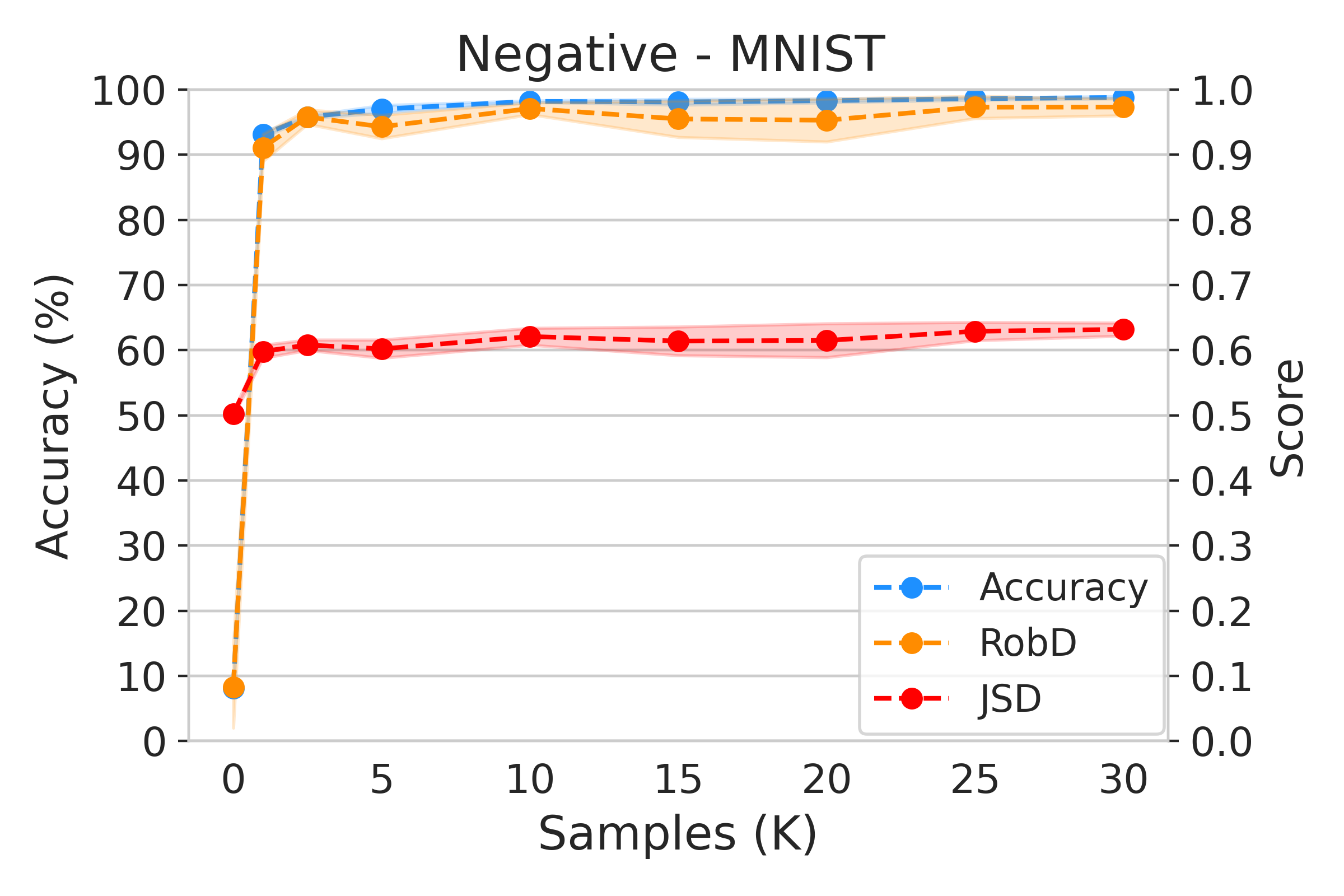}
   \caption{The \emph{RobD} (orange line) and \emph{JSD} (red line) scores between the victim and extracted models throughout the entire extraction procedure (defined by sample sizes, epochs or rounds) on MNIST.}
   \label{fig:mnist_extraction}
\end{figure}

In Fig.~\ref{fig:mnist_extraction}, we further show the evolution of the \emph{RobD} and \emph{JSD} values throughout the entire extraction process of Knockoff, ESA and JBA attacks. We find that both \emph{RobD} (orange line) and \emph{JSD} (red line) values decrease as the extraction progresses, again, except for JBA. This confirms our speculation that, when tested by \dje, a better extracted model will expose more similarities to its victim. By contrast, we also study how these two values change during the training process of the negative model in Fig.~\ref{fig:mnist_extraction}, which shows that the independently trained negative suspect models tend to vary more from the victim model and produce higher \emph{RobD} and \emph{JSD} values.

\begin{tcolorbox}[fonttitle = \bfseries]
  \textbf{Remark 3:} Model extraction attacks are more challenging than finetuning or pruning attacks, however, \deepjudge can still correctly identify those successful extractions. Moreover, the better the extraction, the easier the extracted model will be identified by \dje as a stolen copy.
\end{tcolorbox}

\vspace{-2mm}
\section{Robustness to Adaptive Attackers}
\label{sec:adaptive_exp}
In this section, we explore potential adaptive attacks to \dje based on the adversary's knowledge of \dje: 1) the adversary knows the testing metrics and the test cases, or 2) the adversary only knows the testing metrics. Contrast evaluation of watermarking \& fingerprinting against similar adaptive attacks are in Appendix~\ref{subsec:adapt123}.

\vspace{-2mm}
\subsection{Knowing Both Testing Metrics and Test Cases}
\label{sec:adaptive_attack1}
In this threat model, the adversary has full knowledge of \dje including the testing metrics $\Lambda$ and the secret test cases $T$. We also assume the adversary has a subset of clean data. In \dje, we have two test settings, i.e., white-box testing and black-box testing. The two testings differ in the testing metrics and the generated test cases (see examples in Fig.~\ref{fig:sample}). The black-box test cases are labeled. Therefore, the adversary can mix $T$ into its clean subset to finetune the stolen model to have large testing distances (i.e., black-box testing metrics \emph{RobD} and \emph{JSD}) while maintaining good classification performance. This will fool \dje to identify the stolen model to be significantly different from the victim model. This adaptive attack against black-box testing is denoted by \emph{Adapt-B}. Since the white-box test cases are unlabeled, the adversary can use the predicted labels (by the victim model) as ground-truth and finetunes the stolen model following a similar procedure as \emph{Adapt-B}. This attack against white-box testing is denoted by \emph{Adapt-W}. Note that the suffix `\emph{-B/-W}' marks the target testing setting to attack, while both attacks are white-box adaptive attacks knowing all the information.

\begin{figure}[t]
    \centering
    \includegraphics[width=0.46\linewidth]{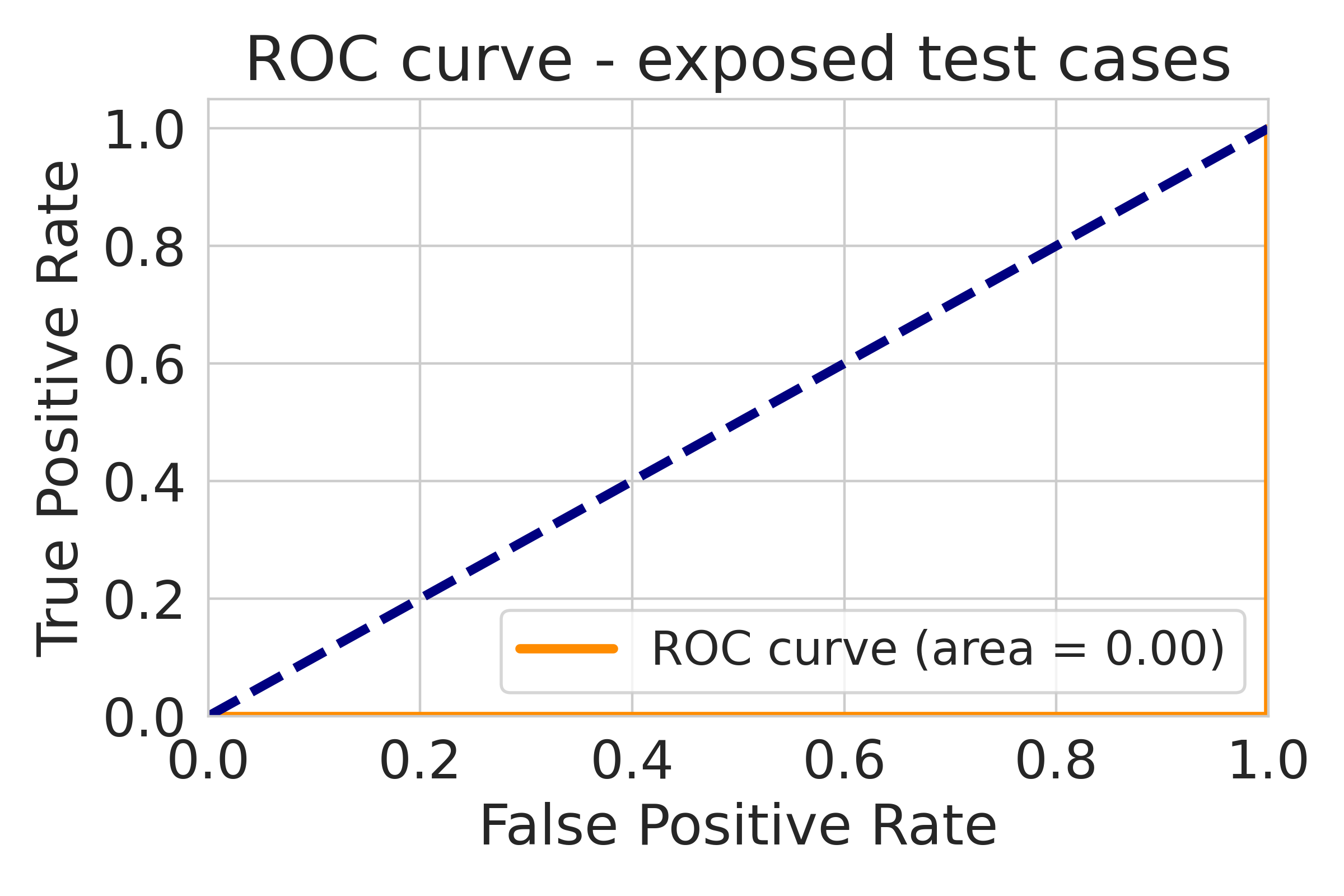}
    \hspace{2mm}
    \includegraphics[width=0.46\linewidth]{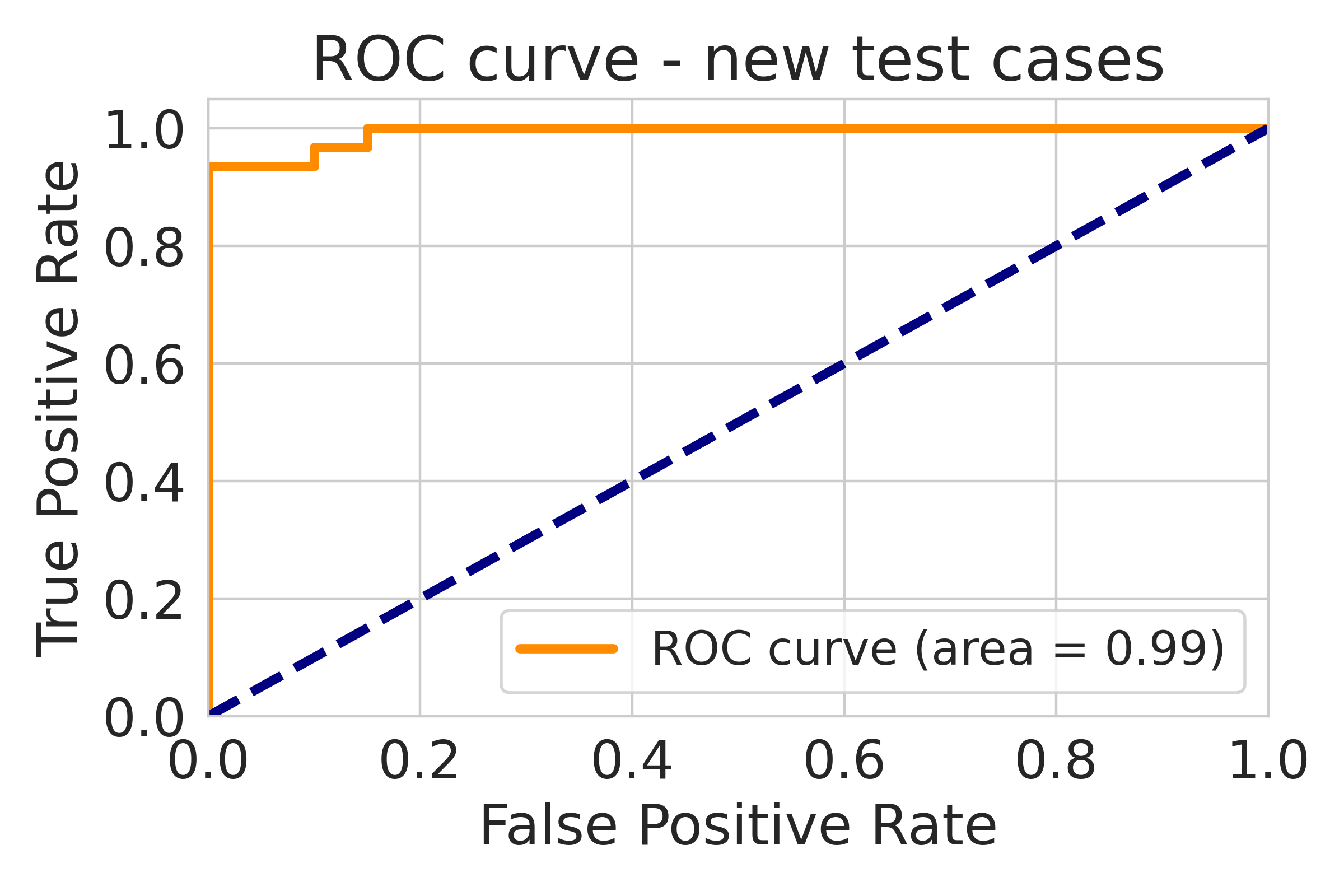}
    \caption{Detection ROC curve of \emph{RobD} with exposed (left) and new (right) test cases against \emph{Adapt-B} attack on CIFAR-10.}
    \label{fig:old}
\end{figure}

The results of \dje using the exposed test cases $T$ are reported in Table \ref{tab:adapt}. It shows that: 1) \dje is robust to \emph{Adapt-W}, which fails to maximize the output distance and activation distance simultaneously nor maintaining the original classification accuracy; 2) though \dje is not robust to \emph{Adapt-B} when the test cases are exposed with labels, it can easily recover the performance with new test cases generated with different seeds (see the ROC curves on the exposed and new test cases in Fig.~\ref{fig:old}); and 3) \dje can still correctly identify the stolen copies by \emph{Adapt-B} when combining black-box and white-box testings (the final judgements are all correct). Comparing the non-trivial effort of retraining/finetuning a model to the efficient generation of new test cases, \dje holds a clear advantage in the arms race against finetuning-based adaptive attacks. 

It is noteworthy that \emph{Adapt-W} did not break all white-box metrics of \dje, since the mechanism of white-box testing is robust. Specifically, black-box testing characterizes the behaviors of the output layer, while white-box testing characterizes the internal behaviors of more shallow layers. Due to the over-parameterization property of DNNs, it is relatively easy to fine-tune the model to overfit the set of black-box test cases, subverting the results of the black-box metrics. However, in white-box testing, changing the activation status of all hidden neurons on the set of white-box test cases is almost impossible without completely retraining the model. Therefore, white-box testing is inherently more robust to adaptive attacks, especially when the test cases are exposed.

\subsection{Knowing Only the Testing Metrics}
\label{sec:adaptive_attack2}
In this threat model, the adversary can still adapt in different ways. We consider two adaptive attacks: \emph{adversarial training} targeting on black-box testing and a general \emph{transfer learning} attack on white-box testing, respectively. 

\subsubsection{Blind adversarial training} 
\label{subsubsec:advtrain}
\begin{figure}[]
    \centering
    \includegraphics[width=0.46\linewidth]{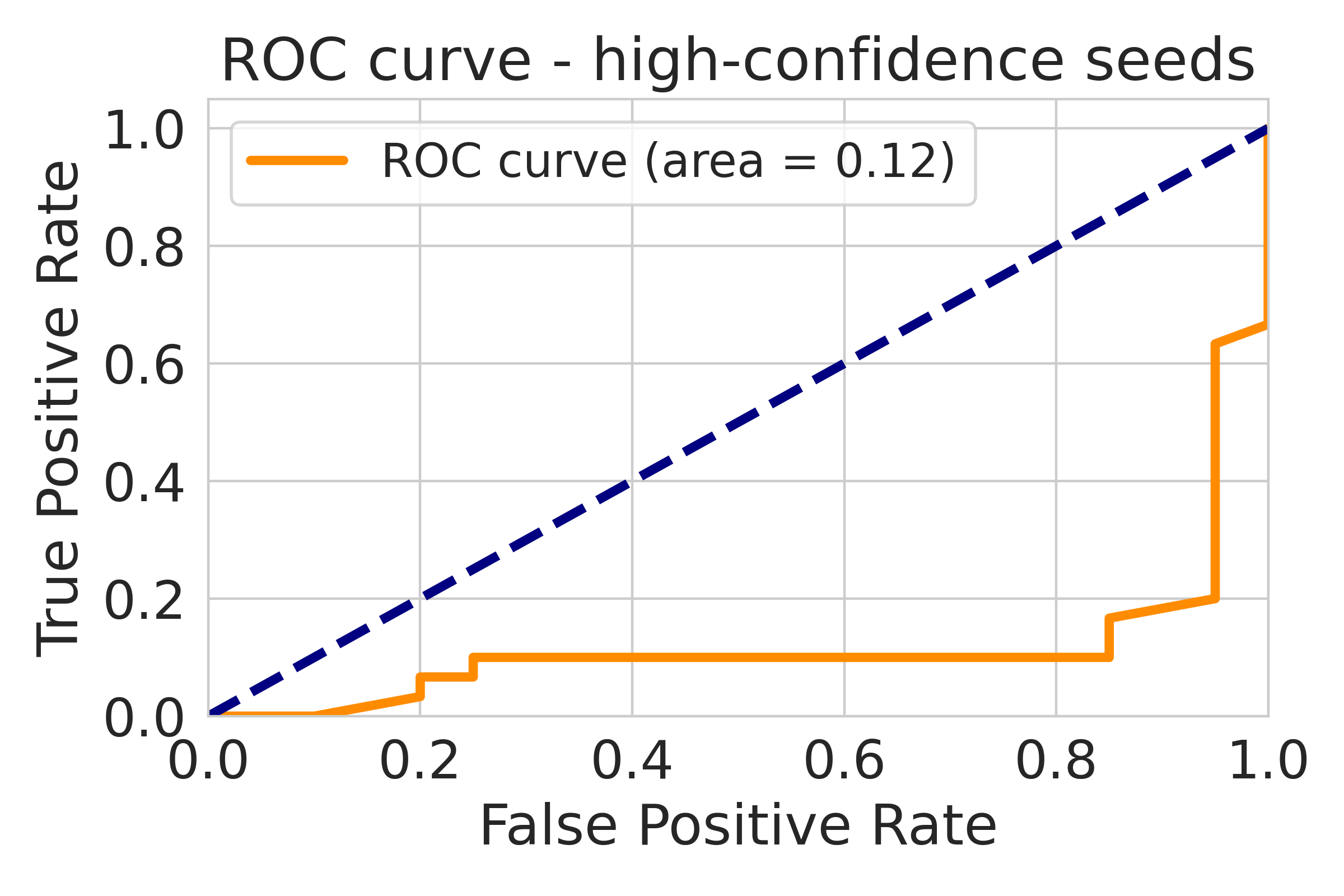}
    \hspace{2mm}
    \includegraphics[width=0.46\linewidth]{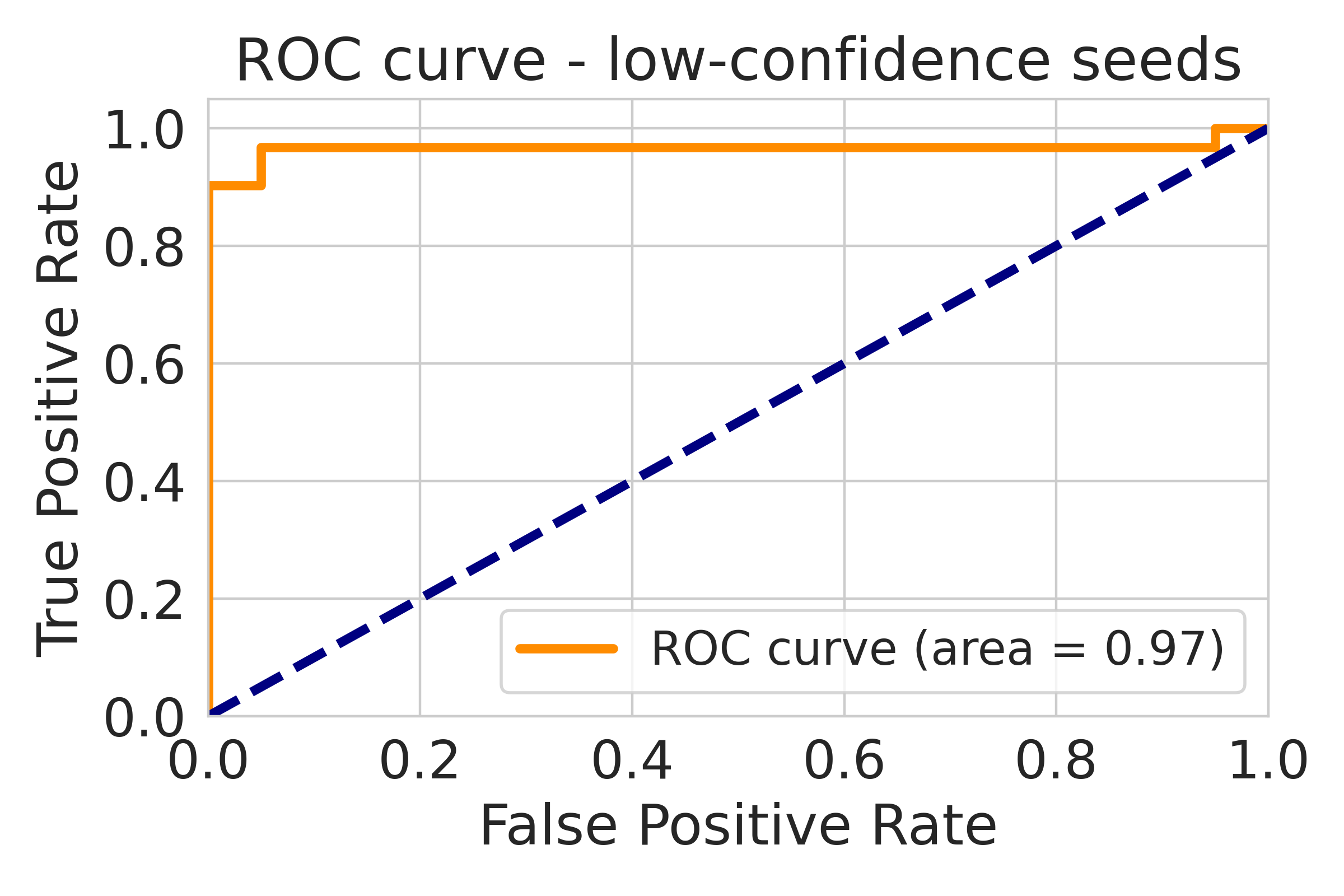}
    \caption{Detection ROC curve of \emph{RobD} with adversarial test cases generated from high-confidence (left) or low-confidence seeds (right) against \emph{{Adv-Train}} attack on CIFAR-10.}
    \label{fig:advtrain}
    \vspace{-0.1in}
\end{figure}

Since our black-box testing mainly relies on probing the decision boundary difference using adversarial test cases, the adversary may utilize adversarial training to improve the robustness of the stolen copy. Given the PGD parameters and a subset of clean data (20\% of the original training data), the adversary iteratively trains the stolen model to smooth the model decision boundaries following \cite{madry2017towards}. This type of adaptive attack is denoted by \emph{Adv-Train}. As Table \ref{tab:adapt} shows, it can indeed circumvent our black-box testing, with a sacrifice of $\sim10\%$ performance (a phenomenon known as accuracy-robustness trade-off \cite{tsipras2018robustness,zhang2019theoretically}). 
However, interestingly, if we replace the high-confidence seeds used in \dje with low-confidence seeds, \dje becomes effective again (as shown in Fig.~\ref{fig:advtrain}). One possible reason is that, compared to high-confidence seeds, these low-confidence seeds are natural boundary (hard) examples that are close to the decision boundary, thus can generate more test cases to cross the adversarially smoothed decision boundary
within certain perturbation budget.
Examples of high/low confidence test seeds are provided in Fig.~\ref{fig:gini-contrast}. It is also worth mentioning that our white-box testing still performs well in this case. Overall, \dje is robust to \emph{Adv-Train} or at least can be made robust by efficiently updating the seeds.


\begin{table*}[]
\footnotesize
   \caption{Performance of \dje against several adaptive attacks on the CIFAR-10 dataset. \emph{Adapt-B}: adaptive attack against black-box testing; \emph{Adapt-W}: adaptive attack against white-box testing; \emph{Adv-Train}: adversarial training, \emph{VTL}: vanilla transfer learning. For each metric, the values below (indicating `copy') or above (indicating `not copy') the threshold $\tau_\lambda$ (the last row) are highlighted in \redbox{red} (copy alert) and \greenbox{green} (no alert) respectively.}
   \label{tab:adapt}
   \centering
   \renewcommand\arraystretch{1.2}
   \setlength\tabcolsep{4pt}

\begin{tabu}{c|c|cccccccc}
\tabucline[1pt]{-}
\multicolumn{2}{c|}{\multirow{2}{*}{\textbf{Model Type}}}     &                                            & \multicolumn{2}{c}{\textbf{Black-box Testing}} & \multicolumn{4}{c}{\textbf{White-box Testing}} \\ \cline{3-10} 
\multicolumn{2}{c|}{}                                                                            & ACC           & \emph{RobD}           & \emph{JSD}           & \emph{NOD}       & \emph{NAD}        & \emph{LOD}        & \emph{LAD}   & \textbf{Copy?}     \\ \tabucline[1pt]{-}
\multirow{4}{*}{\begin{tabular}[c]{@{}c@{}}Positive\\ Suspect\\ Models\end{tabular}} 
& Adapt-B   &  81.4$\pm$0.9\%    & \greenbox{0.985$\pm$0.011}  &   \greenbox{0.665$\pm$0.007}  &   \redbox{0.38$\pm$0.04}    &    \redbox{0.44$\pm$0.15}        &   \redbox{1.12$\pm$0.06}         &  \redbox{0.40$\pm$0.05}   & \textbf{Yes} (4/6)        \\ \cline{2-10} 
& Adapt-W   & 71.9$\pm$1.8\%    & \redbox{0.519$\pm$0.048}    & \redbox{0.372$\pm$0.025}  & \greenbox{3.11$\pm$0.12} & \redbox{1.94$\pm$0.12}  & \greenbox{11.62$\pm$0.54} &  \redbox{1.89$\pm$0.33}  & \textbf{Yes} (4/6)   \\ \cline{2-10} 
& Adv-Train &      74.5$\pm$2.3\%         & \greenbox{0.939$\pm$0.087}      &  \greenbox{0.637$\pm$0.036}             &  \redbox{0.68$\pm$0.11}         &   \redbox{0.79$\pm$0.17}         &    \redbox{1.89$\pm$0.14}        &   \redbox{0.75$\pm$0.08}    & \textbf{Yes} (4/6)      \\ \cline{2-10} 
& VTL  & 93.3$\pm$1.7\%    & $\times$            & $\times$             & \redbox{0.85$\pm$0.23} & \redbox{1.08$\pm$0.14}  & \redbox{2.58$\pm$0.24}  & \redbox{0.64$\pm$0.15}  & \textbf{Yes} (4/4)  \\ \tabucline[1pt]{-}

\multirow{3}{*}{\begin{tabular}[c]{@{}c@{}}Negative\\ Suspect\\ Models\end{tabular}} 
& Neg-1     & 84.2$\pm$0.6\%    & \greenbox{0.920$\pm$0.021}   & \greenbox{0.603$\pm$0.016}   & \greenbox{3.09$\pm$0.30} & \greenbox{10.94$\pm$1.74} & \greenbox{11.85$\pm$1.01} & \greenbox{5.41$\pm$0.67}  & \textbf{No} (0/6)  \\ \cline{2-10} 
& Neg-2     & 84.9$\pm$0.5\%    & \greenbox{0.926$\pm$0.030}    & \greenbox{0.615$\pm$0.021}   & \greenbox{3.21$\pm$0.18} & \greenbox{11.09$\pm$0.71} & \greenbox{12.60$\pm$1.33} & \greenbox{5.37$\pm$0.72} & \textbf{No} (0/6)   \\ \cline{2-10} 
& $\tau_\lambda$         & --             & \textbf{0.816}    & \textbf{0.537}   & \textbf{1.79} & \textbf{6.14} & \textbf{6.89} & \textbf{3.01} & -- \\ \tabucline[1pt]{-}
\end{tabu} 

\end{table*}

\subsubsection{Transfer learning} 
\label{subsubsec:vtl}
The adversary may transfer the stolen copy of the victim model to a new dataset. The adversary exploits the main structure of the victim model as a backbone and adds more layers to it. 
Here, we test a vanilla transfer learning (\emph{VTL}) strategy from the 10-class CIFAR-10 to a 5-class SVHN \cite{netzer2011reading}. The last layer of the CIFAR-10 victim model is first replaced by a new classification layer. We then fine-tune all layers on the subset of SVHN data.
Note that, in this setting, the black-box metrics are no longer feasible since the suspect model has different output dimensions to the victim model, however, the white-box metrics can still be applied since the shallow layers are kept. The results are reported in Table~\ref{tab:adapt}. Remarkably, \dje succeeds in identifying transfer learning attacks with distinctively low testing distances and an $AUC=1$.

In one recent work \cite{DBLP:conf/iclr/MainiYP21}, it was observed that the knowledge of the victim model could be transferred to the stolen models. Dataset Inference (DI) technique was then proposed to probe whether the victim's knowledge (i.e., private training data) is preserved in the suspect model. We believe such knowledge-level testing metrics could also be incorporated into \dje to make it more comprehensive. An analysis of how different levels of transfer learning could affect \dje can be found in Appendix~\ref{discuss}.



\vspace{1mm}
\begin{tcolorbox}[fonttitle = \bfseries]
\textbf{Remark 4:} \dje is fairly robust to adversarial finetuning, adversarial training or transfer learning based adaptive attacks, although sometimes it needs to regenerate the seeds or test cases.
\end{tcolorbox}

\vspace{1mm}
\section{Conclusion}
\label{sec:conclusion}
In this work, we proposed \dje, a novel testing framework for copyright protection of deep learning models. The core of \dje is a family of multi-level testing metrics that characterize different aspects of similarities between the victim model and a suspect model.
Efficient and flexible test case generation methods are also developed in \dje to help boost the discriminating power of the testing metrics.
Compared to watermarking methods, \dje does not need to tamper with the model training process. Compared to fingerprinting methods, it can defend more diverse attacks and is more resistant to adaptive attacks.
\dje is applicable in both black-box and white-box settings against model finetuning, pruning and extraction attacks.
Extensive experiments on multiple benchmark datasets demonstrate the effectiveness and efficiency of \dje. 
We have implemented \dje as a self-contained open-source toolkit.
As a generic testing framework, new testing metrics or test case generation methods can be effortlessly incorporated into \dje to help defend future threats to deep learning copyright protection.

\vspace{1mm}
\section*{Acknowledgement}
We are grateful to the anonymous reviewers and shepherd for their valuable comments. This research was supported by the Key R\&D Program of Zhejiang (2022C01018) and the NSFC Program (62102359, 61833015).

\clearpage
\bibliographystyle{plain}
\bibliography{bare_conf.bib}

\begin{thebibliography}{10}

\bibitem{adi2018turning}
Yossi Adi, Carsten Baum, Moustapha Cisse, Benny Pinkas, and Joseph Keshet.
\newblock Turning your weakness into a strength: Watermarking deep neural
  networks by backdooring.
\newblock In {\em USENIX Security}, pages 1615--1631, 2018.

\bibitem{cao2021ipguard}
Xiaoyu Cao, Jinyuan Jia, and Neil~Zhenqiang Gong.
\newblock {IPGuard}: Protecting intellectual property of deep neural networks
  via fingerprinting the classification boundary.
\newblock In {\em Asia CCS}, pages 14--25, 2021.

\bibitem{carlini2020cryptanalytic}
Nicholas Carlini, Matthew Jagielski, and Ilya Mironov.
\newblock Cryptanalytic extraction of neural network models.
\newblock In {\em CRYPTO}, pages 189--218. Springer, 2020.

\bibitem{carlini2017towards}
Nicholas Carlini and David Wagner.
\newblock Towards evaluating the robustness of neural networks.
\newblock In {\em S\&P}, pages 39--57. IEEE, 2017.

\bibitem{chen2015deepdriving}
Chenyi Chen, Ari Seff, Alain Kornhauser, and Jianxiong Xiao.
\newblock {DeepDriving}: Learning affordance for direct perception in
  autonomous driving.
\newblock In {\em ICCV}, pages 2722--2730, 2015.

\bibitem{choi2017kapre}
Keunwoo Choi, Deokjin Joo, and Juho Kim.
\newblock Kapre: On-gpu audio preprocessing layers for a quick implementation
  of deep neural network models with keras.
\newblock In {\em Machine Learning for Music Discovery Workshop at ICML}, 2017.

\bibitem{collobert2011natural}
Ronan Collobert, Jason Weston, L{\'e}on Bottou, Michael Karlen, Koray
  Kavukcuoglu, and Pavel Kuksa.
\newblock Natural language processing (almost) from scratch.
\newblock {\em Journal of machine learning research}, 12(ARTICLE):2493--2537,
  2011.

\bibitem{correia2018copycat}
Jacson~Rodrigues Correia-Silva, Rodrigo~F Berriel, Claudine Badue, Alberto~F
  de~Souza, and Thiago Oliveira-Santos.
\newblock Copycat {CNN}: Stealing knowledge by persuading confession with
  random non-labeled data.
\newblock In {\em IJCNN}, pages 1--8. IEEE, 2018.

\bibitem{darvish2019deepsigns}
Bita Darvish~Rouhani, Huili Chen, and Farinaz Koushanfar.
\newblock {DeepSigns}: an end-to-end watermarking framework for ownership
  protection of deep neural networks.
\newblock In {\em ASPLOS}, pages 485--497, 2019.

\bibitem{fan2019rethinking}
Lixin Fan, Kam~Woh Ng, and Chee~Seng Chan.
\newblock Rethinking deep neural network ownership verification: Embedding
  passports to defeat ambiguity attacks.
\newblock 2019.

\bibitem{fawzi2017robustness}
Alhussein Fawzi, Seyed-Mohsen Moosavi-Dezfooli, and Pascal Frossard.
\newblock The robustness of deep networks: A geometrical perspective.
\newblock {\em IEEE Signal Processing Magazine}, 34(6):50--62, 2017.

\bibitem{feng2020deepgini}
Yang Feng, Qingkai Shi, Xinyu Gao, Jun Wan, Chunrong Fang, and Zhenyu Chen.
\newblock {DeepGini}: prioritizing massive tests to enhance the robustness of
  deep neural networks.
\newblock In {\em ISSTA}, pages 177--188, 2020.

\bibitem{fuglede2004jensen}
Bent Fuglede and Flemming Topsoe.
\newblock Jensen-shannon divergence and hilbert space embedding.
\newblock In {\em International Symposium on Information Theory}, page~31.
  IEEE, 2004.

\bibitem{goodfellow2014explaining}
Ian~J Goodfellow, Jonathon Shlens, and Christian Szegedy.
\newblock Explaining and harnessing adversarial examples.
\newblock {\em arXiv preprint arXiv:1412.6572}, 2014.

\bibitem{graves2013speech}
Alex Graves, Abdel-rahman Mohamed, and Geoffrey Hinton.
\newblock Speech recognition with deep recurrent neural networks.
\newblock In {\em ICASSP}, pages 6645--6649. IEEE, 2013.

\bibitem{gu2017badnets}
Tianyu Gu, Kang Liu, Brendan Dolan-Gavitt, and Siddharth Garg.
\newblock {BadNets}: Evaluating backdooring attacks on deep neural networks.
\newblock {\em IEEE Access}, 7:47230--47244, 2019.

\bibitem{guo2020hidden}
Shangwei Guo, Tianwei Zhang, Han Qiu, Yi~Zeng, Tao Xiang, and Yang Liu.
\newblock The hidden vulnerability of watermarking for deep neural networks.
\newblock {\em arXiv preprint arXiv:2009.08697}, 2020.

\bibitem{he2016deep}
Kaiming He, Xiangyu Zhang, Shaoqing Ren, and Jian Sun.
\newblock Deep residual learning for image recognition.
\newblock In {\em CVPR}, pages 770--778, 2016.

\bibitem{jagielski2020high}
Matthew Jagielski, Nicholas Carlini, David Berthelot, Alex Kurakin, and Nicolas
  Papernot.
\newblock High accuracy and high fidelity extraction of neural networks.
\newblock In {\em USENIX Security}, pages 1345--1362, 2020.

\bibitem{jia2021entangled}
Hengrui Jia, Christopher~A Choquette-Choo, Varun Chandrasekaran, and Nicolas
  Papernot.
\newblock Entangled watermarks as a defense against model extraction.
\newblock In {\em USENIX Security}, 2021.

\bibitem{juuti2019prada}
Mika Juuti, Sebastian Szyller, Samuel Marchal, and N~Asokan.
\newblock {PRADA}: protecting against dnn model stealing attacks.
\newblock In {\em EuroS\&P}, pages 512--527. IEEE, 2019.

\bibitem{krizhevsky2009learning}
Alex Krizhevsky, Geoffrey Hinton, et~al.
\newblock Learning multiple layers of features from tiny images.
\newblock 2009.

\bibitem{le2020adversarial}
Erwan Le~Merrer, Patrick Perez, and Gilles Tr{\'e}dan.
\newblock Adversarial frontier stitching for remote neural network
  watermarking.
\newblock {\em Neural Computing and Applications}, 32(13):9233--9244, 2020.

\bibitem{lecun2010mnist}
Yann LeCun, Corinna Cortes, and CJ~Burges.
\newblock {MNIST} handwritten digit database.
\newblock 2010.

\bibitem{liu2018fine}
Kang Liu, Brendan Dolan-Gavitt, and Siddharth Garg.
\newblock Fine-pruning: Defending against backdooring attacks on deep neural
  networks.
\newblock In {\em International Symposium on Research in Attacks, Intrusions,
  and Defenses}, pages 273--294. Springer, 2018.

\bibitem{liu2018rethinking}
Zhuang Liu, Mingjie Sun, Tinghui Zhou, Gao Huang, and Trevor Darrell.
\newblock Rethinking the value of network pruning.
\newblock {\em arXiv preprint arXiv:1810.05270}, 2018.

\bibitem{DBLP:conf/iclr/LukasZK21}
Nils Lukas, Yuxuan Zhang, and Florian Kerschbaum.
\newblock Deep neural network fingerprinting by conferrable adversarial
  examples.
\newblock In {\em ICLR}, 2021.

\bibitem{madry2017towards}
Aleksander Madry, Aleksandar Makelov, Ludwig Schmidt, Dimitris Tsipras, and
  Adrian Vladu.
\newblock Towards deep learning models resistant to adversarial attacks.
\newblock {\em arXiv preprint arXiv:1706.06083}, 2017.

\bibitem{DBLP:conf/iclr/MainiYP21}
Pratyush Maini, Mohammad Yaghini, and Nicolas Papernot.
\newblock Dataset inference: Ownership resolution in machine learning.
\newblock In {\em ICLR}, 2021.

\bibitem{mehrabi2019survey}
Ninareh Mehrabi, Fred Morstatter, Nripsuta Saxena, Kristina Lerman, and Aram
  Galstyan.
\newblock A survey on bias and fairness in machine learning.
\newblock {\em arXiv preprint arXiv:1908.09635}, 2019.

\bibitem{netzer2011reading}
Y~Netzer, T~Wang, A~Coates, A~Bissacco, B~Wu, and AY~Ng.
\newblock Reading digits in natural images with unsupervised feature learning.
\newblock In {\em Workshop on deep learning and unsupervised feature learning},
  2011.

\bibitem{orekondy2019knockoff}
Tribhuvanesh Orekondy, Bernt Schiele, and Mario Fritz.
\newblock Knockoff nets: Stealing functionality of black-box models.
\newblock In {\em CVPR}, pages 4954--4963, 2019.

\bibitem{papernot2017practical}
Nicolas Papernot, Patrick McDaniel, Ian Goodfellow, Somesh Jha, Z~Berkay Celik,
  and Ananthram Swami.
\newblock Practical black-box attacks against machine learning.
\newblock In {\em Asia CCS}, pages 506--519, 2017.

\bibitem{pei2017deepxplore}
Kexin Pei, Yinzhi Cao, Junfeng Yang, and Suman Jana.
\newblock {DeepXplore}: Automated whitebox testing of deep learning systems.
\newblock In {\em SOSP}, pages 1--18, 2017.

\bibitem{renda2020comparing}
Alex Renda, Jonathan Frankle, and Michael Carbin.
\newblock Comparing rewinding and fine-tuning in neural network pruning.
\newblock {\em arXiv preprint arXiv:2003.02389}, 2020.

\bibitem{russakovsky2015imagenet}
Olga Russakovsky, Jia Deng, Hao Su, Jonathan Krause, Sanjeev Satheesh, Sean Ma,
  Zhiheng Huang, Andrej Karpathy, Aditya Khosla, Michael Bernstein, et~al.
\newblock Imagenet large scale visual recognition challenge.
\newblock {\em International journal of computer vision}, 115(3):211--252,
  2015.

\bibitem{sharir2020cost}
Or~Sharir, Barak Peleg, and Yoav Shoham.
\newblock The cost of training nlp models: A concise overview.
\newblock {\em arXiv preprint arXiv:2004.08900}, 2020.

\bibitem{tramer2016stealing}
Florian Tram{\`e}r, Fan Zhang, Ari Juels, Michael~K Reiter, and Thomas
  Ristenpart.
\newblock Stealing machine learning models via prediction {APIs}.
\newblock In {\em USENIX Security}, pages 601--618, 2016.

\bibitem{tsipras2018robustness}
Dimitris Tsipras, Shibani Santurkar, Logan Engstrom, Alexander Turner, and
  Aleksander Madry.
\newblock Robustness may be at odds with accuracy.
\newblock {\em arXiv preprint arXiv:1805.12152}, 2018.

\bibitem{uchida2017embedding}
Yusuke Uchida, Yuki Nagai, Shigeyuki Sakazawa, and Shin'ichi Satoh.
\newblock Embedding watermarks into deep neural networks.
\newblock In {\em ICMR}, pages 269--277, 2017.

\bibitem{wang2019neural}
Bolun Wang, Yuanshun Yao, Shawn Shan, Huiying Li, Bimal Viswanath, Haitao
  Zheng, and Ben~Y Zhao.
\newblock Neural cleanse: Identifying and mitigating backdoor attacks in neural
  networks.
\newblock In {\em S\&P}, pages 707--723. IEEE, 2019.

\bibitem{warden2018speech}
Pete Warden.
\newblock Speech commands: A dataset for limited-vocabulary speech recognition.
\newblock {\em arXiv preprint arXiv:1804.03209}, 2018.

\bibitem{wu2020skip}
Dongxian Wu, Yisen Wang, Shu-Tao Xia, James Bailey, and Xingjun Ma.
\newblock Skip connections matter: On the transferability of adversarial
  examples generated with resnets.
\newblock {\em ICLR}, 2020.

\bibitem{yosinski2015understanding}
Jason Yosinski, Jeff Clune, Anh Nguyen, Thomas Fuchs, and Hod Lipson.
\newblock Understanding neural networks through deep visualization.
\newblock {\em arXiv preprint arXiv:1506.06579}, 2015.

\bibitem{yuan2020attack}
Xiaoyong Yuan, Lei Ding, Lan Zhang, Xiaolin Li, and Dapeng Wu.
\newblock Es attack: Model stealing against deep neural networks without data
  hurdles.
\newblock {\em arXiv preprint arXiv:2009.09560}, 2020.

\bibitem{zhang2019theoretically}
Hongyang Zhang, Yaodong Yu, Jiantao Jiao, Eric Xing, Laurent El~Ghaoui, and
  Michael Jordan.
\newblock Theoretically principled trade-off between robustness and accuracy.
\newblock In {\em ICML}, pages 7472--7482. PMLR, 2019.

\bibitem{zhang2018protecting}
Jialong Zhang, Zhongshu Gu, Jiyong Jang, Hui Wu, Marc~Ph Stoecklin, Heqing
  Huang, and Ian Molloy.
\newblock Protecting intellectual property of deep neural networks with
  watermarking.
\newblock In {\em Asia CCS}, pages 159--172, 2018.

\end{thebibliography}

\appendix
\section{Appendix}

\subsection{Details of Datasets and Models}
\label{subsec:datasets}

We use four benchmark datasets from two domains for the evaluation:
\begin{itemize}
\item MNIST \cite{lecun2010mnist}. This is a handwritten digits (from 0 to 9) dataset, consisting of 70,000 images with size $28 \times 28 \times 1$, of which 60,000 and 10,000 are training and test data. 

\item CIFAR-10 \cite{krizhevsky2009learning}. This is a 10-class image classification dataset, consisting of 60,000 images with size $32\times 32\times 3$, of which 50,000 and 10,000 are training and testing data. 

\item ImageNet \cite{russakovsky2015imagenet}. This is a large-scale image dataset containing more than 1.2 million training images of 1,000 categories. It is more challenging due to the higher image resolution $224 \times 224 \times 3$. We randomly sample 100 classes to construct a subset of ImageNet, of which 120,000 are training data and 30,000 are testing data. 

\item Speech Commands \cite{warden2018speech}. This is an audio dataset of 10 single spoken words, consisting of about 40,000 training samples and 4,000 testing samples. We pre-processed the data to obtain a Mel Spectrogram \cite{choi2017kapre}. Each audio sample is transformed into an array of size $120\times 85$. 
\end{itemize}

To explore the scalability of \dje, various model structures are tested as in Table~\ref{tab:setup}. LeNet-5, ResNet-20 and VGG-16 are standard CNN structures, while LSTM(128) is an RNN structure: an LSTM layer with 128 hidden units, followed by three fully-connected layers (128/64/10).


\vspace{-2mm}
\subsection{Seed Selection Strategy}
\label{subsec:gini}
\begin{figure}[t]
  \centering
  \includegraphics[width=0.8\linewidth]{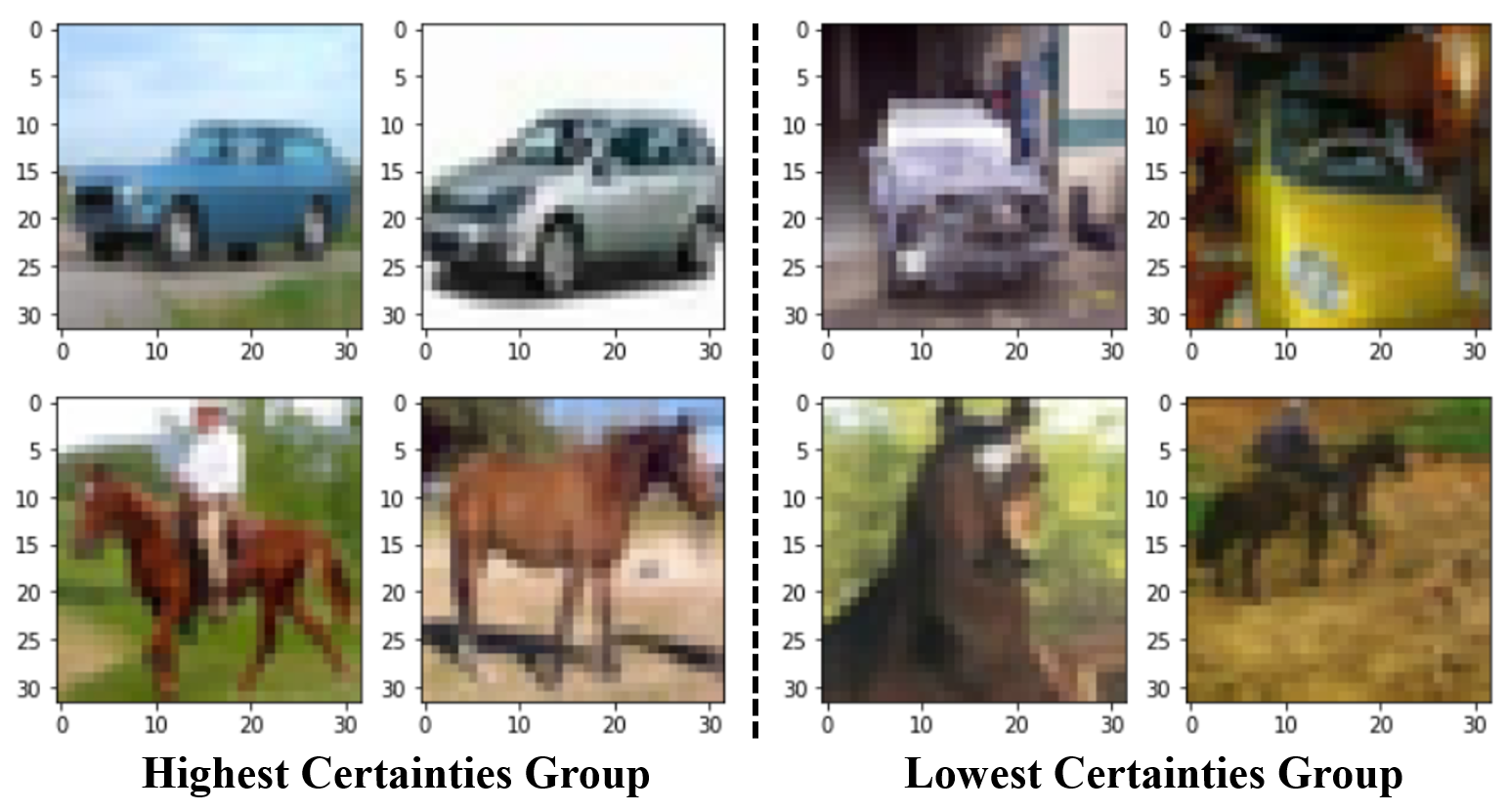}
  \caption{Selected test seeds with the highest or lowest certainty scores. The first row belongs to `automobile' and the second row belongs to `horse'.}
  \label{fig:gini-contrast}
\end{figure}

Seed selection is important for generating high-quality test cases. We use DeepGini \cite{feng2020deepgini} to measure the certainty of each candidate sample. Given the victim model $f$ and a testing dataset $\mathcal{D}$, we first calculate the Certainty Score (CS) for each seed $\vx\in \mathcal{D}$ as: $CS(f^L,\vx)=\sum_i^Cf^L_i(\vx)^2,$ 
then we rank the seed list by the certainty score, and the first part of the seeds of the highest scores (i.e., most certainties) will be chosen for the following generation process. Here, we assume to have two seeds $\{\vx_1, \vx_2\}$ with $CS(f^L,\vx_1)>CS(f^L,\vx_2)$, that means the victim model $f$ is more confident at $\vx_1$, which also means that $\vx_1$ is farther from the decision boundary and easier for classification (see examples in Fig.~\ref{fig:gini-contrast}).

\vspace{-2mm}
\subsection{Data-augmentation}
\label{subsec:aug}
During the finetuning and pruning processes, typical data-augmentation techniques are used to strengthen the attacks except for the SpeechCommands dataset, including random rotation ($10^{\circ}$), random width- and height-shift (both $0.1$).

\vspace{-2mm}
\subsection{Test Case Generation Details and Calibrations} 
\label{subsec:generation}
Specifically, we consider three adversarial attacks for generating \textbf{black-box test cases} (see Section \ref{subsubsec:bound}).

\vspace{0.5mm}
\noindent \textbf{FGSM} \cite{goodfellow2014explaining} perturbs a normal example $\vx$ by one single step of size $\epsilon$ to maximize the model's prediction error with respect to the groundtruth label $y$:$ \vx' = \vx + \epsilon \cdot \sign(\bigtriangledown_{\vx} \mathcal{L}(f^L(\vx), y) )$, where $\sign(\cdot)$ is the sign function, $\mathcal{L}$ is the cross entropy (CE) loss, and $\bigtriangledown_{\vx} \mathcal{L}$ is the gradient of the loss to the input.

\vspace{0.5mm}
\noindent \textbf{PGD} \cite{madry2017towards} is an iterative version of FGSM but with smaller step size: $\vx^k=\Pi_{\epsilon}(\vx^{k-1} + \alpha \cdot \sign(\bigtriangledown_{\vx} \mathcal{L}(f^L(\vx^{k-1}), y) ) )$, where $\vx^k$ is the adversarial example obtained at the $k$-th perturbation step, $\alpha$ is the step size and $\Pi_{\epsilon}$ is a projection (clipping) operation that projects the perturbation back onto the $\epsilon$-ball centered around $\vx$ if it goes beyond. 

\vspace{0.5mm}
\noindent \textbf{CW} \cite{carlini2017towards} generates adversarial examples by solving the optimization problem: $\vx' = \underset{\vx'}{\min} \norm{\vx'-\vx}_{2}^{2} - c\cdot \mathcal{L}(f^L(\vx'), y)$, where $c$ is a hyperparameter balancing the two terms and the pixel values of adversarial example $\vx'$ are bounded to be within a legitimate range, e.g., $[0,1]$ for 0-1 normalized input.

\begin{table}[]
   \renewcommand\arraystretch{1.2}
   \center
    \scriptsize
   \setlength\tabcolsep{3.5pt}
   \caption{The hyper-parameters used in different test case generation strategies.} \label{tab:generation}
   \begin{tabu}{cc|cccc}
   \tabucline[1pt]{-}
   \textbf{Method}       & \textbf{Params}   & \textbf{MNIST}    & \textbf{CIFAR-10}   & \textbf{ImageNet} & \textbf{SpeechCmds} \\ \tabucline[1pt]{-}
   \emph{PGD} \cite{madry2017towards}     & $\epsilon/steps$  & 0.1/10   &  0.03/10     &  0.01/10  & 0.1/10    \\ \hline
   \emph{Alg.~\ref{alg:neuron}}  & $m/iters$   & 3/1k     &    3/1k      &   3/1k  & 1/1k \\ \hline
   \emph{FGSM }\cite{goodfellow2014explaining}     & $\epsilon/steps$  & 0.1/1   &  0.03/1     &  0.01/1   & 0.1/1  \\ \hline
   \emph{CW }\cite{carlini2017towards}       & $c/iters$  & 5/1k   &  5/1k   &  5/1k   & 5/1k \\ \hline
   \emph{IPG }\cite{cao2021ipguard}      & $k/iters$  & 10/1k   & 10/1k  & 10/1k  & 10/1k \\ \tabucline[1pt]{-}
   \end{tabu}
   \end{table}

\begin{table}[t]\centering
   \renewcommand\arraystretch{1.2}
\scriptsize
   \caption{Time cost (seconds) of test cases generation.} \label{tab:time}
   \begin{tabu}{c|cccc}
   \tabucline[1pt]{-}
   \textbf{Method}     & \textbf{MNIST}   & \textbf{CIFAR-10}   & \textbf{ImageNet} & \textbf{SpeechCmds}  \\ \tabucline[1pt]{-}
   \emph{PGD} \cite{madry2017towards}    & 0.3   &  3.7    & 227.6   &  1.2           \\ \hline
   \emph{Alg.~\ref{alg:neuron}}   & 635.3   & 1200.3    & 1280.1      &  4424.6    \\ \tabucline[1pt]{-}
   \end{tabu}
\end{table}

The hyper-parameters used for the generation algorithms on different datasets are summarized in Table~\ref{tab:generation}. Here, we take CIFAR-10 dataset as an example and analyze the influencing factors of the test case generation process.

\vspace{1mm}
\noindent\textbf{Adversarial Examples}. PGD is the default choice for generating adversarial examples in the \textbf{black-box setting}. Here, we further compare PGD with two other methods, FGSM and CW. We use the same selected seeds for the generation. Table~\ref{tab:rob-attacks} shows the results of the \emph{RobD} metric. We observe that the gap in \emph{RobD} values between the positive and negative suspect models is very small when CW is used, which fails to distinguish the two types of models. One reason is that CW attack optimizes adversarial examples for minimal perturbations, which is more sensitive (less robust) to model modifications. FGSM can be regarded as a one-step PGD, which usually has a larger average perturbation than PGD. When the perturbation increases, the \emph{RobD} value of negative suspect models would decrease since adversarial examples with larger perturbations tend to have better transferability \cite{wu2020skip}. It is similar to {PGD$_{3\epsilon}$} when the perturbation bound increases. In general, the absolute \emph{RobD} gap between the positive and negative suspects tested with PGD-generated test cases is larger than that of FGSM and CW. Moreover, PGD is relatively cheaper to calculate than CW, i.e., the time cost of PGD is $100\times$ lower than CW. Overall, PGD is more suitable for fingerprinting the decision boundary with untargeted adversarial examples, as shown in Fig.~\ref{fig:adv}. We will explore more effective metrics and test case generation methods with diverse granularity in future work. 
   
\vspace{0.5mm}
\begin{tcolorbox}[fonttitle = \bfseries]
  \textbf{Remark 5:} Different generation strategies and parameters can impact \dje differently. Overall, PGD is a better choice for characterizing the model's decision boundary.
\end{tcolorbox}

\begin{table*}[t]
      \renewcommand\arraystretch{1.2}
      \centering
      \scriptsize
      \caption{Using different methods to generate adversarial examples for the \emph{RobD} metric evaluation on CIFAR-10 dataset. PGD$_{\epsilon/3}$ : with $\frac{1}{3}\times$  perturbation bound,  PGD$_{3\epsilon}$ : with $3\times$ bound, PGD$_{10s}$: \text{with}~$10\times$ steps. }
      
      \label{tab:rob-attacks}
      \begin{tabu}{cc|cccccc}
\tabucline[1pt]{-}
\multicolumn{2}{c|}{\textbf{Model Type}}                                                                          & \textbf{FGSM}    & \textbf{CW}    & \textbf{PGD}          & PGD$_{\epsilon/3}$   & PGD$_{3\epsilon}$ & PGD$_{10s}$  \\ \tabucline[1pt]{-}
\multicolumn{1}{c|}{\multirow{5}{*}{\begin{tabular}[c]{@{}c@{}}Positive\\ Suspect\\ Models\end{tabular}}} & FT-LL   &  0.024$\pm$0.004   &  0.0$\pm$0.0     & 0.0$\pm$0.0        &  0.034$\pm$0.002       &   0.0$\pm$0.0    &  0$\pm$0.0  \\ \cline{2-8} 
\multicolumn{1}{c|}{}                                                                                     & FT-AL   &  0.261$\pm$0.025   &  0.905$\pm$0.028      & 0.192$\pm$0.028  & 0.733$\pm$0.012        &    0.046$\pm$0.010 &  0.350$\pm$0.027   \\ \cline{2-8}
\multicolumn{1}{c|}{}                                                                                     & RT-AL   &  0.267$\pm$0.025   &  0.917$\pm$0.024      & 0.237$\pm$0.055  & 0.748$\pm$0.046       &   0.073$\pm$0.022      & 0.400$\pm$0.046 \\ \cline{2-8} 
\multicolumn{1}{c|}{}                                                                                     & P-20\% &  0.252$\pm$0.030   &  0.882$\pm$0.038          & 0.155$\pm$0.032  & 0.702$\pm$0.023        &   0.045$\pm$0.020  &  0.299$\pm$0.049  \\ \cline{2-8} 
\multicolumn{1}{c|}{}                                                                                     & P-60\% &  0.293$\pm$0.027  &  0.940$\pm$0.013         & 0.318$\pm$0.036  & 0.792$\pm$0.023        & 0.123$\pm$0.022 &   0.502$\pm$0.031   \\ \tabucline[1pt]{-}

\multicolumn{1}{c|}{\multirow{3}{*}{\begin{tabular}[c]{@{}c@{}}Negative\\ Suspect\\ Models\end{tabular}}} & Neg-1  &  0.662$\pm$0.058    & 0.999$\pm$0.002         & 0.920$\pm$0.021  &  0.989$\pm$0.007     &  0.573$\pm$0.093 & 0.958$\pm$0.013      \\ \cline{2-8} 
\multicolumn{1}{c|}{}                                                                                     & Neg-2  &  0.672$\pm$0.019    & 0.998$\pm$0.003     & 0.926$\pm$0.030  &  0.986$\pm$0.004     & 0.576$\pm$0.030  &  0.948$\pm$0.012    \\ \cline{2-8} 
\multicolumn{1}{c|}{}                                                                                     & $\tau_\lambda$  & \textbf{0.583}    &   \textbf{0.897}     & \textbf{0.816} &     \textbf{0.886}    &   \textbf{0.489}  &  \textbf{0.851}   \\ \tabucline[1pt]{-}
\end{tabu}
\end{table*}

\begin{figure}[t]
 \centering
 \includegraphics[width=0.72\linewidth]{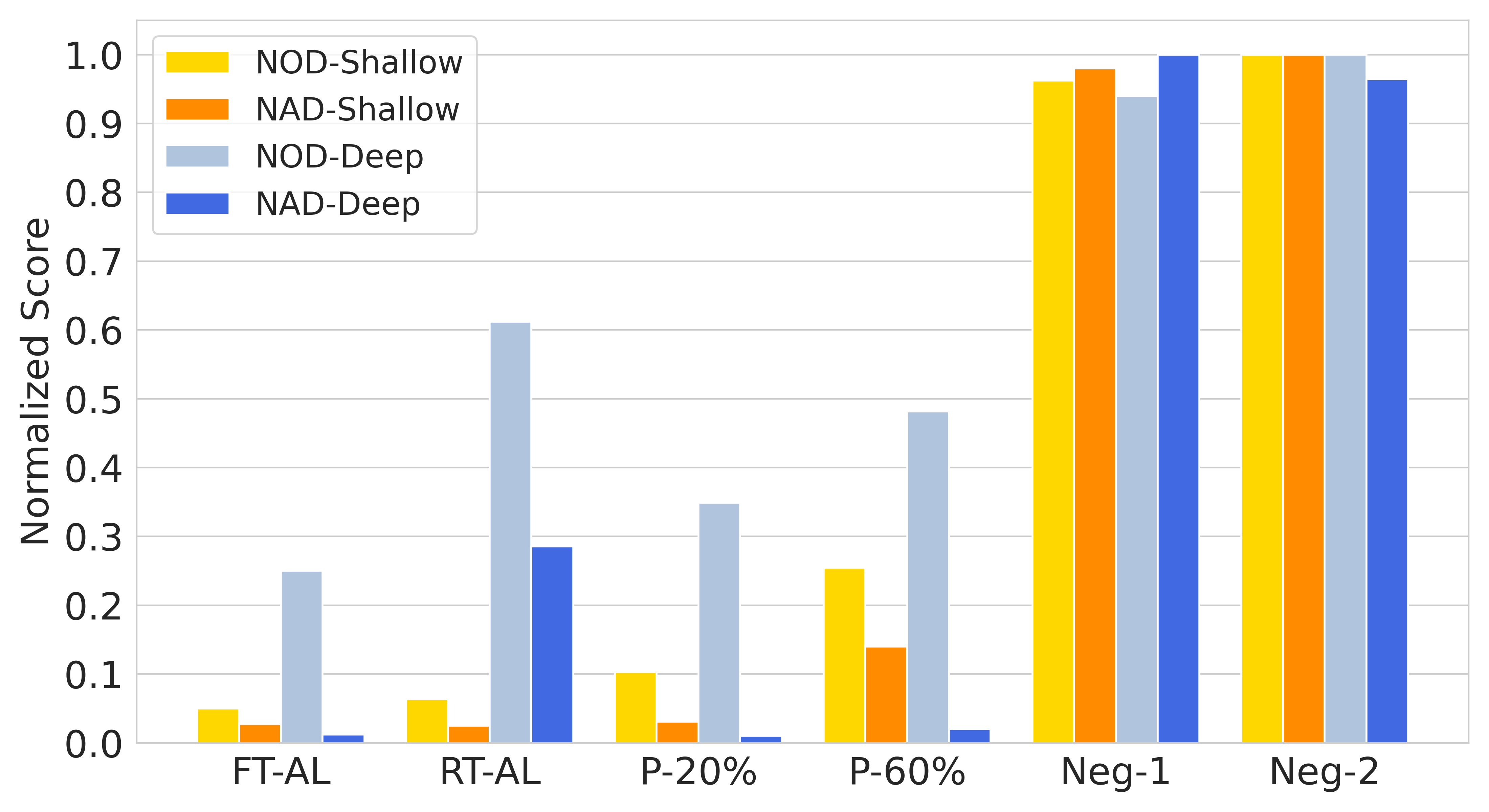}
 \caption{Normalized distance evaluations based on different layers of the CIFAR-10 network: `-Shallow' means the results on the shallow layer, and `-Deep' means the deep layer.}
 \label{fig:hyper-layer}
\end{figure}

\noindent\textbf{Layer Selection.} Layer selection is important when applying \dje in the \textbf{white-box setting} with \emph{NOD} and \emph{NAD}. Here, we evaluate how the choice of layers affects the performance of \dje.
For comparison, we choose a shallow layer and a deep layer of the victim model, and re-generate the test cases respectively for each layer. Fig.~\ref{fig:hyper-layer} shows the results of \emph{NOD} and \emph{NAD} metrics. In general, the \emph{NOD/NAD} difference between the positive and negative suspect models becomes much larger at the shallow layer. The reason is that the shallow layers of a network usually learn the low-level features \cite{yosinski2015understanding}, and they tend to stay the same or at least similar during model finetuning. Particularly, the performance on RT-AL degrades the most when the deep layer is selected, since the parameters of the last layer are re-initialized. Thus, choosing the shallow layers to compute the \emph{NOD} and \emph{NAD} metrics could help the robustness of \dje. Moreover, the time cost of generating and testing with the shallow layer is 10$\times$ less than the deep layer, since most of the back-propagation computations are eliminated.

\vspace{0.5mm}
\begin{tcolorbox}[fonttitle = \bfseries]
  \textbf{Remark 6:} The shallow layers are a better choice for testing metrics \emph{NOD} and \emph{NAD}.
\end{tcolorbox}

\subsection{Defense Baselines}
\label{subsec:watermarking}

\subsubsection{Backdoor-based watermarking (Black-box)} 
\begin{figure}[]
  \centering
  \includegraphics[width=0.28\linewidth]{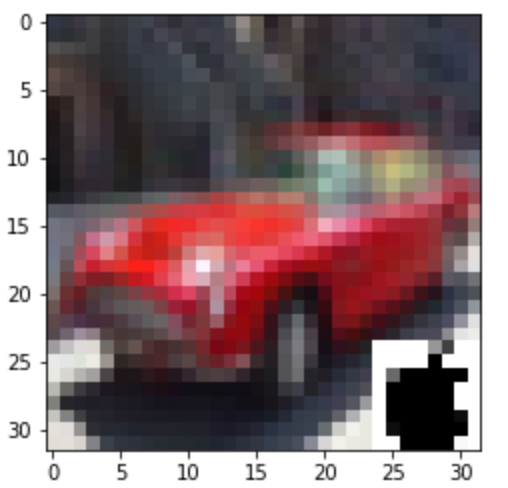}
  \caption{A trigger input example used in backdooring.}
  \label{fig:backdoor}
\end{figure}

\cite{zhang2018protecting} embeds backdoors into the model. In our experiments, we select 500 samples from the training dataset, of which the ground truth labels are ``automobile''. Then we patch an ``apple'' logo at the bottom right corner of each sample and change their labels to ``cat'' (see Fig.~\ref{fig:backdoor}). These trigger examples (i.e., trigger set) are mixed into the clean training dataset to train a watermarked model from scratch. The initial \emph{TSA} of the watermarked model is 100.0\% (on a separate trigger set).

\subsubsection{Signature-based watermarking (White-box)} 
\cite{uchida2017embedding} embeds a $T$-bit vector (i.e., the watermark) $b \in \{0,1\}^T$ into one of the convolutional layers, by adding an additional parameter regularizer into the loss function: $E(w) = E_0(w) + \lambda E_R(w),$ where $E_0(w) $ is the original task loss function, $E_R(w)$ is the regularizer that imposes a certain restriction on the model parameters $w$, and $\lambda$ is a hyper-parameter. In our experiments, $\lambda$ is set to $0.01$, and we embed a 128-bit watermark (generated by the random strategy) into the second convolutional block (Conv-2 group) as recommended in \cite{uchida2017embedding}. The initial \emph{BER} of the watermarked model is 3.13\%.

\subsubsection{Fingerprinting (Black-box)}
IPGuard \cite{cao2021ipguard} proposes a type of adversarial attack that targets on generating adversarial examples $x'$ around the classification boundaries of the victim model, and the matching rate (\emph{MR}) of these key samples is calculated for the verification similar to \cite{zhang2018protecting}. We generate a set of 1,000 adversarial examples following \cite{cao2021ipguard} and the initial \emph{MR} of the victim model on the generated key samples is 100.0\%. 


\subsection{Model Extraction Attacks}
\label{subsec:extraction}

\noindent \textbf{Jacobian-Based Augmentation}. The seeds used for augmentation are all sampled from the testing dataset. We sample 150 seeds for extracting the MNIST victim model, 500 seeds for SpeechCommands, 1,000 seeds for CIFAR-10, and use all other default settings \cite{papernot2017practical}.


\noindent \textbf{Knockoff Nets}.  We use the Fashion-MNIST dataset for extracting the MNIST victim model, an independent speech dataset for SpeechCommands, and CIFAR-100 for CIFAR-10. We use other default hyper-parameter settings of \cite{orekondy2019knockoff}.

\vspace{0.5mm}
\noindent \textbf{ES Attack}. We use the OPT-SYN algorithm \cite{yuan2020attack} to heuristically synthesize the surrogate data. We set the stealing epoch to 50 for MNIST and 400 for CIFAR-10. We failed to extract the SpeechCommands Victim model since the validation accuracy could not exceed 20\%. All other hyper-parameters are the same as in \cite{yuan2020attack}.

\vspace{0.5mm}
\noindent \textbf{$^\ast$Functionality-equivalent Extraction}. Besides the above three extraction attacks, we are also aware of the functionality-equivalent extraction attacks \cite{jagielski2020high, carlini2020cryptanalytic} that attempt to obtain a precise functional approximation of the victim model. For instance, \cite{carlini2020cryptanalytic} proposed a differential attack that could steal the parameters of the victim model up to floating-point precision without the knowledge of training data. We remark that defending this type of attack is a trivial task for \deepjudge as there will be no difference between the extracted model and the victim model in an ideal approximation.


Note that Black-box model extraction is still underexplored, and more extraction attacks may appear in the future. This poses a continuous challenge for deep learning copyright protection. We hope that \dje could evolve with the adversaries by incorporating more advanced testing metrics and test case generation methods, and provide a possibility to fight against this continuing model stealing threat.



\subsection{Adaptive Attacks for Watermarking \& Fingerprinting}
\label{subsec:adapt123}
In addition to Section~\ref{sec:adaptive_exp}, here we conduct an extra evaluation of existing watermarking \cite{uchida2017embedding, zhang2018protecting} and fingerprinting \cite{cao2021ipguard} methods under similar adaptive attack settings. 

\noindent \textbf{Adaptive attacks.} \emph{Adv-Train} and \emph{VTL} are the two adaptive attacks in Table~\ref{tab:adapt}, while the \emph{Adapt-X} attack is specifically designed for each method as follows:
\begin{itemize}
    \item \emph{Adapt-X} for \cite{uchida2017embedding}. Since the embedded watermark (signature) is known, the adversary copies (steals) the victim model then fine-tunes it on a small subset of clean examples while maximizing the embedding loss $E_R(w)$ on the signature.
    \item \emph{Adapt-X} for \cite{zhang2018protecting}. Since the embedded watermark (backdoor) is known, the adversary can follow a similar approach as above to steal the victim model and remove the backdoor watermark with a few backdoor-patched but correctly-labeled examples.
    \item \emph{Adapt-X} for \cite{cao2021ipguard}. Similar to our \emph{Adapt-B} for \dje, the adversary copies the victim model then fine-tunes it on a small subset of clean and correctly-labeled fingerprint examples to circumvent fingerprinting.
\end{itemize}

As the results in Table~\ref{tab:adapt22} show, all three methods are completely broken by the adaptive attacks. \dje is the only method that can survive these attacks and was \emph{partially compromised but not fully broken} (the final judgments are still correct, as shown in the ‘Copy?’ column of Table~\ref{tab:adapt}). This implies that a single metric of watermarking or fingerprinting is not sufficient enough to combat adaptive attacks. By contrast, a testing framework with comprehensive testing metrics and test case generation methods may have the required flexibility to address this challenge. For example, \emph{Adv-Train} may break the black-box testing of \dje but cannot break the white-box testing (see Section~\ref{sec:adaptive_attack2}). Moreover, \dje can quickly recover its performance by switching to a new set of seeds (see Fig.~\ref{fig:advtrain}). 



\begin{table*}[]
\centering
   \renewcommand\arraystretch{1.2}
   \setlength\tabcolsep{4pt}
   
    \scriptsize
   \caption{Performance of existing watermarking and fingerprinting baselines on CIFAR-10 dataset against adaptive attacks: 1) \emph{Adapt-X}, adaptive attack designed specifically against the defense method; 2) \emph{Adv-Train}, blind adversarial training; and 3) \emph{VTL}, vanilla transfer learning. The broken metrics (close to the negatives) are highlighted in \red{red}. \emph{Adapt-X} breaks all three metrics, while \emph{Adv-Train} breaks the adversarial-examples-based fingerprinting.}
    \label{tab:adapt22}
    

\begin{tabu}{c|c|cc|cc|cc}
\tabucline[1pt]{-}
\multicolumn{2}{c|}{\multirow{2}{*}{\textbf{Model Type}}}                                                 & \multicolumn{2}{c|}{\textbf{~~Black-box Watermarking \cite{zhang2018protecting}~~}} & \multicolumn{2}{c|}{\textbf{~~White-box Watermarking \cite{uchida2017embedding}~~}} & \multicolumn{2}{c}{\textbf{~~Black-box Fingerprinting \cite{cao2021ipguard}~~}} \\ \cline{3-8} 
\multicolumn{2}{c|}{}  & ACC   & \emph{TSA}     & ACC    & \emph{BER}  & ACC  & \emph{MR}         \\ \tabucline[1pt]{-}
\multicolumn{2}{c|}{Victim Model}    &  82.9\%    &  100.0\%  &   83.8\%     &     3.13\%    &   84.8\%    & 100.0\%     \\ \tabucline[1pt]{-}
\multirow{3}{*}{\begin{tabular}[c]{@{}c@{}}Positive\\ suspect \\ models \end{tabular}} 
& Adapt-X   &  81.8$\pm$0.8\%  &  \red{0.01$\pm$0.01\%}    &   71.2$\pm$2.6\%    &   \red{46.3$\pm$3.3\%}     &   81.9$\pm$0.2\%     & \red{1.2$\pm$0.8\%}         \\ \cline{2-8} 

& Adv-Train  &  73.8$\pm$1.6\%  &  5.5$\pm$3.2\%    &   73.5$\pm$1.7\%    &   4.0$\pm$0.8\%     &   74.5$\pm$2.3\%     & \red{4.2$\pm$1.5\%}        \\ \cline{2-8} 
& VTL       &  92.2$\pm$1.3\%  &   $\times$     &   91.7$\pm$1.6\%    &  6.1$\pm$1.2\%   & 93.3$\pm$1.7\%      &  $\times$    \\ \tabucline[1pt]{-}
\multirow{2}{*}{\begin{tabular}[c]{@{}c@{}}Negative\\ models\end{tabular}} 
& Neg-1     &  84.2$\pm$0.6\%  &  0.05$\pm$0.04\%     &  84.2$\pm$0.6\%    &  50.3$\pm$4.1\%      &  84.2$\pm$0.6\%  &  2.2$\pm$1.4\%      \\ \cline{2-8} 
& Neg-2     &  84.9$\pm$0.5\%  &    0.03$\pm$0.03\%    &  84.9$\pm$0.5\%   &  50.6$\pm$4.3\%     &  84.9$\pm$0.5\%  & 3.1$\pm$1.2\%     \\ \tabucline[1pt]{-}
\end{tabu}

\end{table*}

\subsection{How Different Levels of Finetuning, Pruning and Transfer Learning Affect \dje?}
\label{discuss}
There is a spectrum of building a new model with access to a victim model, from different ways of finetuning to transfer learning. Different levels of modifications to the victim model would accordingly influence the testing of \dje in different ways. Intuitively, a larger modification would lead to more dissimilarity between the victim and suspect models and a larger metric distance.

\begin{figure}[t]
    \centering
    \includegraphics[width=0.49\linewidth]{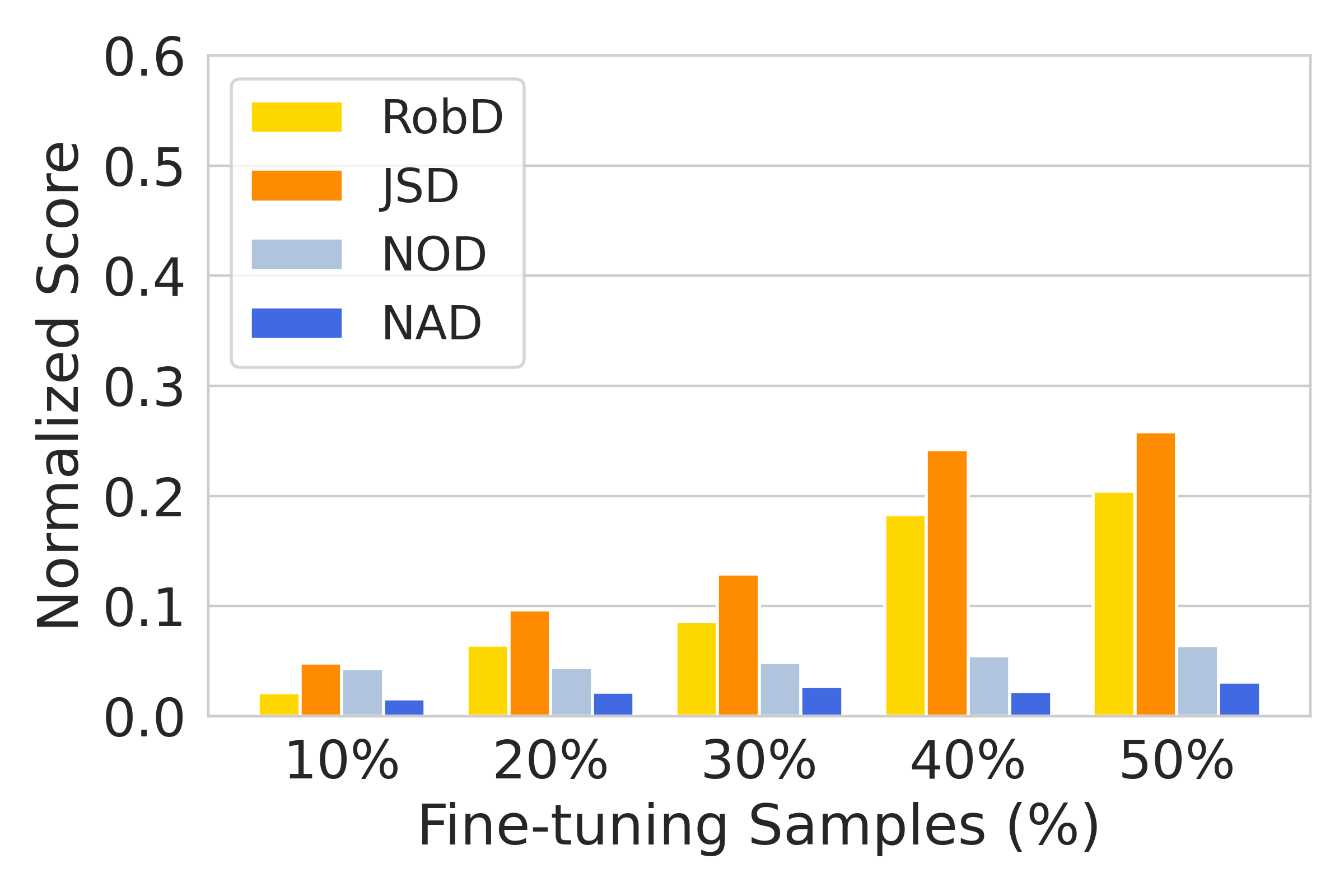}
    \includegraphics[width=0.49\linewidth]{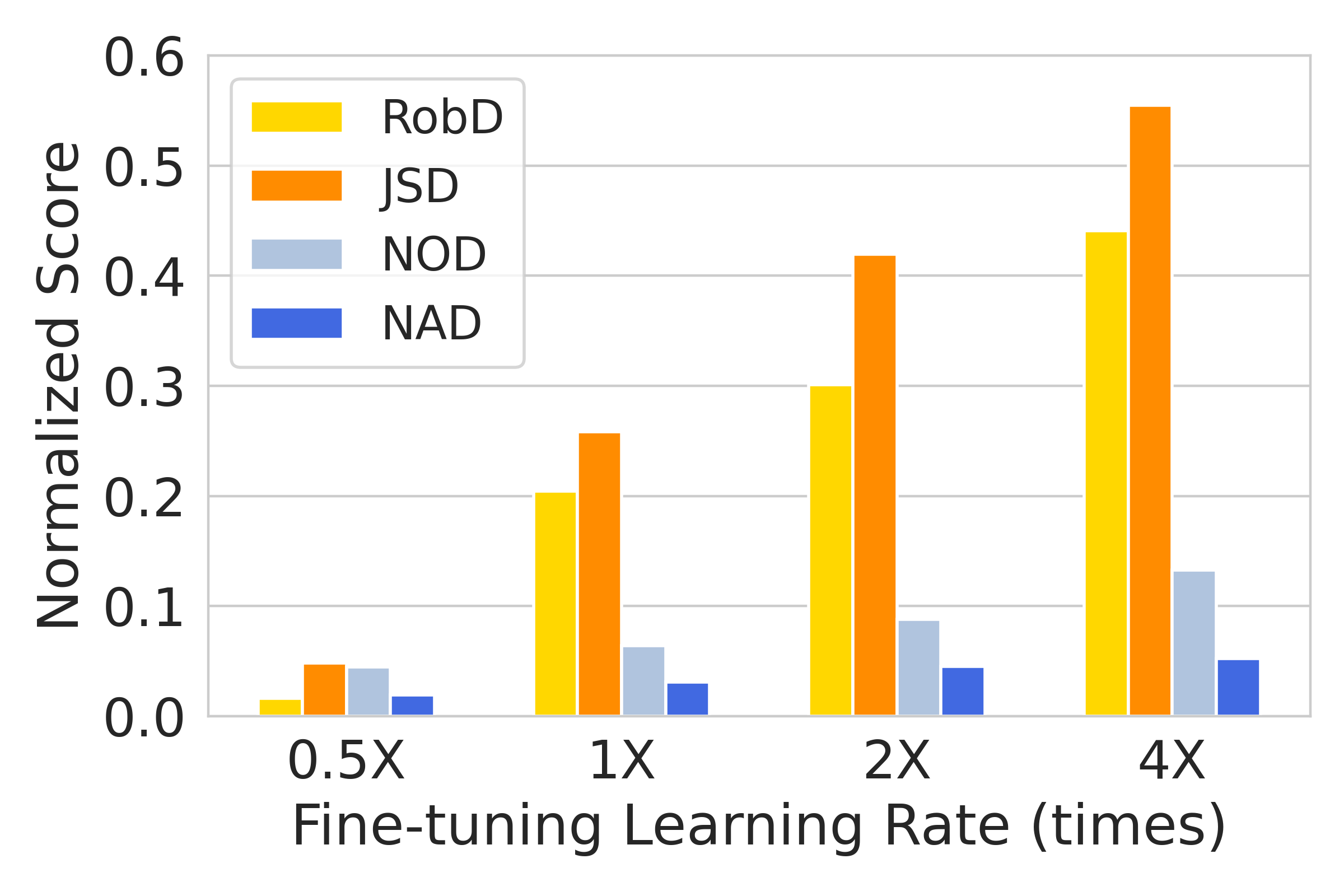}
    \includegraphics[width=0.49\linewidth]{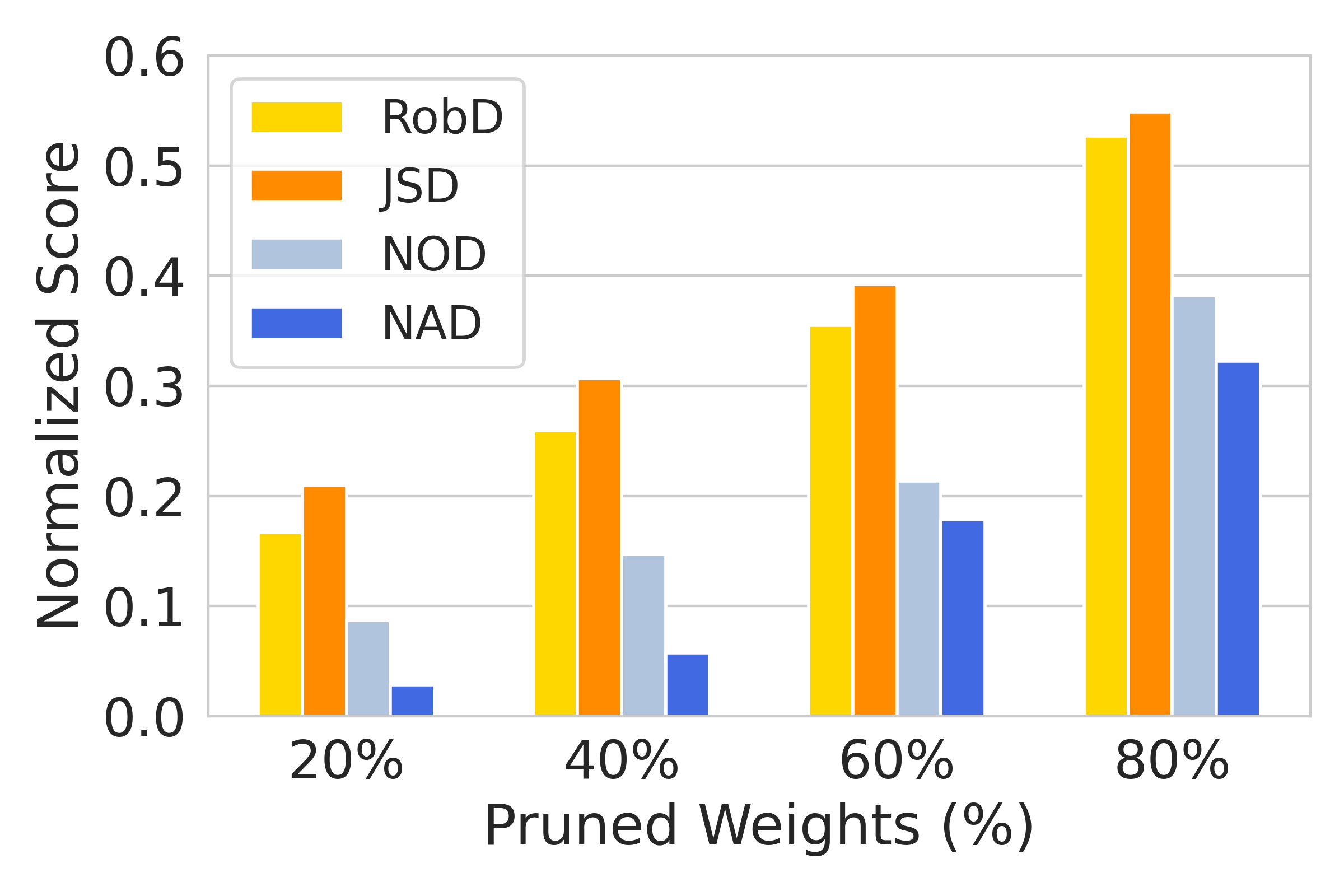}
    \includegraphics[width=0.49\linewidth]{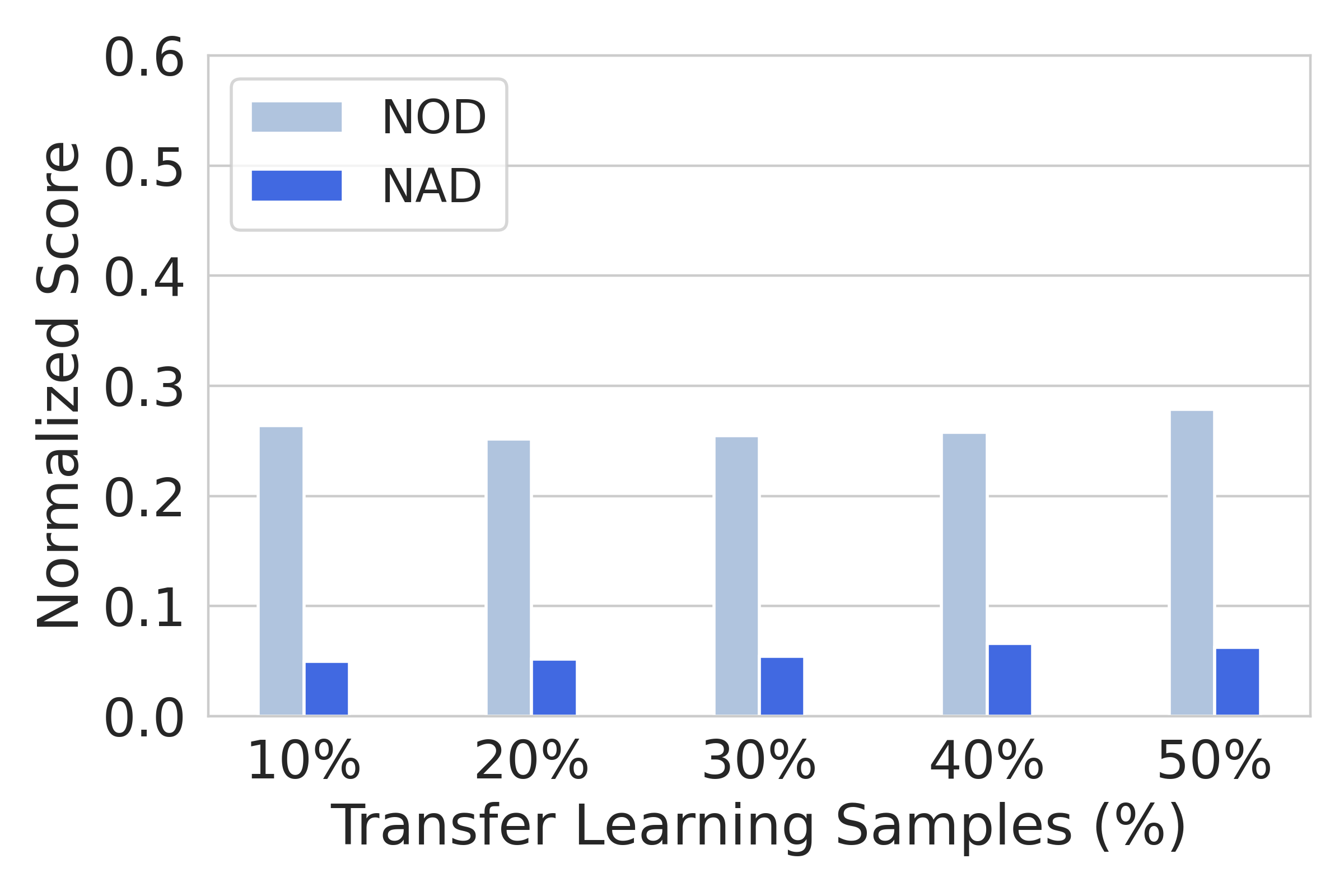}

    \caption{\dje metrics computed at different levels of model modifications on CIFAR-10. `2x' means 2 times the default learning rate.}
    \label{fig:nodnad}
\end{figure}

Here, we test different proportions of training samples and learning rates used for finetuning, proportions of pruned weights (pruning ratios) for pruning, and proportions of samples used for transfer learning (w.r.t. the setting described in Section~\ref{subsubsec:vtl}). 
Fig.~\ref{fig:nodnad} shows the metrics' values at different levels of finetuning, pruning and transfer learning.
At a high level, black-box metrics (i.e., \emph{RobD} and \emph{JSD}) have higher normalized distances than white-box metrics (i.e., \emph{NOD} and \emph{NAD}) on average. This implies that the model's decision boundary is more sensitive to almost all levels of modifications. 
For finetuning, the two black-box metrics (yellow and orange bars) increase significantly with the amount of finetuning samples or amplified learning rate, whereas the two white-box metrics are relatively stable. 
Note that `4x' (4 times the default learning rate) causes a significant drop ($\sim20\%$) in the model accuracy. For pruning, all metrics including the white-box metrics increase with the amount of pruned weights at a much higher rate than finetuning with different sample sizes. This indicates that pruning has more impact on the model than finetuning and will greatly distort the model's internal activations (measured by the two white-box metrics \emph{NOD} and \emph{NAD}). 
Transfer learning has much higher metric values (only white-box metrics are applicable here) than finetuning. This is because the victim model's functionality has been greatly altered by transferring to a new data distribution. However, it seems that the modification caused by transfer learning does not accumulate with more samples, resulting in similar metric values even with 40\% more samples.



In this work, we follow the principle that any derivations from the victim model other than independent training should be treated as having a certain level of copying. However, it can be hard to judge what degree of similarity (or level of modification) should be considered as ``real copying" in real-world scenarios. In \dje, we introduced a good range of testing metrics, hoping to provide more comprehensive evidence for making the final judgement. Moreover, the final judgement mechanism of \dje (Section~\ref{subsec:verification}) can be flexibly adjusted to suit different application needs.




\subsection{Additional Figures}
\label{add-figs}

\begin{figure}[b]
   \centering
   \includegraphics[width=0.99\linewidth]{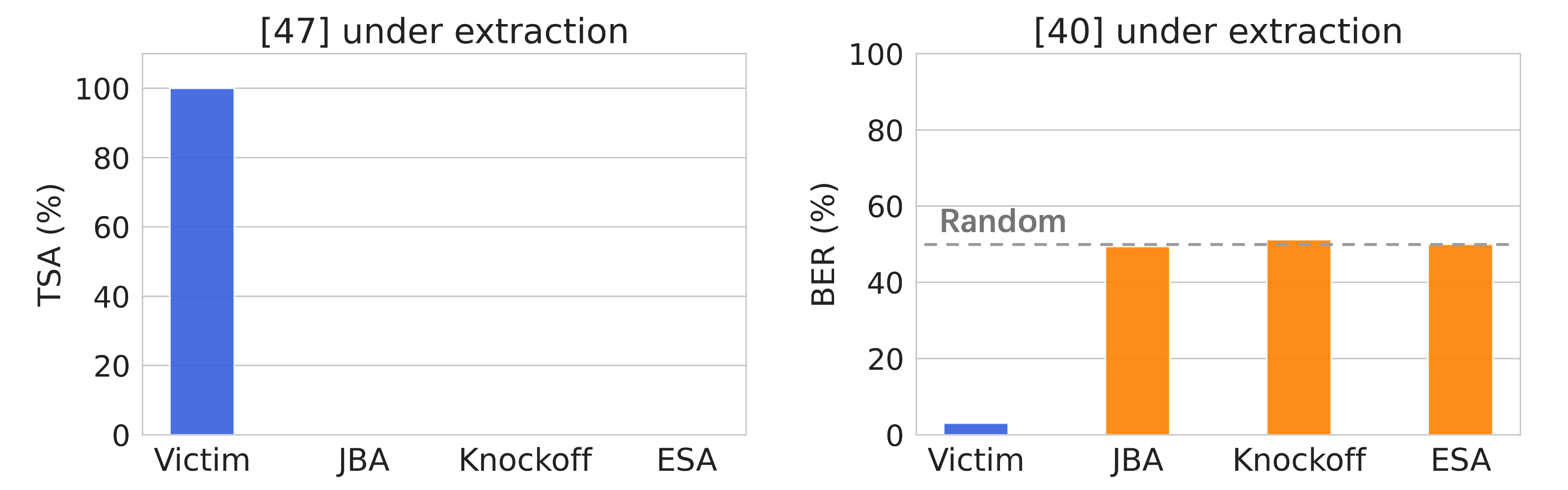}
   \caption{Existing watermarking methods \cite{zhang2018protecting} and \cite{uchida2017embedding} failed to verify the ownership of the extracted models by several extraction attacks.}
   \label{fig:failure}

\end{figure}

\begin{figure*}[t]
  \centering
  \includegraphics[width=0.38\linewidth]{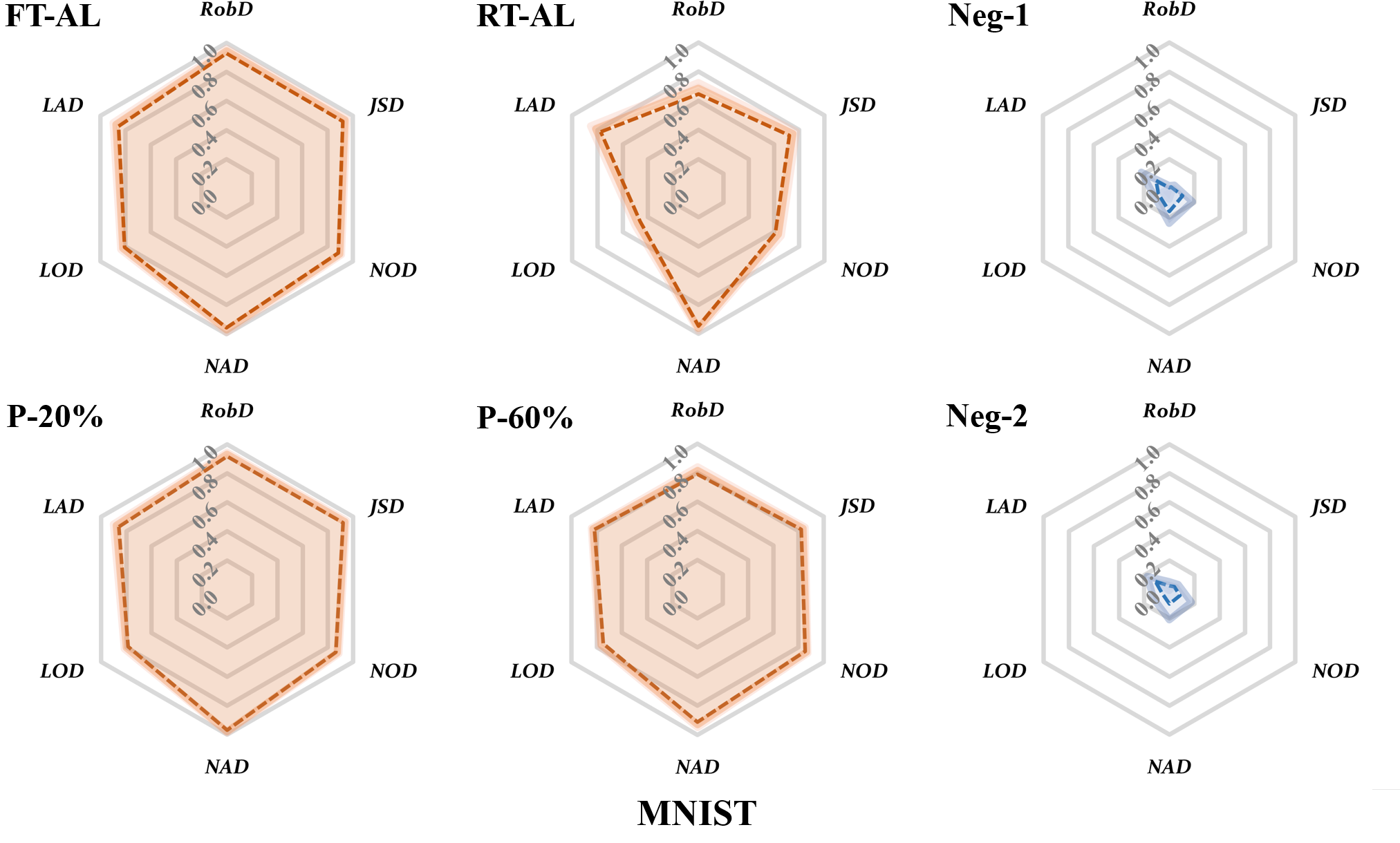}  
  \hspace{10mm}
  \includegraphics[width=0.38\linewidth]{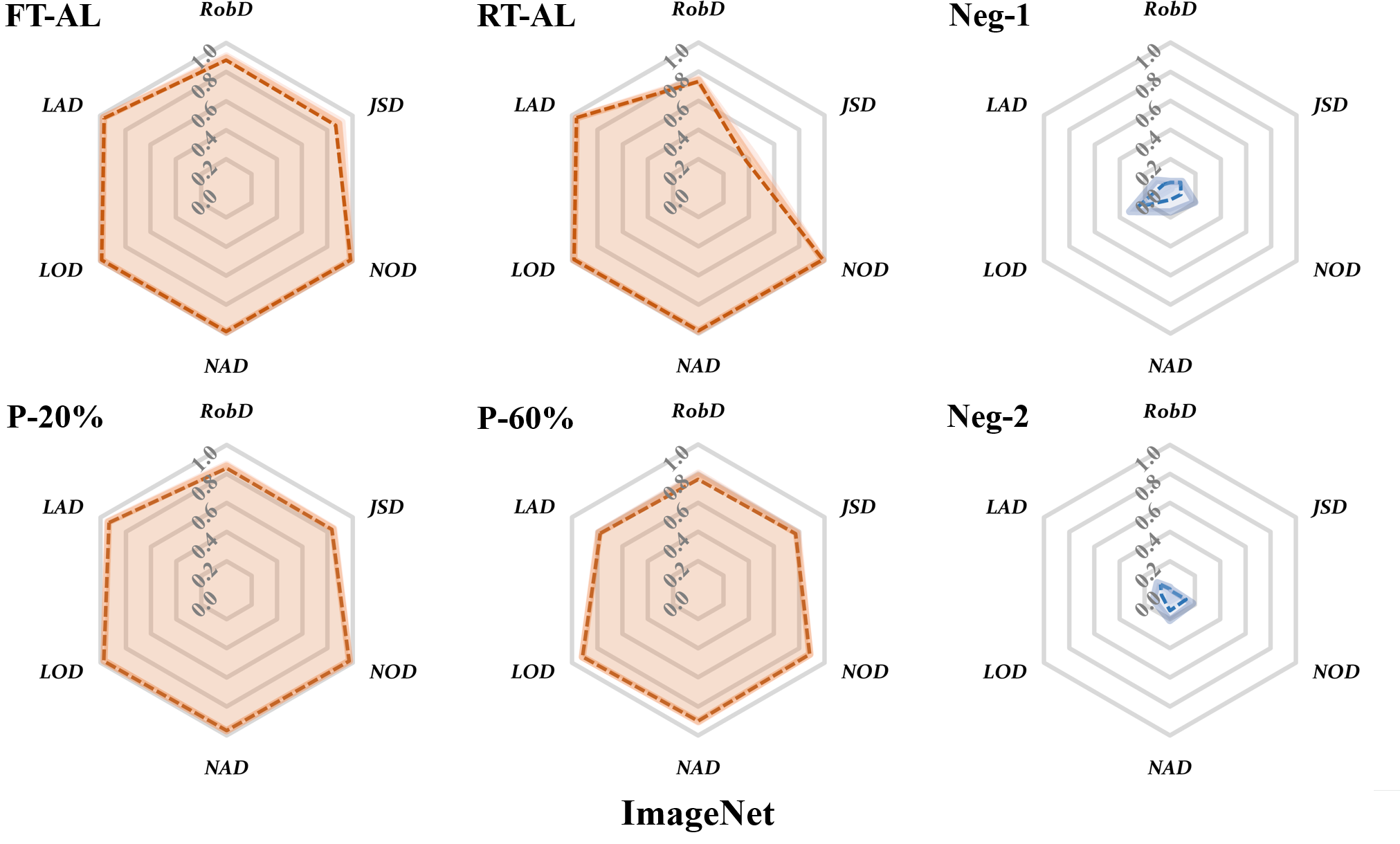}
  \setlength{\abovecaptionskip}{3pt}
  \caption{Similarity evaluation between the victim and suspect models on MNIST (left 3 columns) and ImageNet (right 3 columns). We use the \textcolor{orange}{orange} line for positive suspect models and the \blue{blue} line for negatives.}
  \label{fig:radars2}
\end{figure*}

\begin{figure}[t]
   \centering
   \includegraphics[width=0.49\linewidth]{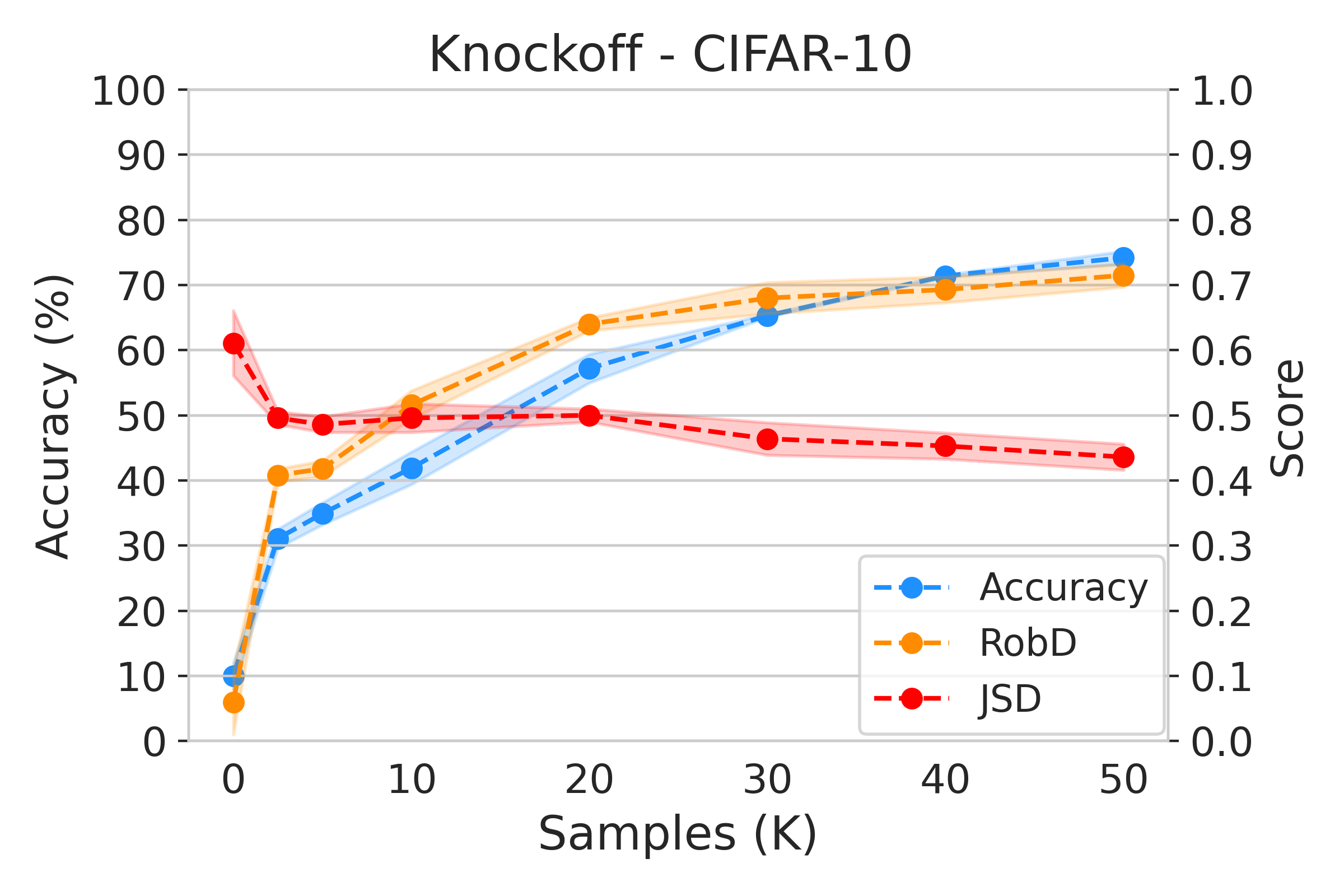}
   \includegraphics[width=0.49\linewidth]{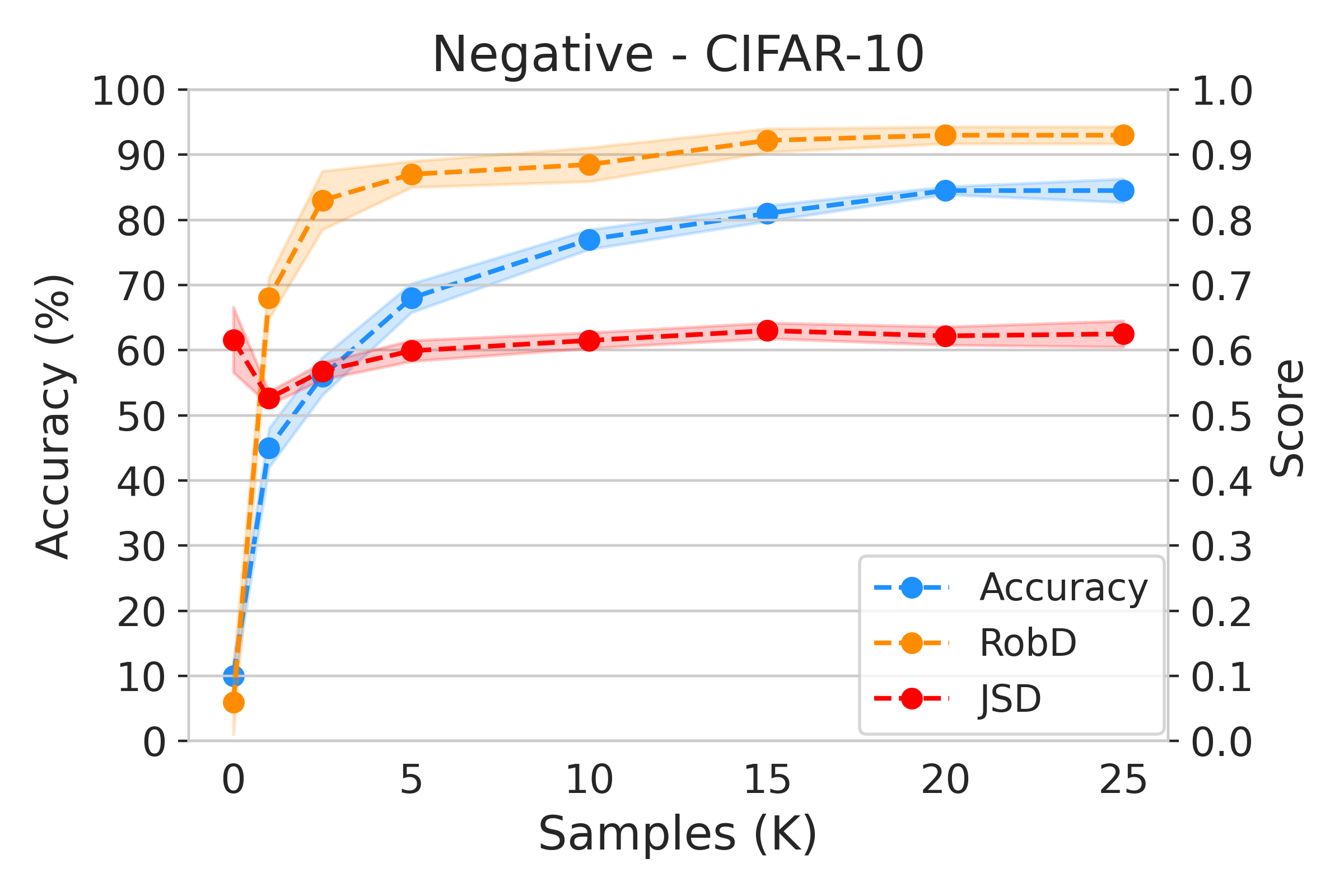}
   \includegraphics[width=0.49\linewidth]{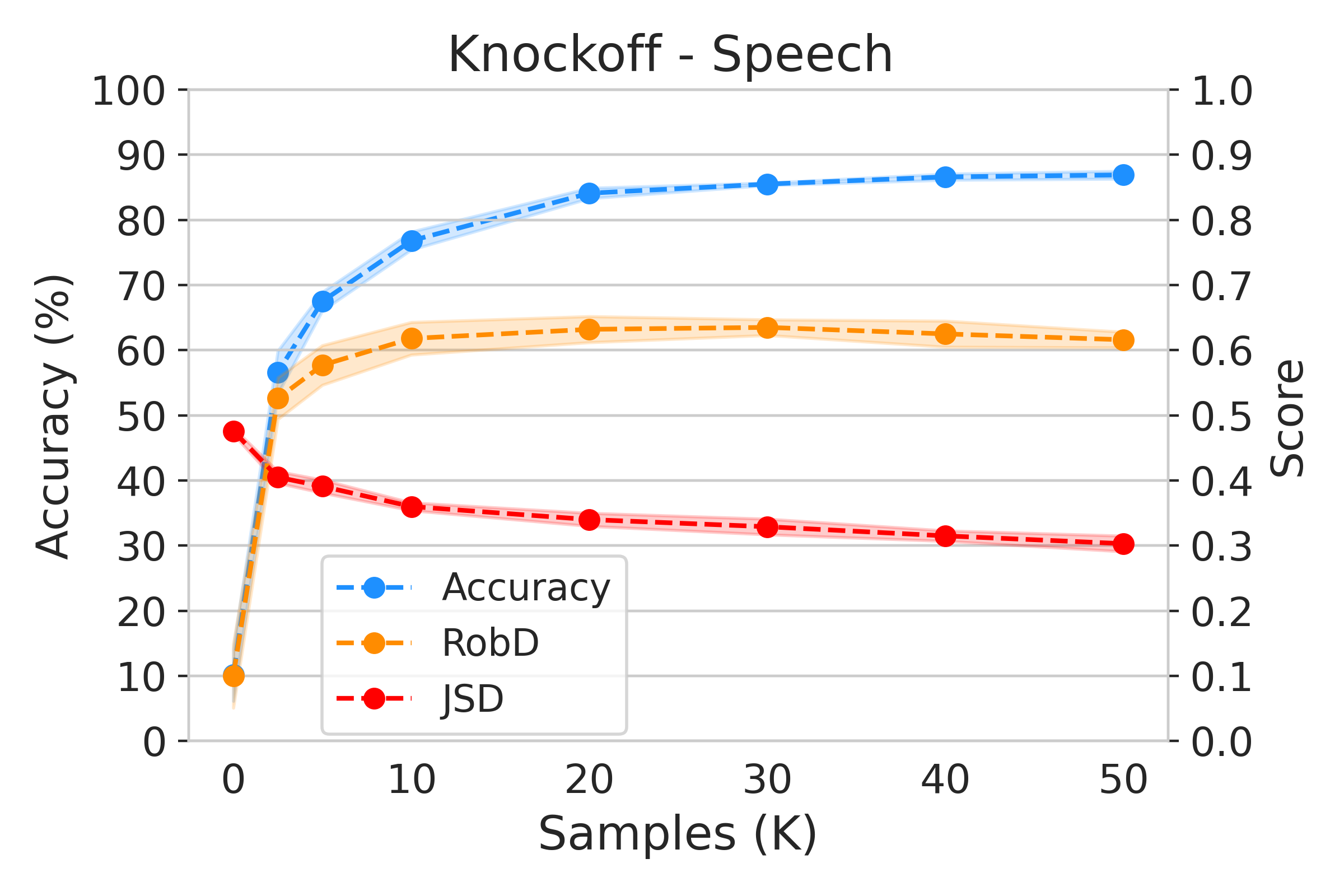}
   \includegraphics[width=0.49\linewidth]{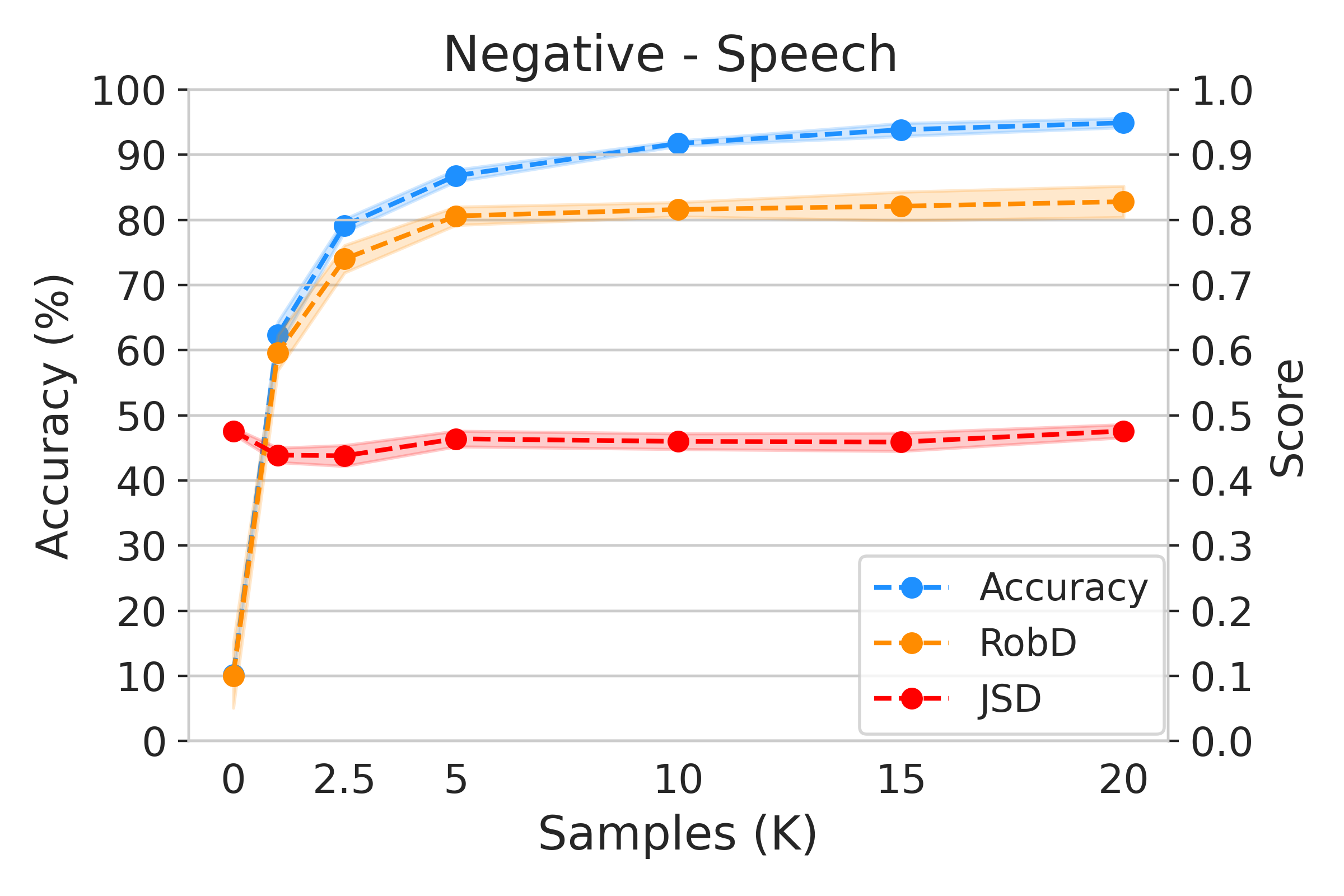}
   \caption{The \emph{RobD/JSD} scores between the CIFAR-10 (first row) and SpeechCommands (second row) victim models and their extracted copies by model extraction attacks.}
   \label{fig:extraction_add}
\end{figure}

\begin{figure}[]
    \centering
    \includegraphics[width=0.8\linewidth]{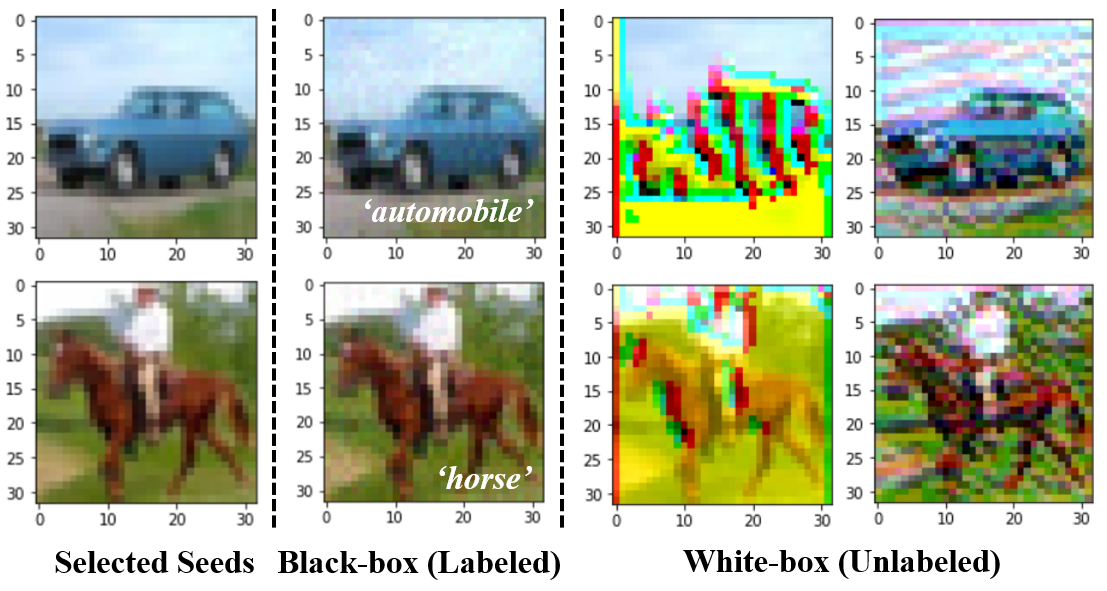}
    \caption{Example test cases generated in black-box and white-box testings. Note that the white-box test cases are not regular images and are \textbf{unlabeled}. }
    \label{fig:sample}
\end{figure}

\clearpage
\newpage

\end{document}